\documentclass{report}

\newcommand{\bmat}{\lt ( \begin{array} }
\newcommand{\emat}{  \end{array} \rt )} 
\newcommand{\oH}{{\ov H}}

\newcommand{\oq}{{\ov \q}}

\newcommand{\oG}{{\ov \G}}

\newcommand{\ED}{\end{document}}
\newcommand{\of}{{\ov f}}

\newcommand{\oY}{{\ov Y}}
\newcommand{\og}{{\ov g}}
\newcommand{\oy}{{\ov \y}}

\newcommand{\ove}{{\ov e}}
\newcommand{\oh}{{\ov h}}

\newcommand{\ok}{{\ov k}}
\newcommand{\oC}{{\ov C}}

\newcommand{\oF}{{\ov F}}
\newcommand{\A}{{\ov A}}

\renewcommand{\a}{\alpha}	
\renewcommand{\b}{\beta}
\newcommand{\g}{\gamma}
\renewcommand{\d}{\delta}
\newcommand{\e}{\epsilon}

\newcommand{\h}{\eta}
\newcommand{\q}{\theta}

\renewcommand{\l}{\lambda}
\newcommand{\m}{\mu}
\newcommand{\n}{\nu}	
\newcommand{\x}{\xi}

\newcommand{\f}{\phi}

\newcommand{\y}{\psi}
\newcommand{\w}{\omega}
\newcommand{\G}{\Gamma}
\newcommand{\D}{\Delta}
	
\renewcommand{\L}{\Lambda}
\newcommand{\X}{\Xi}
\renewcommand{\P}{\Pi}	
\renewcommand{\S}{\Sigma}

\newcommand{\F}{\Phi}

\newcommand{\W}{\Omega}
\newcommand{\la}{\label}
\newcommand{\ci}{\cite}

\newcommand{\ds}{\documentstyle}	
\newcommand{\fr}{\frac}

\newcommand{\na}{\nabla}
\newcommand{\pa}{\partial}
\newcommand{\ov}{\overline}
\newcommand{\be}{\begin{equation}}
\newcommand{\ee}{\end{equation}}
\newcommand{\ba}{\begin{array}} 
\newcommand{\ea}{\end{array}}
\newcommand{\bea}{\begin{eqnarray}}
\newcommand{\eea}{\end{eqnarray}}
\newcommand{\ra}{\rightarrow}

\newcommand{\lra}{\longrightarrow}
\newcommand{\lla}{\longleftarrow}
\newcommand{\Lra}{\Leftrightarrow}
\newcommand{\lt}{\left}
\newcommand{\rt}{\right}

\newcommand{\ben}{\begin{enumerate}}
\newcommand{\een}{\end{enumerate}}
\newcommand{\bitem}{\begin{itemize}}
\newcommand{\eitem}{\end{itemize}}

\begin{document}

\Large

\title{Composite Operators, Supersymmetry Anomalies and Supersymmetry Breaking in the Wess-Zumino Model  }

\author{ John  Dixon\footnote{ jadixon@shaw.ca} \\ Dixon Law Firm\footnote{Fax: (403) 266-1487 } \\1020 Canadian Centre\\
833 - 4th Ave. S. W. \\ Calgary, Alberta \\ Canada T2P 3T5 }
\maketitle

\abstract{   
A possible mechanism for supersymmetry breaking is that it is broken by a host of supersymmetry anomalies that occur in composite operators.   These anomalies are well hidden.  This paper explains where they may be found.

Using a source $\f_{\a}$ to absorb the free spinor index, it is shown that the simplest, potentially anomalous, composite spinor operator has the form: 
\[
{\cal A}^{(0)}_{\f} = 
\int d^4 x \; \f^{\a} f_i^j
\]
\[ \lt \{
\oG^{i} \A_j
C_{\a} 
+
\lt (
\pa_{\a \dot \b} A^i 
+ C_{\a}\oY^{i}_{ \dot \b} \rt )
 \oy_{j}^{\dot \b}
-
\y^i_{\a} 
\oF_{1,j}\rt \} 
\]
and the corresponding supersymmetry anomaly has the form:
\[
{\cal A}^{(1)}_{\f} = 
\int d^4 x \; \f^{\a} C_{\a}
 \lt \{ 
   \fr{1}{2} m {\oh}^{ij}   \oF_{2,ij}
+  \fr{1}{3} {\oh}^{ijk}  \oF_{3,ijk}
\rt \}
\]
where 
\bitem
\item
$m$ is the mass parameter,
\item
 $A^j$ is the scalar field in the chiral multiplet,
\item
 $\y^j_{\a}$ is the spinor field in the chiral multiplet,
\item
$\oG^{i}$ is the Zinn-Justin source for the variation of the scalar field $\d \A_i$,
\item
$\oY^{i \dot \b}$ is the Zinn-Justin source for the variation of the spinor field $\d \oy_{i \dot \b}$,
\eitem
 and the following are composite $\oF$-type invariants (the symmetrizations have `unit weight'):
\[
\oF_{1,j}= - \lt ( 
m {g}_{jl} {A}^l  +
{g}_{jlq} A^l A^q
+
{Y}_j^{  \b} {c}_{\b } 
\rt )
\]
\[
\oF_{2,ij}= 2 \lt ( \A_{(i} \oF_{1,j)} - \oy_{(i} \cdot \oy_{j)}
\rt )
\]
\[
\oF_{3,ijk}= 3 \lt ( \A_{(i} \oF_{2,jk)} - \A_{(i} \oy_{j} \cdot \oy_{k)} \rt )
\]

In accord with the analysis of Becchi, Rouet and Stora, as modified by Zinn-Justin (BRS-ZJ), anomalies ${\cal A}^{(1)}_{\f}$ are characterized by the following anomalous form of the BRS-ZJ identity for the 1PI generating functional ${\cal G}_{1PI, \f}$:
\[
\d {\cal G}_{1PI, \f} = {\cal A}^{(1)}_{\f}
\]
where ${\cal G}_{1PI, \f}$ is calculated with the usual Wess-Zumino action with an inserted spinor composite operator ${\cal A}^{(0)}_{\f}$.

It can be shown that the above expressions for ${\cal A}^{(0)}_{\f}$ and ${\cal A}^{(1)}_{\f}$ are both in  the local cohomology space of the nilpotent BRS-ZJ operator $\d$.  This  is an indication that their matrix elements are physically `on-shell' and have measurable physical effects. 

The simplest interesting case has the following dimensionless parameters:
\bitem
\item 
Parameters in the Action:
\bitem
\item  $\og^{ij}$ for the  mass  term $m \og^{ij} \oF_{ij}$, 
\item $\og^{ijk}$ for  the Yukawa term $ \og^{ijk} \oF_{ijk}$;
\eitem
\item 
Parameters in the Ghost Charge Zero Spinor Invariant ${\cal A}^{(0)}_{\f}$:
\bitem
\item
$f^i_j$    for the spinor cohomology invariant which can develop a supersymmetry anomaly; 
\eitem 
\item
Parameters in the Ghost Charge One Spinor  Invariant ${\cal A}^{(1)}_{\f}$:
\bitem
\item
 $ h_{ij},h_{ijk}$
for the spinor cohomology invariant which characterises  the
 potential supersymmetry anomaly.
\eitem 

\eitem

These parameters are constrained by the cohomology.  To describe the constraints,   the above dimensionless parameters $f,\og,h$  are assembled into a trilinear Generating Function ${\cal T}$:
\[
{\cal T}
=
f_j^i 
\lt ( 2 h_{ik} \og^{jk}  + 6 h_{ikl} \og^{jkl} \rt )
\]
The constraints are that the variation of ${\cal T}$ with respect to each of  the independent parameters in   $f,\og,h$ is zero: 
\[ 
\fr{\pa {\cal T}} {\pa f_i^j }   =     \fr{\pa {\cal T}} {\pa \og^{ij} }  =\fr{\pa {\cal T}} {\pa \og^{ijk} }  =  \fr{\pa {\cal T}} {\pa h_{ij} } =\fr{\pa {\cal T}} {\pa h_{ijk} }  =0
\]

These constraints are derived by a cohomology analysis, together with three assumptions which seem to be consistent with perturbation theory:
\bitem
 \item
No renormalization of the mass and Yukawa coefficients is needed or allowed, in accord with the supersymmetry non-renormalization theorem.
\item
There are no `zero momentum terms' (i.e. terms with only one field or source) in the action.
\item
There are no `zero momentum terms' in the variation $\d {\cal G}_{1PI} $.
\eitem
  If the non-renormalization requirement is relaxed, the constraints are more complicated and certain anomaly possibilities are removed. But, in compensation, allowing renormalization of the mass and Yukawa coefficients also violates supersymmetry. 

The  solutions of these constraints depend on the number of chiral fields $n_{\rm Chiral}$ that the theory contains.   It looks possible that some solutions for $n_{\rm Chiral} \geq 3$ may provide examples of one-loop supersymmetry anomalies.  These anomalies will necessarily give rise to a pattern of supersymmetry breaking that is related to the solution of the constraints.  

The next logical steps after this paper are:
\bitem
\item
to analyze the various solutions of  the constraints and their generalizations, to see if there are any `interesting' solutions; 
\item
to perform one-loop perturbation theory calculations of the potential supersymmetry anomalies that relate to the `interesting' solutions of the constraints; and
\item
to further explore the cohomology of these inserted composite operators; including: 
\bitem
\item
non-zero inserted momentum, which means that
the source  $ \f_{\a}$
becomes a space time dependent antichiral spinor superfield:
 $ \f_{\a} \ra {\hat \f}_{\a}(x)$;
\item
gauge and group coupling; 
\item
spontaneous gauge symmetry breaking; and
\item
kinetic mixing between ${\hat \f}_{\a}$ and the chiral multiplet, analogous to the Higgs-Kibble mixing that gives mass to vector bosons.
\eitem

\eitem

This mechanism for supersymmetry breaking, assuming that interesting solutions exist, and that the anomalies are present, appears to have a good chance of yielding supersymmetry breaking with a zero cosmological constant.  Asuming that supersymmetry itself is not spontaneously broken, which is normally the case, the cosmological constant should remain zero after supersymmetry breaking, even when gauge symmetry is spontaneously broken.}

\chapter{Introduction:}

\la{introchapter}

\section{Some History}

\subsection{The Problem of Supersymmetry Breaking}

Supersymmetry has theoretical appeal for lots of reasons, but we still don't really know whether it has any experimental justification at all.  This was put succinctly as follows in the 2000 Superstring Conference at Ann Arbor, Michigan, as Problem 4, in a series of TEN PROBLEMS \ci{superstrings2000}:

  {\em Is Nature supersymmetric, and if so, how is supersymmetry broken?} 

A possible mechanism for supersymmetry breaking is that it is broken by a host of supersymmetry anomalies that occur in composite operators.    This possibility was discussed about fifteen years ago in  \ci{dixprl} and  \ci{kyoto}. 

The present paper summarizes the simplest aspects of some significant progress relating to that possibility.

This paper takes a new look at the cohomology of the Wess Zumino model, including the Zinn-Justin sources and including composite operators that have Lorentz spinor indices coupled to external sources. 

 A good survey of various approaches in the literature relating to the calculation of cohomology can be found in \ci{barnich}, and a recent calculation of some aspects of supersymmetry cohomology, together with  references to the supersymmetric cohomology literature, can be found in \ci{brandt}.  The spectral sequence method used in this paper was expounded in \ci{specseq} and then used in \ci{holes}  \ci{chirsup} \ci{dixmin} \ci{dixminram} \ci{kyoto}.

\subsection{Chiral Supersymmetry Transformations}

In Table \ref{Purechiral1}, we summarize the Field Transformations  for Pure Chiral Supersymmetry.  These transformations close in the sense that
\be
\d^2 = 0.
\ee

 \begin{table}
 \caption{\Large Chiral Transformations}
\la{Purechiral1}
\vspace{.1in} \framebox{
{\Large $\begin{array}{lll}  
\d A^i&= & 
  \y^{i}_{  \b} {C}^{  \b} 
+ \x^{\g \dot \d} \partial_{\g \dot \d} A^i
\\
\d {\ov A}_i&= & 
 {\ov \y}_{i  \dot \b} {\ov C}^{ \dot  \b} 
+ \x^{\g \dot \d} \partial_{\g \dot \d} {\ov A}_i
\\
 \d \y_{\a}^i &  =& 
\pa_{ \a \dot \b }  A^{i} {\ov C}^{\dot \b}  
+ 
C_{\a}   
F^i
+ \x^{\g \dot \d} \partial_{\g \dot \d}  \y^{i}_{\a  }
\\
\d
 {\ov \y}_{i \dot \a} &  =& 
\pa_{ \a \dot \a }  {\ov A}_{i} {C}^{\a}  
+ 
{\ov C}_{\dot \a}   
{\ov F}_{i}
+ \x^{\g \dot \d} \partial_{\g \dot \d} 
 {\ov \y}_{i \dot \a} 
\\ 
\d F^i &  =&  \pa_{ \a \dot \b }  \y^{i \a}  {\ov C}^{\dot \b}  
 + \x^{\g \dot \d} \partial_{\g \dot \d}  F^i
\\ 
\d \oF_i &  =&  \pa_{ \a \dot \b }  \oy^{\dot \b}_i  C^{\a}
 + \x^{\g \dot \d} \partial_{\g \dot \d}  \oF_i
\\
\d \x_{\a \dot \b} &  =& - C_{\a} \oC_{\dot \b}
\\
\end{array}$} }
\end{table}

\subsection{Results from Fifteen Years Ago}

It was  noted and discussed  in  \ci{holes} \ci{chirsup} \ci{dixprl} \ci{dixmin} \ci{dixminram} \ci{kyoto} that local integrated polynomials of the form
\be
{\cal A}^{(1)}_{\f}=
\int d^4 x \; \f^{\a} \oh^i \oF_i C_{\a}
\ee
or the complex conjugates
\be
{\ov {\cal A}}^{(1)}_{{\ov \f}}=
\int d^4 x \; {\ov \f}^{\dot \a}  h_i F^i \oC_{\dot \a} 
\ee
generate cohomology classes of the operator defined by Table \ref{Purechiral1}. 

Here we have inserted a spinor $\f_{\a}$ and a vector $\oh^i$ to absorb the spinor and chiral field index labels. $\f_{\a}$ and    $\oh^i$   are initially assumed to be invariant under supersymmetry and constant over spacetime.

This means that: 
\bitem
\item
$  {\cal A}^{(1)}_{  \f}$ is a cocycle of  $\d$ (i.e. it is in the kernel of $\d$), and 
\item
$  {\cal A}^{(1)}_{{  \f}}$ is not a coboundary  of  $\d$ (i.e.   it is not of the form ${\cal A}^{(1)}_{\f}= \d {\cal P}^{(0)}_{\f}$ for any local integrated ${\cal P}^{(0)}_{\f}$). 
\eitem
We can see that $ {\cal A}^{(1)}_{{  \f}}$ is a cocycle easily :
\be
\d {\cal A}^{(1)}_{\f}=
\d \int d^4 x \; \f^{\a} \oh^i \oF_i C_{\a}
\ee
\be
= - \int d^4 x \; \f^{\a} \oh^i \lt (
\pa_{ \g \dot \b } C^{\g} \oy^{\dot \b}_i  
 + \x^{\g \dot \d} \partial_{\g \dot \d}  \oF_i
\rt ) C_{\a}= 0
\ee
and also we can check that there are no local polynomials ${\cal P}^{(0)}_{\f}$ such that 
${\cal A}^{(1)}_{\f}=
\d \int d^4 x \; \f^{\a} 
   {\cal P}^{(0)}_{\f}$.  This is easy to see for this lowest example. The 
only possibility is: 
\[
\d \int d^4 x\;
\f^{\a} q_i \y^i_{\a}
=
 - \int d^4 x\;
\f^{\a}q_i F^i C_{\a}
\]
\be
\neq
 \int d^4 x\;
\f^{\a} \oh^i \oF_i C_{\a} = {\cal A}^{(1)}_{\f}\ee
and it does not work,  because $\oF$ is not equal to its complex conjugate $F$.

The higher analogues have the same properties.
The next one is:
\be
{\cal A}^{'(1)}_{\f}= \int d^4 x \; 
\f^{\a} \oh^{ij}  \lt (
\A_{i} \oF_{j} +\A_{j} \oF_{i} -   \oy^{\dot \b}_i 
\oy_{j \dot \b} 
\rt )
C_{\a}
\ee
So we say that all of these ${\cal A}^{(1)}_{\f}$   are in the cohomology space of the nilpotent operator $\d$. Note that all of these ${\cal A}^{(1)}_{\f}$ have ghost charge one.

\subsection{Unanswered Questions that lead to full BRS-ZJ Cohomology}

Generally we expect that when there are cocycles in the cohomology space like these, then there should be a one-loop anomaly in perturbation theory relating to them \ci{brs}. However it has not been easy to see what the anomalous operator would be.

The following questions have remained unanswered:
\bitem
\item
Are there any operators that have this kind of anomaly?  
\item
If there are any,  what do these operators look like?
\eitem

To determine the answer to this question we need to find the cohomology of the theory including Zinn-Justin's sources. This, plus a perturbation theory calculation, could assure us that we really have an anomaly, whereas the above results are only a suggestion of that. 

 This has proved to be a difficult problem for several reasons:
\bitem
\item

 In the first place, it has appeared that supersymmetry anomalies are unlikely because superspace perturbation theory, if it is valid, seems to guarantee that one will generate only superfields in any calculation, and that 
would mean that supersymmetry anomalies must be hard to generate. 
However this is somewhat unclear, because non-locality in superspace could generate anomalies. 
\item
In the second place, the full cohomology of the theory, including the Zinn-Justin sources, is needed to even know where to look for anomalies.  This is a very tricky problem by itself.
\eitem

The claim made here is that a correct formulation of the BRS-ZJ identity can only be done by eliminating the auxiliary field so as to write down a `Physical Formulation' of the BRS-ZJ identity.  It looks difficult, or perhaps impossible, to do this in superspace.  In fact it appears below that a correct formulation  of the BRS-ZJ identity can only be done by giving up manifest supersymmetry. 

The cohomology in general is complicated. It generates interesting and complicated `constraint equations' which are discussed below.  In the present paper we formulate the cohomology for the simplest cases, sufficiently to identify some relatively simple operators that can have the anomalies.  The general cohomology is not needed yet. 

There are some puzzling issues relating to the combination of the non-renormalization theorems of supersymmetry and the cohomology. There are also some important and tricky issues relating to `zero momentum' terms.  These are discussed below.

The next logical step after this paper is to solve the constraint equations for some more challenging cases, and then to actually calculate these anomalies in perturbation theory. The details of this will be discussed below.

\section{Summary of this Paper}

Chapter \ref{introchapter} reviews some of the history of this topic and then presents a derivation of the BRS identity in Zinn-Justin's form.  This is trickier than it looks at first sight.  The formulation of the BRS-ZJ identity is done in a form that is not manifestly supersymmetric, to avoid certain problems resulting from the existence of the auxiliary field. Those problems are discussed in Subsection \ref{auxilproblem}.

The result, (in Section \ref{physzj}), will be called the Physical Formulation of the BRS-ZJ identity.   Tables \ref{physicaltable}  and \ref{compterms} record the resulting non-linear nilpotent operator $\d$ that follows from the Physical Formulation of the BRS-ZJ identity.

Chapter \ref{cohomchapter} introduces some results using the spectral sequence method. The issue of `zero momentum' terms is discussed.  These results are used throughout the paper. A more detailed paper on the spectral sequence, including supersymmetric gauge theory, is in the course of preparation.

Chapter \ref{diagramchasechapter} uses the results of Chapter \ref{cohomchapter} to analyze some simple cases by the method of `Diagram Chasing'.

Chapter \ref{superfieldschapter} puts the results of 
Chapter  \ref{diagramchasechapter} into a relatively simple form by enumerating the supermultiplets that are discoverable from the results of Chapter \ref{diagramchasechapter}. 

Chapter \ref{insertionchapter} then discusses the effect of placing the new results of Chapter \ref{superfieldschapter} into the original action as inserted composite operators. 
New cohomology problems arise.  A revised spectral sequence is put into place.

Chapter \ref{diagramchasewithinsertchapter} analyzes the new cohomology problems that arise from the quadratic nature of the BRS-ZJ identity when the operators referred to in Chapter \ref{insertionchapter} are inserted.

Chapter \ref{notsimpleconstraintchapter}
 discusses some aspects of the solution of the simplest constraint equations which arise from the considerations above.

Chapter \ref{remarkschapter}
summarizes the results, adds some more general cohomology results, and also discusses the issue of a symmetry breaking term and vacuum expectation values of the scalar field.

Chapter \ref{conclusionchapter} includes an example of the kind of Feynman diagrams that  need to be calculated to determine whether the supersymmetry anomalies have non-zero coefficients or not. Section \ref{speculationsection} contains speculations on the general significance of  the computation of these kind of anomalies and the possibilities of supersymmetry breaking by this mechanism. Section \ref{speculationsection}   includes some remarks about the Standard Supersymmetric Model and other problems that would be affected by this mechanism.
Section \ref{envoisection} summarizes the various steps needed to calculate the simplest possible supersymmetry anomaly in a composite operator in pure chiral supersymmetry.

A Table of Contents and List of Tables are included at the end for ease of reference.  

\section{Conventions, Signs, Factors and Errors}

This paper uses the same conventions as \ci{dixmin}. In a paper of this size, it is not very easy to get all the signs and factors correct everywhere.  Also, after a certain point, finding the last such errors is not a good return on time and effort, except when the signs and factors actually matter.  The spectral sequence is very forgiving in this regard, which encourages a certain laxity. For example, the spectral sequence (see below) tells us that expressions such as 
\be
( C \x^2 C) f^{ij}_k \A_i \A_j \y^k \in  E_{\infty}
\ee
give rise to cohomology.  But the verification of the actual signs and factors of the detailed cohomology expression itself requires a lot of work. The same thing happens in Superspace.  This work is only really justified when one is calculating things that depend on the signs and factors.  Then of course the signs and factors really needs to be right. 

The result is that if the reader thinks that he or she detects errors and inconsistencies in signs and factors, the reader is probably right.   Realistically speaking, there must be some of them still in the paper.  I intend to try to collect all the errors together and publish a final corrected version of this paper in due course.

\section{Action and BRS Identity for massive interacting chiral supersymmetry including Zinn-Justin's sources}

\la{brssection}

The most obvious way to formulate the BRS-ZJ identity is to start with the following action 
\ci{WZ} \ci{ZJ}.

\be
{\cal A}_{\rm Guess} 
=
{\cal A}_{\rm WZ} 
+
{\cal A}_{\rm ZJ} 
+
{\cal A}_{\rm Sources} 
\ee
where the Wess-Zumino action is \ci{WZ}:
\[
{\cal A}_{\rm WZ} 
= \int d^4 x \;\lt \{
F^i \oF_i   -
\y^{i }_{\a  }     \pa^{\a \dot \b  }   
{\ov \y}_{i \dot \b}
+
\fr{1}{2} 
\pa_{ \a \dot \b  }   A^{i}    \pa^{\a \dot \b  } {\ov  A}_{k} 
\rt.
\]
\[
-   m^2 g_{i} F^i
 -
   m g_{ij} \lt (
F^{i} A^j  
- \fr{1}{2} \y^{i \a} \y^j_{\a} 
\rt ) 
- *
\]
\be
\lt.
-
g_{ijk}\lt (
   F^{i}   A^j A^k    -    \y^{i \a} \y^j_{\a} A^k   
\rt ) - * 
\rt \}
\ee
For generality in this first look, we have included the possibility that the term $   m^2 g_{i} F^i $  is present.  If this term cannot be eliminated by a shift $A^i \ra A^i + m v^i$ where the Vacum Expectation Value is $<A^i > = m v^i$, then  supersymmetry is spontaneously broken.  From now on we shall usually set $g_i = 0$, for reasons that will be explored in Section 
\ref{isolampsection}. Some more discussion of this issue can be found in Section \ref{vevsection}.

  The Zinn Justin action is \ci{ZJ}:
\[
{\cal A}_{\rm ZJ} 
= \int d^4  x\;
\lt \{
\G_i    \y^{i}_{  \b} { C}^{  \b} 
+
 Y_{i}^{ \a}
\lt (
   \pa_{ \a \dot \b }  A^{i} {\ov C}^{\dot \b}   
+
   F^{i} C_{\a}
\rt )
+*
\rt.
\]
\[
+
 \L_{i}
\lt (
   \pa_{ \a \dot \b }  \y^{i \a} {\ov C}^{\dot \b}   
\rt )
+ *
\]
\[
+
\x^{\g \dot \d} \lt ( 
 Y_{i}^{ \a}  \pa_{\g \dot \d}  \y^{i}_{ \a}
-
 \G_{i}   \pa_{\g \dot \d}  A^{i} 
-
 \L_{i}   \pa_{\g \dot \d}  F^{i} 
\rt ) +*
\]
\be
-
X^{\g \dot \d} C_{\g} {\ov C}_{\dot \d} 
\ee
and the Source term is:
\be
{\cal A}_{\rm Sources} 
=
\int d^4 x
\left \{
{\breve A}_i A^i
+
{\breve F}_i F^i
+ {\breve \y^{\a}}_i
\y_{\a}^i
+ *
 \right \}
\ee

There is a problem with this `Guess',  which we shall now discuss.

\subsection{First Guess at Generating Functional}

Standard arguments tell us that we should define the generating functionals \ci{ZJ}:

\[
{\cal G}_{\rm Disconnected;\; Guess} =
e^{i{\cal G}_{\rm Connected; \; Guess}} =
\]
\be
\la{disconn1}
\P_{x,\a,{\dot \a},i} 
\int \;
\d A^i
\d {\ov A}_i
\d \y_{\a}^i
\d {\ov \y}_{i \dot \a}
\d F^i
\d {\ov F}_i
e^{i {\cal A}_{\rm Guess}}
\ee

Now we note that if we perform any infinitesimal transformation on the integration variables, the integral is invariant if we include the appropriate Jacobian.

If we perform the following special infinitesimal transformation on the integration variables:
\be
X \ra X + \e \d X
\ee
where $X = A^i,\y^i_{
a},F^i $, $\e$ is an anticommuting quantity and $\d$ is the variation defined above in Table \ref{Purechiral1}, we note that the variation of the Action is zero  except for the Source parts.  
In this case, the Jacobian of the transformation is unity.

Then this yields the identity:
\[
\int d^4 x
\left \{
{\breve A}_i \fr{\d {\cal G}_{\rm Connected; \; Guess}}{\d \G_i} 
+
{\breve F}_i \fr{\d {\cal G}_{\rm Connected; \; Guess}}{\d \L_i} 
\rt.
\]
\[
\lt.
+ {\breve \y^{\a}}_i
\fr{\d {\cal G}_{\rm Connected; \; Guess}}{\d  Y^{\a}_i}
+
{\breve {\ov A}}^i
 \fr{\d {\cal G}_{\rm Connected; \; Guess}}{\d {\ov \G}^i }
\rt.
\]
\[
\lt.
+
{\breve {\ov F}}^i
 \fr{\d {\cal G}_{\rm Connected; \; Guess}}{\d {\ov \L}^i }
 + {\breve {\ov \y}}^{i \dot \a}
 \fr{\d {\cal G}_{\rm Connected; \; Guess}}{\d  {\ov Y}^{i \dot \a}}
\right \}
\]
\be
+ C^{\a} {\ov C}^{ \dot \b}   \fr{\pa {\cal G}_{\rm Connected; \; Guess}}{\pa  \x^{\a \dot \b} }
 = 0
\ee

\subsection{A Problem with the auxiliary and the formulation of the BRS Identity}

\la{auxilproblem}

Zinn Justin's formulation of the BRS identity, (for example see Subsection \ref{physzj} below), requires  a Legendre transform of each field source. 

A combinatorial argument \ci{ZJ} shows that, when the relevant source gives rise to a Propagator leg attached to the rest of the diagram, a Legendre transform takes the connected Green's functional ${\cal G}_{\rm Connected}$ 
to the 1PI functional ${\cal G}_{\rm 1PI}$. 

The relation between them should be of the form:

\[
{\cal G}_{\rm Connected; \; Guess} \stackrel{??}{=}
{\cal G}_{\rm 1PI; \; Guess} +
\]
\be
\int d^4 x
\left \{
{\breve A}_i A^i
+
{\breve F}_i F^i
+ {\breve \y^{\a}}_i
\y_{\a}^i
+ *
 \right \}
\ee

This makes sense for the physical fields $A^i$ and $\y^i_{\a}$, but there seems to be a problem for the auxiliary field $F^i$.

\begin{picture}(400,120)
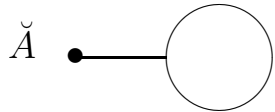
\put(40,30){\Large \parbox{4 in}{Figure 1:    ${\breve A}_i A^i$  generates a $\fr{1}{k^2 + m^2 }$  propagator leg.  The circle represents the rest of a diagram.} }
%
%
%
\put(144,75){\line(1,0){36}}
\put(142,72){$\bullet$}
\put(120,75){{${\breve A}$}}
%
%
%
%
\put(200,75){\circle{40}}
%
%
%
\end{picture}

\begin{picture}(400,120)
\put(40,30){\Large \parbox{4 in}{Figure 2:    ${\breve \y}_{i \a} \y^{i \a}$  generates a $\fr{k_{\a \dot \b}}{k^2 + m^2 }$  propagator leg} }
%
%
%
\put(144,75){\line(1,0){36}}
\put(142,72){$\bullet$}
\put(120,75){{${\breve \y}_{i \a}$}}
%
%
%
%
\put(200,75){\circle{40}}
%
%
%
\end{picture}

\begin{picture}(400,120)
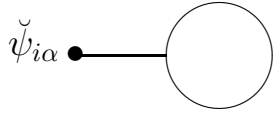
\put(40,30){\Large \parbox{4 in}{Figure 3:    ${\breve \y}_{i \a} \y^{i \a}$  also generates a $\fr{m \e_{\a   \b}}{k^2 + m^2 }$  propagator leg} }
%
%
%
\put(144,75){\line(1,0){36}}
\put(142,72){$\bullet$}
\put(120,75){{${\breve \y}_{i \a}$}}
%
%
%
%
\put(200,75){\circle{40}}
%
%
%
\end{picture}

 It is natural to integrate the auxiliary field at some point.  This can be done exactly by completing the square,  and we will assume that this has been done.  Because the field equation of the auxiliary contains no derivatives, this integration gives rise to a local term of the form:
\[
\int d^4 x \; \lt \{
- \lt ( 
{\breve {\ov F}}^i  +
m {\ov g}^{ij} {\ov A}_j  +
 {\ov g}^{ijk} {\ov A}_j  {\ov A}_k  +
{\ov Y}^{i \dot \b} {\ov C}_{\dot \b } 
\rt )
\rt.
\]
\be
\lt.
\lt ( {\breve {F}}_i   +
 m {g}_{il} {A}^l  +
{g}_{ilq} A^l A^q
+
{Y}_i^{  \b} {c}_{\b } 
\rt )
\rt \}
\ee

The ${\breve F}_i$ part of this expression  has two parts that give rise to Feynman diagrams.  First there is a part that generates a propagator leg:

\begin{picture}(400,120)
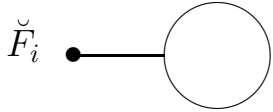
\put(40,30){\Large \parbox{4 in}{Figure 4:    ${\breve F}_i  m   {\ov g}^{ik} 
{\ov A}_k   $ 
  generates a $\fr{m }{k^2 + m^2 }$  propagator leg }}
%
%
%
\put(144,75){\line(1,0){36}}
\put(142,72){$\bullet$}
\put(120,75){{${\breve F}_{i}$}}
%
%
%
%
\put(200,75){\circle{40}}
%
%
%
\end{picture}

 Then there is a  composite operator part, which generates no propagator leg:

\begin{picture}(400,120)
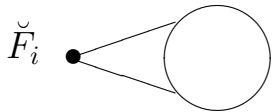
\put(40,20){\Large \parbox{4 in}{Figure 5:      ${\breve F}_i   {\ov g}^{ikl} 
{\ov A}_k {\ov A}_l  $ does not generate  a $\fr{1}{k^2 + m^2}$  propagator leg.  It generates part of a closed loop or else two legs.}}

%
%
%
\put(142,72){$\bullet$}
\put(120,75){{${\breve F}_i$}}
\put(144,75){\line( 3, -1){40}}
\put(144,75){\line( 3,1){40}}
%
%
%
%
\put(200,75){\circle{40}}
%
%
%
\end{picture} 

So it appears that the Legendre transform may have difficulties for the auxiliary field, if one integrates the auxiliary field as one needs to do at some point. 

This seems to spell trouble for superspace perturbation theory too, since it uses the auxiliary field in a way that is analogous to the above.  However this is not completely clear, since in superspace we do not integrate the auxiliary field separately \ci{superspace}.

Does superspace Feynman perturbation theory suffer from a problem with the BRS identity? The sources in Superspace perturbation theory \ci{superspace} 
give rise to superpropagator legs.  Are these  different from propagator legs for the above reason?

The results of this paper, set out below, seem to indicate that this is a real and important issue. But rather than try to understand this question in some complicated way, the problem will be avoided by an alternative formulation of the BRS-ZJ identity. Are the two formulations equivalent? 

  If they are not equivalent, then the formulation below is more likely to be correct, because it certainly does not have the problem mentioned above.  That would mean that superspace perturbation theory, or at least the naive BRS identity in superspace perturbation theory, is incorrect in some subtle way.  

This `guess' method and the equivalent `Hamiltonian plus Antifield' approach both suffer from this `auxiliary--Legendre' problem.  This somewhat dubious approach has been assumed to be correct for a long time by the present author\ci{kyoto} as well as many other people. Probably the best way to decide whether there is really a problem is to find out if there are anomalies and then return to see if the above problems can be fixed somehow. 

As will be seen below, the anomaly possibilities are very closely connected with this very issue.

\section{`Physical' Formulation of BRS-ZJ Identity}

\la{physzj} 

The auxiliary field issues above will be avoided by defining an alternative set of Green's functions: 
\be
{\cal G}_{\rm Disconnected.} =
e^{i {\cal G}_{\rm Connected.} } =
\ee
\be
\la{disconn2}
\P_{x} 
\int \;
\d A^i
\d {\ov A}_i
\d \y_{\a}^i
\d {\ov \y}_{i \dot \a}
\d F^i
\d {\ov F}_i
e^{i 
{\cal A}_{ \rm  New } }
\ee
where we simply remove the $\L_i$ and the ${\breve F}_i$ terms:
\be
{\cal A}_{\rm New} 
= {\cal A}_{\rm WZ} 
+
{\cal A}_{\rm ZJ;\; Physical}
+
{\cal A}_{\rm Sources;\; Physical}
\ee
where
\[
{\cal A}_{\rm ZJ;\; Physical}
 =
 \int d^4  x\;
\lt \{
\G_i    \y^{i}_{  \b} { C}^{  \b} 
+
 Y_{i}^{ \a}
\lt (
   \pa_{ \a \dot \b }  A^{i} {\ov C}^{\dot \b}   
+
   F^{i} C_{\a}
\rt )
\rt.
\]
\be
\lt.
+
\x^{\g \dot \d} \lt ( 
 Y_{i}^{ \a}  \pa_{\g \dot \d}  \y^{i}_{ \a}
-
 \G_{i}   \pa_{\g \dot \d}  A^{i} 
\rt ) 
\rt \} 
-
X^{\g \dot \d} C_{\g} {\ov C}_{\dot \d} 
\ee
and
\be
 {\cal A}_{\rm Sources;\; Physical}
=
\int d^4 x
\left \{
{\breve A}_i A^i
+ {\breve \y^{\a}}_i
\y_{\a}^i
+ *
 \right \}
\ee

The same reasoning as before yields the new identity:
\[
\int d^4 x
\left \{
{\breve A}_i \fr{\d {\cal G}_{\rm Connected}}{\d \G_i} 
+ {\breve \y^{\a}}_i
\fr{\d {\cal G}_{\rm Connected}}{\d  Y^{\a}_i}
\rt.
\]
\[
\lt.
+
{\breve {\ov A}}^i
 \fr{\d {\cal G}_{\rm Connected}}{\d {\ov \G}^i }
 + {\breve {\ov \y}}^{i \dot \a}
 \fr{\d {\cal G}_{\rm Connected}}{\d  {\ov Y}^{i \dot \a}}
\right \}
\]
\be
\la{correctone} + C^{\a} {\ov C}^{ \dot \b}   \fr{\pa {\cal G}_{\rm Connected}}{\pa  \x^{\a \dot \b} }
 = 0
\ee

\subsection{Action for massive interacting chiral supersymmetry after integration of auxiliary $F^i$}

By performing the integration of $F^i$ and $\oF_i$, which can be done by completing the square since there is no kinetic term for the auxiliary fields, this can be written:

\be
{\cal G}_{\rm Disconnected.} =
e^{i {\cal G}_{\rm Connected.} } =
\ee
\be
\la{disconn3}
=
\P_{x} 
\int \;
\d A^i
\d {\ov A}_i
\d \y_{\a}^i
\d {\ov \y}_{i \dot \a}
e^{i \lt \{
{\cal A}_{\rm Physical }+ {\cal A}_{\rm Sources;\; Physical}
\rt \} }
\ee
where
\[
{\cal A}_{\rm Physical}
\]
\[
= \int d^4 x \;\lt \{
- \lt (    m {\ov g}^{ij} {\ov A}_j  +
 {\ov g}^{ijk} {\ov A}_j  {\ov A}_k  +
{\ov Y}^{i \dot \b} {\ov C}_{\dot \b } 
\rt )
\rt.
\]
\[
\lt (
 m {g}_{il} {A}^l  +
{g}_{ilq} A^l A^q
+
{Y}_i^{  \b} {c}_{\b } 
\rt )
  -
\y^{i }_{\a  }     \pa^{\a \dot \b  }   
{\ov \y}_{i \dot \b}
\]
\[
- \fr{1}{2} m g_{ij} \y^{i \a} \y^j_{\a}   - g_{ijk}  \y^{i \a} \y^j_{\a} A^k   
\]
\[
- \fr{1}{2} m {\ov g}^{ij} {\ov \y}_{i }^{\dot \a} {\ov \y}_{j  \dot \a} -
 {\ov g}^{ijk} {\ov \y}_{i }^{\dot \a} {\ov \y}_{j  \dot \a} {\ov A}_k 
\]
\[
+
\fr{1}{2} 
\pa_{ \a \dot \b  }   A^{i}    \pa^{\a \dot \b  } {\ov  A}_{k} 
+
\G_i    \y^{i}_{  \b} {  c}^{  \b} 
+
{\ov \G}^i  {\ov \y}_{i \dot  \b} {\ov C}^{\dot   \b}  
\]
\[
+
 Y_{i}^{ \a}   \pa_{ \a \dot \b }  A^{j} {\ov C}^{\dot \b}   
+
 \ov{Y}^{i \dot \b}  \pa_{ \a \dot \b }  {\ov A}_{j} 
{  c}^{  \a}  
\]
\[
\lt.
+
\x^{\g \dot \d} \lt ( 
 Y_{i}^{ \a}  \pa_{\g \dot \d}  \y^{i}_{ \a}
+
 {\ov Y}^{i \dot \b}  \pa_{\g \dot \d}{\ov \y}_{i \dot \b}  
-
 \G_{i}   \pa_{\g \dot \d}  A^{i} 
-
 {\ov \G}^{i  }  \pa_{\g \dot \d}{\ov A}_{i }  
\rt )
\rt \}
\]
\be
-
X^{\g \dot \d} C_{\g} {\ov C}_{\dot \d} 
\la{Aphysical}
\ee

\subsection{BRS-ZJ identity in the Physical Formulation }

\la{brsinphyspform}

A valid Legendre transform now takes the connected Green's functional to the 1PI functional.  For the Physical fields $\y$ and $A$, the relevant source always gives rise to a propagator leg attached to the rest of the diagram.

The Legendre transform is of the form:

\be
{\cal G}_{\rm Connected.} =
{\cal G}_{\rm 1PI} +
\int d^4 x
\left \{
{\breve A}_i A^i
+
 {\breve \y^{\a}}_i
\y_{\a}^i
+ *
 \right \}
\ee
where
\be
\fr{\d {\cal G}_{\rm Connected.}}{\d {\breve A}_i} = A^i
\ee
\be
\fr{\d {\cal G}_{\rm Connected.}}{
 {\breve \y^{\a}}_i
} = \y^i_{\a}
\ee
\be
\fr{\d {\cal G}_{\rm 1PI}}{\d {A}^i} = {\breve A}_i
\ee
\be
\fr{\d {\cal G}_{\rm 1PI}}{\d \y^i_{\a}
} = 
 {\breve \y^{\a}}_i
\ee
and then the identity above in equation (\ref{correctone}) is equivalent to:

\[
\int d^4 x \: \lt \{
\fr{\d {\cal G}_{\rm 1PI} }{\d \G_i} 
\fr{\d {\cal G}_{\rm 1PI} }{\d A^i} 
+
\fr{\d {\cal G}_{\rm 1PI} }{\d {\ov \G}^i} 
\fr{\d {\cal G}_{\rm 1PI} }{\d {\ov A}_i} 
\rt.
\]
\be
\lt.
+
\fr{\d {\cal G}_{\rm 1PI} }{\d {\ov \y}_{i \dot \b}} 
\fr{\d {\cal G}_{\rm 1PI} }{\d {\ov Y}^{i \dot \b} } 
+
\fr{\d {\cal G}_{\rm 1PI} }{\d {  \y}^i_{   \b}} 
\fr{\d {\cal G}_{\rm 1PI} }{\d {  Y}_i^{   \b} } 
\rt \}
+
\fr{\pa {\cal G}_{\rm 1PI} }{\pa { \x}^{\a \dot    \b}} 
\fr{\pa {\cal G}_{\rm 1PI} }{\pa { X}_{\a \dot    \b}} 
=0
\ee
which we will abbreviate to
\be
\la{star1PI}
{\cal G}_{\rm 1PI}* 
{\cal G}_{\rm 1PI}
=0
\ee
Here we can use the loop expansion:
\be
 {\cal G}_{\rm 1PI} =  {\cal A}_{\rm Physical} 
+
 {\cal G}_{\rm 1PI-One \; Loop}
+
 {\cal G}_{\rm 1PI-Two \; Loop}
+\cdots
\la{loopexp}
\ee

Note that `Manifest supersymmetry' is lost by this formulation. 
This is clear because 
\bitem
\item
we have eliminated the ${\breve F}_i$ part of the source superfield,
\item 
we have eliminated the $\L_i$ part of the Zinn-Justin superfield, 
\item
 we have integrated the $F^i$ part of the matter superfields to all orders, 
while leaving the $A^i$ and $\y^i_{\a}$ parts to be integrated in the loop expansion. 
\eitem

Note that we have the following identity from zero loops:
\be
\la{star1PIphys}
{\cal A}_{\rm Physical} * 
{\cal A}_{\rm Physical} =0
\ee

\subsection{Boundary Operator $\d$}

\la{boundaryopinphyspform}

Now we have a new nilpotent operator that is the `square root' of the BRS-ZJ identity:

\[
\d  = 
\int d^4 x \; 
\left \{
\fr{\d {\cal A}_{\rm Physical} }{\d A^i }  
\fr{\d   }{\d \G_i } 
+
\fr{\d {\cal A}_{\rm Physical} }{\d \G_i } 
\fr{\d   }{\d A^i }  
\rt.
\]
\[
\left.
+
\fr{\d {\cal A}_{\rm Physical} }{\d {\ov A}_i } 
\fr{\d  }{\d {\ov \G}^i } 
+
\fr{\d {\cal A}_{\rm Physical} }{\d {\ov \G}^i } 
\fr{\d  }{\d {\ov A}_i } 
\rt.
\]
\[
\left.
+
\fr{\d {\cal A}_{\rm Physical} }{\d \y^{i \a} } 
 \fr{\d   }{\d  Y_{i  \a}}
+
 \fr{\d {\cal A}_{\rm Physical} }{\d  Y_{i  \a}}
\fr{\d   }{\d \y^{i \a} } 
\rt.
\]
\[
\left.
+
\fr{\d {\cal A}_{\rm Physical} }{\d {\ov \y}_{i}^{\dot  \a} }  
\fr{\d  }{\d  {\ov Y}^{i}_{ \dot  \a}}
+
\fr{\d {\cal A}_{\rm Physical} }{\d  {\ov Y}^{i}_{ \dot  \a}}
\fr{\d  }{\d {\ov \y}_{i}^{\dot  \a} }  
\right \}
\]
\be
+
\fr{\pa {\cal A}_{\rm Physical} }{\pa X_{\a \dot \b  }  }  
\fr{\pa  }{\pa  \x^{\a \dot \b  }  } 
+
\fr{\pa {\cal A}_{\rm Physical} }{\pa  \x^{\a \dot \b  }  }  
\fr{\pa }{\pa X_{\a \dot \b  }  }  
\ee 
Its explicit form is summarized in 
Table \ref{physicaltable}, which uses composite terms defined in Table \ref{compterms}.

The equation
\be
\d^2 = 0
\ee
follows from equation (\ref{star1PIphys}), as does the equation:
\be
\d {\cal A}_{\rm Physical}
 = 0
\ee

\begin{table}
\caption{\Large   Transformations  for the Physical Formulation of the BRS-ZJ Identity}
\la{physicaltable}
\vspace{.1in}
\framebox{{\Large $\begin{array}{lll}  
\\
\d A^i&= & 
\fr{\d {\cal A}}{\d \G_i} 
=  \y^{i}_{  \b} {}^{  \b} 
+ \x^{\g \dot \d} \partial_{\g \dot \d} A^i
\\
\d {\ov A}_i&= & 
\fr{\d {\cal A}}{\d {\ov \G}^i} 
=  {\ov \y}_{i  \dot \b} {\ov C}^{ \dot  \b} 
+ \x^{\g \dot \d} \partial_{\g \dot \d} {\ov A}_i
\\

\d \y_{\a}^i &  =& 
\fr{\d {\cal A}}{\d {  Y}_i^{   \a} } = 
\pa_{ \a \dot \b }  A^{i} {\ov C}^{\dot \b}  
+ 
C_{\a}   
F_1^i
+ \x^{\g \dot \d} \partial_{\g \dot \d}  \y^{i}_{\a  }
\\

\d
 {\ov \y}_{i \dot \a} &  =& 
\fr{\d {\cal A}}{\d { {\ov Y}}^{i \dot   \a} } = 
\pa_{ \a \dot \a }  {\ov A}_{i} {C}^{\a}  
+ 
{\ov C}_{\dot \a}   
{\ov F}_{1,i}
+ \x^{\g \dot \d} \partial_{\g \dot \d} 
 {\ov \y}_{i \dot \a} 
\\
\d \G_i 
&= &
 \fr{\d {\cal A}}{\d A^i} 
=
 - \fr{1}{2} \pa_{ \a \dot \b  }       \pa^{ \a \dot \b  }        {\ov  A}_{i} 
  +   m {g}_{iq} F_1^q  + g_{ijk} { F}_2^{jk}
\\
&&
\pa_{ \a \dot \b } Y_{i}^{ \a}    {\ov C}^{\dot \b}   
+ \x^{\g \dot \d} \partial_{\g \dot \d} \G_i
\\

\d {\ov \G}^i 
&= & \fr{\d {\cal A}}{\d {\ov A}_i} 
=
- \fr{1}{2} \pa_{ \a \dot \b  }       \pa^{ \a \dot \b  }        { A}^{i} 
+  m {\ov g}^{ij} {\ov  F}_{1, k}  
+  {\ov g}^{ijk}     {\ov  F}_{2,j k} 
\\
&&
- \pa_{ \a \dot \b } {\ov Y}^{ i \dot \b}    {C}^{\a}   
+ \x^{\g \dot \d} \partial_{\g \dot \d} 
 {\ov \G}^i
\\
\d Y_{i}^{ \a} 
&=&\fr{\d {\cal A}}{\d {  \y}^i_{   \a}} 
= 
-
  \pa^{\a \dot \b  }   
{\ov \y}_{i   \dot \b}
+  m {g}_{iq}   
\y^{q \a} 
\\
&&
 +
2 g_{ijk}  \y^{j \a} A^k    
-
\G_i  
 {C}^{  \a}
+ \x^{\g \dot \d} \partial_{\g \dot \d}  Y_{i}^{ \a}
\\
\d 
{\ov Y}^{i \dot \a} 
&=&\fr{\d {\cal A}}{\d {\ov \y}_i^{ \dot \a} 
} 
= 
-
  \pa^{\b \dot \a  }   
{ \y}^i_{ \b}
+  m {\ov g}^{ik}   
{\ov \y}_{k}^{\dot  \a} 
\\
&&
+
2 {\ov g}^{ijk} {\ov \y}_{j}^{\dot  \a} 
{\ov A}_k  
-
{\ov \G}^i  
 {\ov C}^{\dot  \a}
+ \x^{\g \dot \d} \partial_{\g \dot \d}  
{\ov Y}^{i \dot \a} 
\\
 \d F_1^i 
&=&
  \pa_{\a \dot \b}   \y^{i \a} {\ov C}^{\dot \b} 
+ \x^{\g \dot \d} \partial_{\g \dot \d}  F_1^i 
\\
 \d F_2^{ij} &=& \pa_{\a \dot \b} \lt ( A^i  \y^{j \a} + A^j  \y^{i \a} \rt )  {\ov C}^{\dot \b} 
+ \x^{\g \dot \d} \partial_{\g \dot \d}  F_1^{ij} 
\\
 \d F_3^{ijk} &=& 2 \pa^{\a \dot \b} \lt ( A^{i}   A^{j} \y^{k}_{\a}  + A^{j}   A^{k} \y^{i }_{\a}  + A^{k}   A^{i} \y^{j}_{\a} \rt )  {\ov C}_{\dot \b} 
+ \x^{\g \dot \d} \partial_{\g \dot \d}  F_1^{ijk} 
\\
\d \x_{\a \dot \b} 
&=& \fr{\pa {\cal A}}{\pa { X}^{\a \dot    \b}} 
=
 -   C_{\a} {\ov C}_{\dot \b}
\\
\d X_{\a \dot \b} 
&=& \fr{\pa {\cal A}}{\pa { \x}^{\a \dot    \b}} 
= 
\int d^4 x \; \X_{\a \dot \b} 
\\
\d C_{\a}
&=&
0
\\
\d  {\ov C}_{\dot \b}
&=&
0

\end{array}$}} 
\end{table}

\begin{table}
\caption{\Large  Composite Terms $F$  for   Chiral Supersymmetry}
\la{compterms}
\vspace{.1in}
\framebox{{\Large $\begin{array}{lll}  
\\
 F_1^i 
&=&
- \lt ( 
m {\ov g}^{ij} {\ov A}_j  +
 {\ov g}^{ijk} {\ov A}_j  {\ov A}_k +
{\ov Y}^{i \dot \b} {\ov C}_{\dot \b } 
\rt )
\\
F_2^{ij}&= & 
     A^{i}   
F_1^j
 +     A^{j}   
F_1^i
- \y^{i \a} 
\y^{j}_{ \a}  
 \\
F_3^{(ijk)}
&= & 
 A^{i}  A^{j}  F_1^k
+ A^{j}  A^{k}  F_1^i
+ A^{k}  A^{i}  F_1^j
\\&&
 - 
 \y^{i \a} \y^{j}_{ \a}  A^k
 - 
 \y^{j \a} \y^{k}_{ \a}  A^i
 - 
 \y^{k \a} \y^{i}_{ \a}  A^j
\\
\\
{\ov F}_{1,i} 
&=&
-\lt ( 
m {g}_{il} {A}^l  +
{g}_{ilq} A^l A^q
+
{Y}_i^{  \b} {C}_{\b } 
\rt )
\\
{\ov F}_{2,ij} 
&=&
\A_i \oF_{1,j} 
+\A_j \oF_{1,i} 
-
\oy_i^{\dot \b}  
\oy_{j,\dot \b}  
\\
&=&
-\A_j \lt (  
 m {g}_{il} {A}^l  +
{g}_{ilq} A^l A^q 
+
{Y}_i^{  \b} {C}_{\b } 
\rt )
\\
&&
-
\A_i \lt ( 
m {g}_{jl} {A}^l  +
{g}_{jlq} A^l A^q
+
{Y}_j^{  \b} {C}_{\b } 
\rt )
-
\oy_i^{\dot \b}  
\oy_{j,\dot \b}  
\\
\\
\X_{\g \dot \d}
&=&
 Y_{i}^{ \a}  \pa_{\g \dot \d}  \y^{i}_{ \a}
+
 {\ov Y}^{i \dot \b}  \pa_{\g \dot \d}{\ov \y}_{i \dot \b}  
-
 \G_{i}   \pa_{\g \dot \d}  A^{i} 
-
 {\ov \G}^{i  }  \pa_{\g \dot \d}{\ov A}_{i }  
\\

\end{array}$}} 
\end{table}



\chapter{
Cohomology including Zinn-Justin's Sources and the Spectral Sequence}

\la{cohomchapter}

\section{BRS identity at One Loop}

\la{brsidesection}

Here is the naive BRS Identity for ${\cal G}_{\rm 1PI-One \; Loop}
$ which follows from the identities 
(\ref{loopexp}) and (\ref{star1PI}),
using the $\d$ defined in 
Table \ref{physicaltable}:
\be
\d 
{\cal G}_{\rm 1PI-One \; Loop}
 = 0
\ee
This is expected to be true, except if there are linear (or greater) divergences in the Feynman integrals. If there are linear (or greater) divergences, then we expect that 
\be
\d {\cal G}_{\rm 1PI-One \; Loop}
 = {\cal A}^{(1)}
\la{brsvar}
\ee
The right hand side here comes from the ambiguities associated with linear divergences in the momentum loop integrals \ci{brs}. ${\cal A}^{(1)}$ necessarily is the integral of a local polynomial (LIP) in the fields, the sources and the operator $\pa_{\a \dot \b}$. Moreover as calculated from the diagrams, each term in it must be proportional to at least one momentum.  We can eliminate some of those terms however using ghost charge zero counterterms proportional to Zinn Sources which bring in the equation of motion. That is how terms without momenta could arise. 

The possible anomalies are the local integrated polynomials (LIP)  in the local cohomology space with ghost charge one.  This means that an anomaly satisfies the relation
\be {\cal A}^{(1)}\in {\cal H}^{(1)}_{\rm LIP}
\ee
where
\be
{\cal H}_{\rm LIP} 
=
\fr{\{ {\cal Z}_{\rm LIP}: \d {\cal Z}_{\rm LIP} 
=0 \}}{ \{
{\cal B}_{\rm LIP}: {\cal B}_{\rm LIP} 
= \d {\cal P}_{\rm LIP} \} }
\ee
This equation has the form
\be
{\rm Cohomology} = 
\fr{\rm Cocycles}{\rm Coboundaries}
\ee

Clearly an anomaly is only defined up to a coboundary by this definition.

\section{Some Useful Properties}

See Tables \ref{parametertable}, \ref{fieldparametertable} and \ref{Zinnparametertable} for a summary of the parameters and fields.  This also includes some notation for Supersymmetric Gauge theory.
We will use the abbreviations in Table \ref{Abbreviations}.  The acronyms in Table \ref{acronymstable} can help us to distinguish the various possibilities.   We shall frequently use the acronyms SSSM, LIP, ZMOZ,  ZMOF  and ZMT.

\begin{table}[hptb]
\caption{ Table of Parameters with some Properties}
\la{parametertable}

\vspace{.1in}
\framebox{ 
{\large $\begin{array}{lrrlll} 
 
{\rm Parameter} & {\rm Dim.} & N_{\rm Form} & \mbox{Conjugate} &\mbox{S-partner}&\mbox{Comment }\\

C_{\a} & - \fr{1}{2}  & 1 & {\ov C}_{\dot \b} & \x_{\a \dot \b} &  \mbox{Super Ghost} \\

\x_{\a \dot \b} & - 1  & 1 & \x_{\b \dot \a} & c_{\a} ,{\ov c}_{\dot \b}  &  \mbox{Translation Ghost} \\

m  &  1  & 0 & m &  &  \mbox{Mass Parameter} \\

\pa_{\a \dot \b}   &  1  & 0 & \pa_{\b \dot \a}  &  &  \mbox{Derivative} \\

x_{\a \dot \b}   & - 1  & 0 & x_{\b \dot \a}  &  &  \mbox{Spacetime} \\

k_{\a \dot \b}   &  1  & 0 & k_{\b \dot \a}  &  &  \mbox{Momentum} \\

\int d^4 x   &  -4   & 4 &  \int d^4 x     &  &  \mbox{Space Integral} \\

\int d^4 k   &  4   & -4 &  \int d^4 k     &  &  \mbox{Momentum Integral} \\

\d   &  0   & 1 &  \d      &  &  \mbox{BRS Operator} \\

\d^{\dag}   & 0 & -1 &  \d^{\dag}&  &  \mbox{Adjoint BRS Operator} \\

\end{array}$ }}
\end{table}

\begin{table}[hptb]
\caption{ Table of  Fields in the General Model   with some Properties}
\la{fieldparametertable}
\vspace{.1in}
\framebox{ 
{\Large $\begin{array}{lrrlll} 
 
{\rm Field} & {\rm Dim.} & N_{\rm Form} & 
\mbox{Conjugate} &\mbox{S-partner}&\mbox{Comment and Mass}\\
A^i & 1 & 0 & {\ov A}_i&\y^i_{\a}&\mbox{Scalar} \\
{\y }^i_{\a} & \fr{3}{2} & 0 & {\ov {  \y}}_{i \dot \a}& A^i & \mbox{Spinor} \\

V^a_{\a \dot \b}  & 1 & 0 &V^a_{\b \dot \a}   &\l^a_{\a},{\ov {  \l}}^a_{\dot \b}&  \mbox{Gauge Vector} \\
{\l }^a_{\a} & \fr{3}{2} & 0 & {\ov {  \l}}^a_{\dot \a}& V^a_{\a \dot \b}  &\mbox{Gaugino} \\

\w^a & 0 & 1 & \w^a &  &  \mbox{Gauge Ghost}  \\

\h^a & 2 & -1 & \h^a &  &  \mbox{Gauge AntiGhost} \\
 
\end{array}$ }}
\end{table}

\begin{table}[hptb]
\caption{ Table of  Zinn Sources in the General Model with some Properties}
\la{Zinnparametertable}

\vspace{.1in}
\framebox{ 
{\large $\begin{array}{lllllll}  
{\rm Field} & {\rm Dim.} & N_{\rm Form} & \mbox{Conjugate} &\mbox{Superpartner} & \mbox{Source for}\\
\G_i & 3 & -1 & {\ov \G}^i&Y_{i \a }& \d A^i \\
Y_i^{\a} & \fr{5}{2} & -1 & {\ov Y}^{i \dot \a} &\G_i & \d \y^i_{\a} \\

\S^a_{\a \dot \b}  & 3 & -1 & \S^a_{\b \dot \a}   &L^a_{\a},{\ov L}^a_{\dot \b} & \d V^{a \a \dot \b}\\

L^{a \a} & \fr{5}{2} & -1 & {\ov L}^{a \dot \a} &\S^a_{\a \dot \b}& \d \l^a_{\a} \\

W^a & 4 & -2  & W^a &  &  \d \w^a \\

H^a & 2 & 0 & H^a &  & \d \h^a \\
 
X_{\a \dot \b} &   1  & -2 & X_{\b \dot \a} & &  c_{\a} {\ov c}_{\dot \b} = \d \x^{\a \dot \b} 
 \\
\x_{\a \dot \b} &   -1  & 1 & \x_{\b \dot \a} & C_{\a} {\ov C}_{\dot \b}   &  \X^{\a \dot \b} = \d X^{\a \dot \b}  
 \\

\end{array}$ }}
\end{table}

\begin{table}[hptb]
\caption{ Table of Abbreviations}
\vspace{.1in}
\framebox{ 
{\large $\begin{array}{llrllll}  
\la{Abbreviations}

{\rm Abbreviation} & {\rm Meaning} & {\rm Dim.} & N_{\rm Form} & \mbox{Conjugate} \\

( \x^2 )_{\a \b}
  &  \x_{\b }^{\;\; \dot \b} \x_{\a \dot \b} 
& -2 & 2 & ( \x^2 )_{\dot \a \dot \b}\\  

(\x^4)   &  ( \x^2 )_{\a \b}
( \x^2 )^{\a \b}
& -4 & 4 & 
(\x^4) 
\\

(C \x^2 C) & C^{\a} ( \x^2 )_{\a \b} C^{\b} & -3 & 4 &  
({\ov C} \x^2 {\ov C})& \\

(\f C) & \f^{\a} C_{\a }   & 0 & 1&  
({\ov \f}   {\ov C}) & \\

(C \x {\ov C}) & C^{\a}  \x_{\a \dot \b} {\ov C}^{\dot \b} & -2 & 3 &  (C \x {\ov C}) \\

(C \x {\ov \y}_i) & C^{\a}  \x_{\a \dot \b} {\ov \y}_i^{\dot \b} & 0 & 2 & - (\y^i \x {\ov C}) \\

( C \pa A^i \oC )& C^{\a}  \pa_{\a \dot \b} A^i
\oC_{\dot \b}  & 1 & 2 &  ({\ov C} \pa \A_i C)  
   \\
(\x \pa ) &
\e^{\dot \g \dot \d} 
 \partial_{\g \dot \d}
 & 0 & 1 &  (\x \pa ) \\
F_{\a \b} &
\e^{\dot \g \dot \d} 
\lt ( \partial_{\a \dot \g} V_{\b \dot \d}
+
 \partial_{\b \dot \g} V_{\a \dot \d}
\rt )
 & 2 & 0 &  \oF_{\dot \a \dot \b} \\
\\

\end{array}$ }}
\end{table}

\begin{table}[hptb]
\caption{\large Some Concepts}

\la{acronymstable}
\framebox{
{\large $\begin{array}{llll}  

 \mbox{Acronym}  & \mbox{Meaning}  & \mbox{Discussion}  
  \\   
\\

 \mbox{LIP} & \mbox{Local Integrated Polynomial}  & \mbox{Sections \ref{isolampsection}, \ref{brsidesection}
 }  
  \\   
 \mbox{SS} & \mbox{Supersymmetry}  & 
   \\   
\mbox{SSBL} & \mbox{Spontaneous SS Breaking of Local SS}  & \mbox{Section \ref{isolampsection}  }  
  \\   

 \mbox{SSBG} & \mbox{Spontaneous Gauge Breaking of Gauge SS}  & \mbox{Section \ref{isolampsection}  }  
  \\   
 \mbox{SSBR} & \mbox{Spontaneous SS Breaking of Rigid SS}  & \mbox{Section \ref{vevsection}  }  
  \\   
 \mbox{SSSM} & \mbox{SS  Standard Model }  & \mbox{Section \ref{SSSMsection} }  
  \\   

 \mbox{ZMOF} & \mbox{Zero Momentum One Field}  & \mbox{Section \ref{isolampsection} }  
\\

 \mbox{ZMOZ} & \mbox{Zero Momentum One Zinn}  & \mbox{Section \ref{isolampsection} }  
  \\   

 \mbox{ZMT} & \mbox{ ZMOF  or ZMOZ}  & \mbox{Section \ref{isolampsection} }  
  \\   

\end{array}$ }}
\la{acroTable}
\end{table}

\section{Synopsis of Results from the Spectral Sequence}

This Chapter will summarize very briefly the results that have been obtained using the spectral sequence technique on the operator defined above in Table  \ref{physicaltable}, which uses the composite terms defined in Table \ref{compterms}. This is not meant to be a complete or introductory treatment of the spectral sequence for this case.  In this paper, it will be assumed that the reader is familiar with references 
\ci{holes} \ci{specseq} \ci{chirsup} \ci{dixprl} \ci{dixmin} \ci{dixminram} \ci{kyoto}.  A more complete and more readable treatment is being prepared.  There are plenty of unsolved problems. 
 One of the difficulties is that the treatment needs to be very general to treat the spectral sequence properly, and here we are only interested in the simplest cases.

This Chapter and  Chapters \ref{diagramchasechapter} and \ref{insertdiagramchasechapter} all use the spectral sequence  and Fock Space techniques developed in the above references.
The other chapters are fairly free from the use of these techniques, and can be read fairly independently of the spectral sequence techniques. 

Some results here that relate to the cohomology of supersymmetric gauge theory are also briefly mentioned.  A paper on that topic is also in preparation. 

\section{Isomorphisms, Looking under the Lamp and Zero Momentum Terms}
\la{isolampsection}
As mentioned above, the cohomology that is relevant here to the anomalies and invariants is 
\be
{\cal H}_{\rm LIP} 
=
\fr{\{ {\cal Z}_{\rm LIP}: \d {\cal Z}_{\rm LIP} 
=0 \}}{ \{
{\cal B}_{\rm LIP}: {\cal B}_{\rm LIP} 
= \d {\cal P}_{\rm LIP} \} }
\ee
The cohomology that is calculable with the Fock space techniques in the references is:
\be
{\cal H}_{\rm Fock} 
=
\fr{ \{
{\cal Z}_{\rm Fock}: \d {\cal Z}_{\rm Fock}
=0 \}}{ \{
{\cal B}_{\rm Fock}: {\cal B}_{\rm Fock} 
= \d {\cal P}_{\rm Fock} \} }
\ee
and with the spectral sequence techniques we can calculate a third cohomology space that satisfies
\be
{\cal H}_{\rm Fock} 
\approx E_{\infty} 
\ee
In the references it was shown that one could describe the space of interest by a two step process
\be
E_{\infty} 
\ra 
{\cal H}_{\rm Fock} 
\ra
{\cal H}_{\rm LIP} 
\ee

In previous work, for the examples looked at, ${\cal H}_{\rm LIP} 
$ and ${\cal H}_{\rm Fock} 
$ could be shown to be isomorphic. 

 Here that is not quite the case. We need to be a little more careful. 

We need to divide the parameters into two classes:
\ben 
\item
{\em Fields}: These are $A,\y,Y,\G,\A,\oy,\oY,\oG$.  
Fields  have the property that 
$\pa_{\a \dot \b}$ does not yield zero when applied to a Field.

\item
{\em Constants}: These are $m,\x,C,\oC, X$. Constants have the property that 
$\pa_{\a \dot \b}$ yields zero when applied to a Constant.  

\een

A simple isomorphism holds only when
there are no  terms in $\d$ that convert a single Field into a product of Constants.  We will call such terms Zero Momentum One Field  (ZMOF)  terms.  They originate from  Zero Momentum One Zinn  (ZMOZ) terms in the action.  Here are some examples of  ZMOF  and ZMOZ terms, together with comments on them:
\ben
\item
For example, when there is a global symmetry (as distinguished from a local gauged symmetry) that is spontaneously broken, one gets a  ZMOF  term:
\[
\d_{\rm Gold} \ni \int d^4 x \;
m G^{ai} \w^a \fr{\d}{\d A^i(x)}
\]
\be
 = 
m G^{ai} \w^a \int d^4 x \;\fr{\d}{\d A^i(x)}
\ee
For any such term there is of course a ZMOZ in the action.  In this case it is
\[
{\cal A}_{\rm Gold}  \ni 
\int d^4 x \;
\G_i(x) m \w^a  G^{ai} 
\]
\be
=
 m \w^a  G^{ai}  \int d^4 x \;
\G_i(x)
\ee
This is a ZMOZ term.  If we make $\w^a$ into a field $\w^a(x)$:
\be
\d_{\rm Gold,Local} \ni  \int d^4 x \;
m G^{ai} \w^a(x) \fr{\d}{\d A^i(x)}
\ee
then this term is no longer a  ZMOF  term, and  
the term in the action becomes bilinear.  In this case it is
\[
{\cal A}  \ni 
 \int d^4 x \;\G_i(x) m \w^a(x)  G^{ai}
\]
\be
=
m  G^{ai}
 \int d^4 x \;\G_i(x)  \w^a(x) \ee

Closure of this algebra results in super Yang Mills with spontaneous  breaking of gauge symmetry.

\item
For a second example, when there is a spontaneous breaking of rigid supersymmetry (SSBR), one gets a term:
\[
\d_{\rm SBSS} \ni  \int d^4 x \;
m^2 \og^i C_{\a} \fr{\d}{\d \y_{\a}^i(x)}
\]
\be
=
m^2 \og^i C_{\a} \int d^4 x \;
\fr{\d}{\d \y_{\a}^i(x)}
\ee
This is a  ZMOF  term.  If we make $C_{\a}$ into a field $C_{\a}(x)$, then this becomes
\be
\d_{\rm SBSS, Local} \ni \int d^4 x \;
m^2 \og^i C_{\a}(x) \fr{\d}{\d \y_{\a}^i(x)}
\ee
and the term is no longer a ZMOF   term.  Closure of this algebra results in supergravity with spontaneous  breaking of local supersymmetry.
\item
For a third example, we will see that it is possible to insert operators such that one generates a term:
\[
\d_{\rm Spinor} \ni  \int d^4 x \;
\f^{\a}  C_{\a}  m f_i  \fr{\d}{\d \A_i(x)}
\]
\be
=
\f^{\a}  C_{\a}  m f_i  \int d^4 x \;
 \fr{\d}{\d \A_i(x)}
\ee
This is a ZMOF   term.  If we make $\f_{\a}$ into a field $\f_{\a}(x)$, then this becomes
\be
\d_{\rm Spinor, Local} \ni  \int d^4 x \;
\f^{\a}(x)  C_{\a} m f_i  \fr{\d}{\d \A_i(x)}
\ee
and the term is no longer a ZMOF   term.  Closure of this algebra results in 
a new action in which there is an antichiral superfield coupled to and mixed with chiral matter.  This is treated in Chapter \ref{superfieldschapter}.  The kinetic and kinetic mixing and ghost terms for these $\f_{\a}$ terms need to be examined. 
\item
For a fourth example, we will see that it is possible to insert operators such that one generates a term:
\[
\d_{\f}  \ni  \int d^4 x \;
\f^{\a}  C_{\a}   f^i_j \A_i(x)  \fr{\d}{\d \A_j (x)}
\]
\be
= \f^{\a}  C_{\a}   f^i_j  \int d^4 x \;
\A_i(x)  \fr{\d}{\d \A_j (x)}
\ee
This term is not  a ZMOF   term.  We can leave $\f_{\a}$ a constant and calculate the cohomology here, and that has the advantage of simplicity.  This is our main concentration in this paper.  This example may be sufficient to demonstrate the existence of supersymmetry anomalies. 

\item
For a fifth example, we will see that a possible anomaly appears to have terms like:
\be
{\cal A}_{\f}^1 \ni 
\int d^4 x \;
\f^{\a}  C_{\a}  m^2 \oh^i \oF_{1,i}
\ee
This term is another kind of ZMOZ term. Part of it is
\be
{\cal A}_{\f}^1 \ni 
\int d^4 x \;
\f^{\a}  C_{\a}  m^2 \oh^i Y^{\b}_i c_{\b} 
=
\f^{\a}  C_{\a}  m^2 \oh^i  c_{\b} 
\int d^4 x \; Y^{\b}_i(x)
\ee

A term like this cannot arise as the variation $\d {\cal G}_{1PI}$ from any term that is bilinear in the fields unless $\d$  contains a ZMT term. It could arise from the variation $\d {\cal G}_{1PI}$ of 
a term that is linear in a field.  But such a term in  $ {\cal G}_{1PI}$ does not appear likely to happen, since such a term necessarily is proportional to a momentum, and the relevant field would also have zero momentum, which means that the term must be zero.   So we do not expect to generate these kinds of terms, and shall also label them ZMOZ or ZMOF    terms. Care with respect to them is needed to avoid confusion however. 

\een

If there are such ZMOF   terms in $\d$ then 
${\cal H}_{\rm LIP}$
and ${\cal H}_{\rm Fock}$ are not simply related.  I do not know how to calculate ${\cal H}_{\rm LIP}$ for such cases. Also these kinds of actions appear uninteresting compared to the related theories with local symmetries.  So we shall largely ignore them.  One can calculate ${\cal H}_{\rm Fock}$ for such cases, but its usefulness is not obvious.

If ZMOF   terms are absent, then the isomorphism is effectively:
\be
{\cal H}_{\rm LIP} [   N_{\rm Field} \neq 0] 
\approx
{\cal H}_{\rm Fock} [  N_{\rm Field} \neq 0 ] 
\la{isonoZMT}
\ee

So for the computations that we will look at, we can and will calculate the cohomology using these methods for the case where Spontaneous SS Breaking of Rigid SS (SSBR) is absent.

The isomorphism in Equation (\ref{isonoZMT}) can be demonstrated by using  the correspondence
\be
\d_{\rm Fock} = 
\d_{\rm Integral} -
C_{\a} \oC_{\dot \b} \x_{\a \dot \b}^{\dag} 
+ \x^{\a \dot \b} \pa_{\a \dot \b} 
\ee
and the grading
\be
N_{\rm Fock} = N_{c} - N_{\G} - N_{Y} + * 
\ee
which satisfies
\be
[ N_{\rm Fock}, \x^{\a \dot \b} \pa_{\a \dot \b}] = 0
\ee
\be
[ N_{\rm Fock}, \d_{\rm Integral} ] = \d_{\rm Integral}
\ee
\be
[ N_{\rm Fock}, C_{\a} \oC_{\dot \b} \x_{\a \dot \b}^{\dag} 
 ] = 2 C_{\a} \oC_{\dot \b} \x_{\a \dot \b}^{\dag} 
\ee

This demonstration depends on the identity
\be
[\pa_{\m}^{\dag}, \pa_{\n} ] = 
\d_{\n}^{\m} N_{\rm Fields}
\ee
as is explained in the references   \ci{holes} \ci{specseq} \ci{chirsup} \ci{dixprl} \ci{dixmin} \ci{dixminram} \ci{kyoto}.

{\em For this reason, and for reasons that relate to actual calculations in perturbation theory,  $\d$ for which ZMT  are absent yield comprehensible cohomology.  If ZMT  are present, we are tempted to remove them or else we could promote some constant to a field and find the relevant new $\d$ without a ZMT. If ZMT are excluded one can ignore the pure constant terms.}

The term 
$\X_{\a \dot \b} \fr{\pa}{\pa X_{\a \dot \b} }$ has been ignored in the above.  It needs a separate treatment.  It does not seem to relate to any interesting problem at this stage.  It is treated below in subsection \ref{deltaxisubsection}. It is not a ZMT  because it involves more than one Field or Zinn in every term.

\section{Grading and Some details about $d_r$ }

\subsection{Grading}

We will use the following grading to generate the spectral sequence here. 

\[
{N}_{\rm  Grading} 
= 
{N}_c 
+
{ {N}}_{\ov C}
+
2  N_{\x}
+4  N_{m}
+ 
 4
N_{\rm Fields}
+
4 N_{\rm Zinn}
\]
\be
+
9 {N}_X
+
6 {N}_{\f;\fr{1}{2}}
\la{NewGrading}
\ee

The operators $d_r$ of the resulting spectral sequence are summarized in Tables \ref{specseqsum} and \ref{specseqsum2}.  We anticipate some of the results of Chapter \ref{insertionchapter} in this Table by referring to the $d_7$ operators which arises from those considerations.

This Table \ref{specseqsum} represents the result of a large amount of trial and error.  Some gradings yield insight and some do not.  Once a grading yields insight, it takes a long time to figure out the $d_r$ that arise from it, and the role they play.  So this Table, and its correctness, are not at all obvious. The use of the Table in the next chapter is probably the best illustration of its origin and usefulness.

\begin{table}[hptb]
\caption{ Operators  for the Spectral Sequence for Rigid Supersymmetry  }
\la{specseqsum}
\vspace{.1in}
\framebox{ 
{\large $\begin{array}{lll }  

{\rm Operator} & {\rm Form} 
\\
d_0 \equiv \d_0 & d_{\rm Kinetic} +d_{\rm Structure}
\\
d_1 & \P_1 \lt \{
C^{\a} \na_{\a} + {\ov C}^{\dot \b}  {\ov \na}_{\dot \b} 
\rt \}
\P_1
\\
d_2  & \P_2 \lt \{ \x \pa -
\lt (C \na + {\ov C} {\ov \na} \rt )
\fr{\d_{\rm Str}^{\dag}}{\D_{\rm Str}} 
\lt (C \na + {\ov C} {\ov \na} \rt )
 \rt \} \P_2
\\
d_3 & d_{\rm Gaugino} + d_{\X}
\\
d_4 = d_{\rm Gauge} & \P_4 \lt \{    \w^a \lt (
m G^{a i}+ T^{a i}_j A^j \rt )   A^{i \dag} + *
- \fr{1}{2} f^{abc} \w^a   \w^b   \w^{c \dag}
\rt \} \P_4
\\
d_5 = d_{5,\rm MY} & \P_5 \lt \{
 C_{\a}
\lt ( m {\ov g}^{ij} {\ov A}_j 
+ {\ov g}^{ijk} \A_j \A_k 
\rt )
\y_{\a}^{i \dag}   + * \rt \} \P_5
\\
d_6 = d_{6,\rm MY} & \P_6 
\lt \{
( C \x \oC^{\dag} )
\lt ( m {\ov g}^{ij} {\ov A}_j 
+ {\ov g}^{ijk} \A_j \A_k 
\rt )
A^{i \dag}   + * \rt \} \P_6
\\
 d_{7,\fr{1}{2}} & \P_7 
\lt \{
( \f C) 
f^j_i \lt (  \A_j 
 \A_{i}^{ \dag} + \y^i_{\b} 
 \y_{j \b}^{ \dag}  \rt ) + * \rt \} 
\P_7 \mbox{ for Dim $\f_{\a} = \fr{1}{2}$ Case}
\\

\\
\end{array}$ }}
\end{table}

\begin{table}[hptb]
\caption{ Supplemental Information for
Operators  for the Spectral Sequence for Rigid Supersymmetry  }
\la{specseqsum2}
\vspace{.1in}
\framebox{ 
{\large $\begin{array}{lll }  

{\rm Operator} & {\rm Form} 
\\
 \d_{\rm Kinetic} 
& \int d^4 x \; \lt \{
\pa_{\a \dot \b} {\ov \y}_i^{\dot \b} \fr{\d}{\d Y_{i \a} } + (\pa \pa) {\ov A}_i^{\dot \b} \fr{\d}{\d \G_{i} } + *
\rt \}
\\
\d_{\rm Structure}\equiv \d_{\rm Str}& C_{\a} {\ov C}_{\dot \b} \x_{\a \dot \b}^{\dag}
\\
\na_{\a} & \y^i_{\a} A^{i \dag} + {\ov A}_{i \dot \b \a} \oy_{i \dot \b}^{\dag} + \cdots
\\
{\ov \na}_{\dot \a} & {\ov \y}_{i \dot \a} \A_{i}^{ \dag} + {A}^i_{\b   \dot \a} \y_{\b}^{i \dag} + \cdots
\\
d_{\rm Gaugino} & \P_3 
\lt \{ 
(  C  \x  {\ov \l}^a )\w^{a \dag}
+
(  \l^a \x  {\ov C} )\w^{a \dag}
 \rt \}  \P_3
\\
d_{\X} & \P_3 \P_{ C \x {\ov C} }( C^{\a}  \x_{\a \dot \b}
 {\ov C}^{\dot \b} )
 \lt \{ {\ov A}_i    \pa_{\g \dot \d} A^i  -  \pa_{\g \dot \d} {\ov A}_i    A^i - \y^i_{\g  } {\ov \y}_{i  \dot \d}  
\rt \}   X_{\g \dot \d}^{\dag}\P_3

\\
d_5^{\dag} = d_{\rm MY} & \P_5 \lt \{
\y_{\a}^{i}  
\lt ( g_{ij} {\ov A}_j^{\dag} m^{\dag}
+ g_{ijk} \A_j^{\dag} \A_k^{\dag}  
\rt )
 C_{\a}^{\dag}
 + * \rt \} \P_5
\\
d_6^{\dag} = d_{6,\rm MY}^{\dag} & \P_6 
\lt \{
A^{i }  
\lt ( g_{ij} {\ov A}_j^{\dag} m^{\dag}
+ g_{ijk} \A_j^{\dag} \A_k^{\dag}  
\rt )
( C^{\dag} \x^{\dag} \oC )
 + * \rt \} \P_6
\\
 
 d_{7,\fr{1}{2}}^{\dag} & \P_7 
\lt \{ \of^j_i \lt (  \A_j 
 \A_{i}^{ \dag} + \y^i_{\b} 
 \y_{j \b}^{ \dag}  \rt )  
  ( \f C)^{\dag}    + * \rt \} 
\P_7 \mbox{ for Dim $\f_{\a} = \fr{1}{2}$ Case}
\\

\\
\end{array}$ }}
\end{table}

\subsection{Stucture and Kinetic Terms: The operator $\d_0$ and the Laplacian $\D_0$}
\la{delta0disc}
\be
\d_0 
= 
\d_{\rm Structure} +\d_{\rm Kinetic} 
\ee
where
\be
\d_{\rm Structure} =
 C_{\a} {\ov C}_{ \dot \b}  \x_{\a \dot \b}^{\dag}
\ee
and
\be
\d_{\rm Kinetic}  = \int d^4 x 
  \pa_{\a \dot \b} {\ov \y}^{\dot \b}_{i} \fr{\d}{\d Y_{i \a}}  
+ *
\ee 
\be
+\int d^4 x \;  
\pa_{\a \dot \b} \pa^{\a \dot \b}  {\ov A}_i
\fr{\d}{\d \G_{i }}   
+ *
\ee

 These first two operators give rise to the following  Laplacian operators:
\be
\D_0 = \D_{\rm Structure} +\D_{\rm Kinetic} 
\ee
Note that the two suboperators commute:
\be
[ \D_{\rm Structure} ,\D_{\rm Kinetic} ] =0
\ee
where
\be
\D_{\rm Structure} = \lt ( \d_{\rm Structure}+
 \d_{\rm Structure}^{\dag} \rt )^2 
\ee
\be
\D_{\rm Kinetic} = \lt ( \d_{\rm Kinetic}+ \d_{\rm Kinetic}^{\dag} \rt )^2 
\ee

The Laplacian $ \D_{\rm Kinetic} $ has a simple structure when it is expressed in terms of the variables
\be
\pa_{\a_1 \dot \b_1} \cdots \pa_{\a_n \dot \b_n} A^i \equiv
A^i_{\a_1 \dot \b_1,  \cdots \a_n \dot \b_n}, \; {\rm etc.}
\ee
It is easy to see that $E_1=\ker \D_0$ is a function only of the 
the Physical Fields as tabulated in Table \ref{PhysFieldsTable}.  
A similar analysis shows that supersymmetric gauge theory yields
the other results in Table \ref{PhysFieldsTable}.  
These fields are obtained from the above by symmetrization in all the undotted and all the dotted indices:
\be
A^i_{\a_1 \dot \b_1,  \cdots \a_n \dot \b_n} \ra 
A^i_{\a_1 \cdots \a_n, \dot \b_1 \cdots \dot \b_n} 
\equiv 
A^i_{(\a_1 \cdots \a_n), (\dot \b_1 \cdots \dot \b_n)} 
\la{physfield}
\ee
In other words, contractions such as 
\be
A^i_{a \dot \b , \g \dot \d} \e^{\dot \b \dot \d}
\ee
are not permitted in $E_1$. This makes the work after $E_1$ much easier.

$\D_{\rm Structure}$ has a unique form special to supersymmetry.  Its cohomology was worked out in \ci{dixmin}, and we shall use those results here.  In conjunction with the action of 
$\D_{\rm Kinetic} $, they give rise to the following form. 

\[
E_1 = \ker \D_0 = \ker \D_{\rm Structure} \cap \ker \D_{\rm Kinetic} 
\]
\[
=
{\cal P}[\A, \oy, A, \y, C]
\oplus
{\ov {\cal P}}[ A, \y, \A, \oy, \oC]
\]
\[
\oplus
{\cal Q}_{\dot \b}[\A, \oy, A, \y, C]
 (\x C)^{\dot \b}
\oplus
{\ov {\cal Q}}_{\b}[ A, \y, \A, \oy, \oC]
 (\x \oC)^{\b}
\]
\[
\oplus
{\cal R} [\A, \oy, A, \y, C]
 (C\x^2 C)
\oplus
{\ov {\cal R}} [A, \y, \A, \oy, \oC]
 (\oC \x^2 \oC)
\]
\be
\la{E1eq} 
\oplus
{\cal S}_0 [\A, \oy, A, \y]
 (C\x  \oC)
\ee
where $A,\y,\A,\oy$ are the special symmetrized `Physical Fields' such as those in Equation (\ref{physfield})
that are referred to in Table \ref{PhysFieldsTable}.

In Table \ref{E1Table} we use the notation: 
\be
{\cal P} = \sum_{n=0}^{\infty} {\cal P}_n ; {\rm etc.}
\ee
where
\be
N_c {\cal P}_n = n {\cal P}_n ; {\rm etc.}
\ee

Table \ref{E1Table} further subdivides these polynomials for use in Tables \ref{d1d1dagtable}, \ref{d2table}
and \ref{d2dagtable}.

We do not attempt to use all these results here.  The purpose of including them is to establish some notation for the problem.   Our approach in this paper is more pedestrian.  We will simply use the operators above as maps in the `Diagram Chases' contained in the next chapters.

\begin{table}[hptb]
\caption{  `Physical' Fields and Parameters in $E_1$. Some gauge theory is included here for future reference.  
The gauge results are not discussed in the text and will be presented in a future paper. }

\vspace{.1in}
\framebox{ 
{\large $\begin{array}{lllllll}  
\la{PhysFieldsTable}
\\
\mbox{`Physical' }  {\ov {\cal F}} \mbox { Fields   }  & \mbox{Dim} &  \mbox{Grass}& \mbox{Comments}  \\   
\\

 {\ov A}_{i  \dot \b_1 \cdots \b_q ,\a_1 \cdots \a_q}; q=0,1\ldots
& q+1& \mbox{Even}&\mbox{Scalar} \\

  \y^i_{\a \a_1 \cdots \a_q ,\dot \b_1 \cdots \b_q}; q=0,1 \ldots
& q+\fr{3}{2} & \mbox{Odd}& \mbox{Spinor}\\

 F^a_{\a \b \a_1 \cdots \a_q \dot \b_1 \cdots \b_q}; q=0,1\ldots
&q+2 & \mbox{Even}& \mbox{Gauge Field Strength}\\

  {\ov \l}^a_{\dot \b \dot \b_1 \cdots \b_q,   \a_1 \cdots \a_q }; q=0,1\ldots
&q+\fr{3}{2} & \mbox{Odd}& \mbox{Gaugino}\\
\\

\mbox{`Physical' } { {\cal F}} \mbox { Fields   }  & \mbox{Dim}& \mbox{Grass}& \mbox{Comments}  \\   
\\

 A^i_{\a_1 \cdots \a_q, \dot \b_1 \cdots \b_q}; q=0,1\ldots
&q+1 & \mbox{Even} & \mbox{Scalar}\\

  {\ov \y}_{i \dot \b \dot \b_1 \cdots \b_q , \a_1 \cdots \a_q }; q=0,1 \ldots
&q+\fr{3}{2} & \mbox{Odd}& \mbox{Spinor}
\\
  {\ov F}^a_{\dot \a \dot \b \dot \b_1 \cdots \b_q,  \a_1 \cdots \a_q}; q=0,1\ldots
&q+2 & \mbox{Even}& \mbox{Field Strength}
\\
  \l^a_{\a \a_1 \cdots \a_q \dot \b_1 \cdots \b_q}; q=0,1\ldots
&q+\fr{3}{2} & \mbox{Odd}& \mbox{Gaugino}
\\
\\

\mbox{`Physical'   Parameters   }  & \mbox{Dim} & \mbox{Grass}& \mbox{Comments}  \\ \\  
\w^a = {\ov \w}^a
& 0 & \mbox{Odd}&\mbox{Gauge Ghost}\\
  \f_{\a}  &  \fr{3}{2}, \fr{1}{2} & \mbox{Even}&\mbox{Source } 
\\
  {\ov \f}_{\dot \a}
& \fr{3}{2}, \fr{1}{2}\cdots & \mbox{Even}&\mbox{Source } 
\\
  C_{\a} 
& - \fr{1}{2} & \mbox{Even}&\mbox{Super Ghost}
\\
  \x_{\a \dot \b}& -1  & \mbox{Even}&\mbox{Translation Ghost}
\\
  {\ov C}_{\dot \a}
& -\fr{1}{2}& \mbox{Even}&\mbox{Super Ghost}
\\
  m = {\ov m}
& 1  & \mbox{Even}& \mbox{Mass Parameter}\\

\end{array}$ }}
\end{table}

\subsection{The Supertranslation Operator $\d_1$}

The next operator in the spectral sequence obeys a rather simple algebra, but we shall not use that algebra here, since it would  make this paper even longer than it is without adding much of present use. For this paper, we shall need  only the first few terms of $\na_{\a}$ as expressed in Table \ref{specseqsum}.
The operator $d_1$ has the form
\be
d_1 =
\P_1 
\lt \{
C^{\a} \na_{\a} + {\ov C}^{\dot \b}  {\ov \na}_{\dot \b} 
\rt \}
\P_1
\ee
where
\be
\na_{\a}
\equiv 
 \int d^4 x \;
\lt \{
 {  \y}^{i}_{  \a} 
\fr{\d}{\d A^{i} } 
+
 \pa_{\a \dot  \b} \A_i
\fr{\d}{\d {\oy}_{i \dot \b} } 
\rt \}
\ee

The equations resulting from this nilpotent operator can be separated into the Regular part, which is complicated but soluble, and the Irregular part which is very complicated and unsolved.
  Unfortunately, the two are quite mixed up as is expressed in Table \ref{d1d1dagtable}.  Moreover the most complicated bits are the ones with low dimensions that are our natural interest at first. 
This complexity is the reason that we adopt the more pedestrian technique used in Chapter \ref{diagramchasechapter}
 for the simple cases. 

\subsection{ The Translation Operator $\d_2$ }
\be
\d_{2 }  =\d_{ {\rm Translation}}  =  \x^{\a \dot \b}  \pa_{\a \dot \b}
\ee

Next we need to deal with the rather complicated operator $d_2$ which turns out to have the form:
\be
d_2 = \P_1 \lt ( \d_{2, {\rm Translation}} -
\d_{1, {\rm Super}}
\fr {\d_{\rm Struc}^{\dag}}{\D_{\rm Struc} }
\d_{1, {\rm Super}}
\rt )
\P_2
\ee

Its action on the various parts of Equation  (\ref{E1eq}) is summarized in Tables \ref{d2table}
 and  
\ref{d2dagtable}. It appears again briefly in some of the 
Tables in Chapter 
\ref{diagramchasechapter}.

\subsection{ The $\d_{\X}$ and $\d_{\rm 
gaugino}$ Operators }

\la{deltaxisubsection}

Supersymmetric gauge theory involves the operator $d_{\rm gaugino}$  referred to in Table \ref{specseqsum}.  We shall return to it in a future paper on supergauge cohomology.  The point of including references to gauge theory here is that they add very little to the complication in the 
Tables of this Chapter. However the formulation of the gauge theory is quite involved. 

 The operator $\d_{\X}$
 removes some very specific items from $E_3$. 
We do not need to worry about it for the results in this paper.  It has the form
\[
d_{\X}= \P_3 \d_{\X} \P_3 =
\]
\be
 \P_3
( C^{\a}  \x_{\a \dot \b}
 {\ov C}^{\dot \b} )
 \lt \{ {\ov A}_i    \pa_{\g \dot \d} A^i  -  \pa_{\g \dot \d} {\ov A}_i    A^i - \y^i_{\g  } {\ov \y}_{i  \dot \d}  
\rt \} \fr{\pa} X_{\g \dot \d}^{\dag}\P_3 
 \ee
 It  performs the following mapping from the ghost charge minus two and minus one sectors, namely
\[ X_{\g \dot \d} \in E_3
\stackrel{d_3}{\lra } 
\]
\be
( C^{\a}  \x_{\a \dot \b}
 {\ov C}^{\dot \b} )
 \lt \{ {\ov A}_i    \pa_{\g \dot \d} A^i  -  \pa_{\g \dot \d} {\ov A}_i    A^i - \y^i_{\g  } {\ov \y}_{i  \dot \d} \rt \}
\in E_3
\ee

\subsection{The Lie Algebra operator $\d_{\rm Gauge}$}
 This will appear when we include  Gauge Theory.  Separating out this operator is one of the motivations for the choice of grading that we have made. Its form is indicated in  Table \ref{specseqsum}. We will not use it in this paper.

\subsection{The Mass-Yukawa Operator $\d_{5, {\rm MY}}$}
\be
d_{5, {\rm MY}}  = 
\int d^4 x \;
 \lt (
m {\ov g}^{ij}  {\ov A}_{j} 
 + \og^{ijk}  {\ov A}_{j} \A_k
\rt )
 { C}_{ \a} 
\fr{\d}{\d { \y}^{i}_{\a} } 
+ *
\ee

This generates a number of constraints, which depend on the details of the   mass and Yukawa terms $g_{i}$ $g_{ij}$ and $g_{ijk}$.  

This operator is crucial to the present paper, and its effect is our main concern here. It gives rise to the Constraint Equations.

\subsection{The Operator $d_{6, {\rm MY}}$ }

This arises as follows:
\be
d_{6, {\rm MY}} = \P_6 \d_{5}  \fr{\d_{0 }^{\dag}}{\D_{0 } } \d_1 \P_6 + * 
\approx 
\ee
\be
\P_6 ( C \x {\ov C}^{\dag} ) 
\lt \{
m  {\ov g}^{ij} {\ov A}_j  
+  {\ov g}^{ijk} {\ov A}_j {\ov A}_k
\rt \}
 A^{i \dag} \P_6 + *
\ee
This operator performs the following kinds of mappings between vectors in the   space $E_6$:
\be
(C \x {\ov C} ) A^i \stackrel{d_{6, {\rm MY}}}{\lra} 
(C \x^2 C) \lt \{
m  {\ov g}^{ij} {\ov A}_j 
+  {\ov g}^{ijk} {\ov A}_j {\ov A}_k
\rt \}
\ee
\be
(C \x {\ov C} ) t^k_i A^i {\ov A}_k \stackrel{d_{6, {\rm MY}}}{\lra} 
\ee
\be
  t^k_i  
(C \x^2 C)  \lt \{
m  {\ov g}^{ij} {\ov A}_j 
+  {\ov g}^{ijk} {\ov A}_j {\ov A}_k
\rt \}
 {\ov A}_k 
\ee
\be
+ t^k_i  
({\ov C} \x^2 {\ov C} )  
   A^i 
\lt (
m g_{kl} A^l 
+ g_{klq} A^l A^q
\rt )
\ee
The matrix here is hermitian, which means that:
\be
 (t^k_i)^*  \equiv  {\ov t}^i_k  =  t_k^i  
\ee

We shall also frequently use this operator in this paper.

\subsection{The Operators $d_{7}$ }
The  operator $d_7$ will be discussed in Chapter 
\ref{insertionchapter}.

\begin{table}[hptb]
\caption{\large Terms   in $E_1$.   $E_1$ is the sum of all the Terms shown as in Equation \ref{E1eq} . The Terms are arbitrary \underline{Lorentz scalar} functions of all the `Physical' Fields and the $\f$ Sources  except that the $C$, $\x_{\a \dot \b} $ and ${\ov C}$  dependence is restricted as shown in Equation \ref{E1eq}.    The $\f$ sources are included to absorb and symmetrize uncontracted spinor indices. }

\vspace{.1in}
\framebox{ 
{\large $\begin{array}{llll}  

 \mbox{Terms in } E_1 & \mbox{ Properties of Terms}    \\   

{\cal P}_{n }, n \geq 0  & N_c {\cal P}_{n } =n {\cal P}_{n }
   \\

{\ov {\cal P}}_{n }& N_c {\ov {\cal P}}_{n } =0
   \\

{\cal Q}_n = ( \x C)_{\dot \b} {\cal Q}^{\dot \b}_{n }, n \geq 0& N_c {\cal Q}^{\dot \b}_{n } =n {\cal Q}^{\dot \b}_{n }  \\

{\ov {\cal Q}}

_n = ( \x {\ov C})_{  \b} {\ov {\cal Q}}^{ \b}_{n }& N_c {\ov {\cal Q}}^{ \b}_{n } =0 
\\

( C \x {\ov C}){\cal S}_{0 }& N_c {\cal S}_{0 } =0  \\

& {\cal S}_{0 }= {\ov {\cal S}}_{0 }  &  \\

( C \x^2 C) {\cal R}_{n }, n \geq 0& N_c {\cal R}_{n } =n {\cal R}_{n } \\

( {\ov C} \x^2 {\ov C}) {\ov {\cal R}}_{n }& N_c {\ov {\cal R}}_{n } = 0  \\

\end{array}$ }}
\la{E1Table}
\end{table}

\begin{table}[hptb]
\caption{  Equations from $d_1$ and $d_1^{\dag}$ }

\vspace{.1in}
\framebox{ 
{\large $\begin{array}{llllll}  

\la{d1d1dagtable}
  {\rm Term} &   \mbox{First Equation}&  \mbox{Second Equation}&  \mbox{\rm Reg or Irreg?}     \\
 {\cal P}_0 & \na_{\a} {\cal P}_0 
& {\ov \na}_{\dot \a} {\cal P}_0 = 0
& \mbox{\rm Irregular}   \\
 {\cal P}_1  \equiv {\cal P}_{\a} C^{\a} &  \na_{ ( \a} {\cal P}_{\b )} =0 & \na_{\a}^{\dag} {\cal P}_{\a} + 
 {\ov \na}_{\dot \a}^{\dag} {\ov {\cal P}}_{\dot \a} = 0 
& \mbox{\rm Irregular}   \\
{\cal P}_{n \geq 2}  & ( C \na ) {\cal P}_{n} =0 & ( C \na)^{\dag} {\cal P}_{n} = 0  
& \mbox{\rm Regular}   \\
{\cal Q}_{0}^{\dot \b}  &    \na_{\a} {\ov {\cal Q}}^{\a} 
+
  {\ov \na}_{\dot \b} Q^{\dot \b} =0
&    \na_{\a} {  {\cal Q}}_{\dot \b} = 0
& \mbox{\rm Irregular} \\
{\cal Q}_{1}^{\dot \b}, {\cal S}_0  &    \na_{(\a} { {\cal Q}}_{\b)}^{\dot \g} =0
&    \na_{\a}^{\dag} {\cal S}_0 +  {\ov \na}_{\dot \b}^{\dag} {\ov {\cal Q}}^{\a }_{\dot \b} = 0 
& \mbox{\rm Irregular}    \\
{\cal Q}_{n \geq 2}^{\dot \b}  & ( C \na ) {\cal Q}_{n}^{\dot \b} =0 & ( C \na)^{\dag} {\cal Q}_{n}^{\dot \b} = 0 
& \mbox{\rm Regular}    \\

{\cal R}_{0}  &   \na_{\a} {\cal R}_{0} =0 &    
& \mbox{\rm Irregular}   \\

{\cal R}_{n \geq 1}  & ( C \na ) {\cal R}_{n} =0 & ( C \na)^{\dag} {\cal R}_{n} = 0   
& \mbox{\rm Regular} \\

\end{array}$ }}
\end{table}

\begin{table}[hptb]
\caption{ Equations from $d_2$   }
\vspace{.1in}
\framebox{ 
{\large $\begin{array}{llllll}  

\la{d2table}
 {\rm Term} &  d_2  \mbox{ Equation}&  \mbox{ Maps To }
&\mbox{\rm Reg or Irreg?}    \\   
{\cal P}_1 & \lt ( \pa_{  \a \dot \b } + {  \na}_{  \a} {\ov \na}_{\dot \b}\rt )
{\cal P}^{\a }  + (\na)^2 {\ov {\cal P}}_{\dot \b} =0  & {\cal Q}_{0,\dot \b}   
& \mbox{\rm Irregular}    \\

{\cal P}_{n \geq 2}  &  C_{\a}^{\dag} \pa_{  \a \dot \b }  {\cal P}_{n } =0  & {\cal Q}_{n-1,\dot \b}    
& \mbox{\rm Regular}     \\

{\cal S}_0, {\cal Q}_1 &             (\na)^2 {\cal S}_0 + \lt ( \pa + \na {\ov \na} \rt )_{\a \dot \b}  {\cal Q}_1^{\a \dot \b}   =0
  &  {\cal R}_{0}  
& \mbox{\rm Irregular}    \\

{\cal Q}_{n \geq 2}  &  C_{\a}^{\dag} \pa_{  \a \dot \b }  {\cal Q}^{\dot \b}_{n} =0  &  {\cal R}_{n-1}   
& \mbox{\rm Regular}    \\

\end{array}$ }}
\end{table}

\begin{table}[hptb]
\caption{ Equations from  $d_2^{\dag}$  }
\vspace{.1in}
\framebox{ 
{\large $\begin{array}{llllll}  
\la{d2dagtable}
 {\rm Term} &    \mbox{First Equation} &  \mbox{ Maps To } 
& \mbox{\rm Reg or Irreg?}       \\   

{\cal Q}_{0,\dot \b}  &     C_{\a} \pa_{  \a \dot \b }^{\dag}   {\cal Q}_{0,\dot \b} =0         
  & {\cal P}_1   & \mbox{\rm Irregular}      \\

{\cal Q}_{1,\dot \b}  &     C_{\a} \pa_{  \a \dot \b }^{\dag}   {\cal Q}_{1,\dot \b} =0         
  & {\cal P}_2    & \mbox{\rm Irregular}     \\

{\cal Q}_{n \geq 2,\dot \b}  &     C_{\a} \pa_{  \a \dot \b }^{\dag}   {\cal Q}_{n,\dot \b} =0         
  & {\cal P}_{n+1}  & \mbox{\rm Regular}     \\

{\cal R}_0 &      \pa_{  \a \dot \b }^{\dag}   {\cal R}_{0 } =0  &  
{\cal Q}_1 & \mbox{\rm Irregular}    \\

{\cal R}_0 &      (\na)^{2 \dag} {\cal R}_0 + {\ov \na}^{2 \dag} {\ov {\cal R}}_0= 0         &  {\cal S}_0 & \mbox{\rm Irregular}    \\

{\cal R}_{n \geq 1}  &   C_{\a}^{\dag} \pa_{  \a \dot \b }^{\dag}  {\cal R}_{n } =0  & {\cal Q}_{n+1 }  & \mbox{\rm Regular}    \\

\end{array}$ }}
\end{table}


\chapter{Diagram Chasing for Six Easy Examples}

{\Large
\la{diagramchasechapter}

\section{Diagram Chasing}

It is simplest to think of the action as an object with Form number four, and potential anomalies as objects with Form number five.  Form number can be defined by
\be
N_{\rm Form} =  N_{\x} + (N_c  - N_Y - N_{\G} +*)
\ee
and the ghost number is
\be
N_{\rm Ghost } =  N_{\rm Form } -4
\ee

We shall now look at the mappings from $d_r$ for the six easiest cases. Table \ref{specseqsum} will be used to determine the cohomology by elimination at each order of $E_r$.

 The first three examples are scalars and the last three examples are spinors.

This is a tricky business.  Ideally one would have a general solution and deduce the results here in a systematic way.  That is not easy.

The diagram chasing method here is fairly straightforward.
 But it is also mechanical and laborious.  It has the advantage that no general solution of the problem is required to work out these low dimensional examples.  The general solution will be easier to find after one understands the simple examples of course. So that is a hope for the future. In Chapter \ref{remarkschapter}
 we will return to some general remarks about the solutions.

\section{Diagram Chasing Rules}

Here are the rules for diagram chasing:
\ben
\item
Write down all possible terms in $E_1$ with a given dimension and other quantum numbers.   The operators $\d$  and $d_r$ conserve the same quantum numbers as the action. These quantum numbers and the relevant fields are in Tables  \ref{Abbreviations} and \ref{PhysFieldsTable}. 
\item
Because we restrict ourselves to theories which do not have ZMT (See Section \ref{isolampsection}), we can, in principle, ignore all terms that do not have fields in them, and all terms that have only one field in them.  
\item
However, we will analyze some of these cases because they illustrate principles in a simple way, and they are also interesting to use for the construction of the relevant local theory in which a constant is made into a field as discussed in Section \ref{isolampsection}.
  
\item
The structure of $E_1$ is well defined from Equation (\ref{E1eq}), together with Table \ref{PhysFieldsTable}.
\item
Apply the operators $d_r, r \geq 1$ to these in the order $d_1,d_2\cdots$. These operators are contained in Table  \ref{specseqsum}. If $d_r$ connects two terms, then neither of them survive to $E_{r+1}= \ker d_r \cap \ker d_r^{\dag} $. 
\item
Once all possible $d_r$ have been used, the terms that are left are in $E_{\infty}$.
\item
Using the quantum numbers, one then looks for a mapping to ${\cal H}_{\rm Fock}$, which is isomorphic to $E_{\infty}$. Again, this mapping preserves the quantum numbers that are conserved by the action and $\d$. Then one looks for a mapping  ${\cal H}_{\rm Fock} \ra {\cal H}_{\rm Integral}$, taking into account that these spaces are isomorphic as discussed above in Section \ref{isolampsection}.
\item
In practice it is difficult, at least in a complicated case like the present one, to be sure that one has not missed a $d_r$.  The result for ${\cal H}$ needs to be checked by hand and in this way one gradually comes to understand the intricacies of the problem, and finds all the $d_r$.  This is how Table \ref{specseqsum} was constructed. I believe it is now correct but it is too early to try to construct a proof of that.

\item
For the present the use of the spectral sequence is more of an art than a science for this problem. However it seems to be a useful art.  The present problem, at least at the present state of understanding,  could not be resolved without it, and that is why it is used in this paper.  Eventually it should be possible to establish the complete cohomology using it.  Perhaps there is a simpler method analogous to superspace yet to be found.
\een

\section{Scalar Cohomology}

\la{ScalarCohomologySection}

\subsection{Scalar Cohomology with Dimension  $-2$}

This is the simplest case of diagram chasing.  The scenario is summarized in Tables 
\ref{minustwoscalars}
and \ref{minustwoscalars2}.

Table 
\ref{minustwoscalars} summarizes the effects of all possible
 $d_r$ on Scalars in $E_r$ that have ${\rm Dimension}  = -2$.
These have  $N_{\rm Form} = 4,5$ as shown.  This Table  shows that   $(C \x^2 C)  {\ov k}^{i}   \A_i $ survives to $E_{\infty}$.

To understand how things work, note, for example, that 
\[ d_1 ( C \x^2 C)   k_i  A^i
=
\P_1( C \na + \oC {\ov \na}) (C \x^2 C)   k_i  A^i
\]
\be
  = 
(C \x^2 C)   k_i  (C \y^i)
\ee
and that
\be \P_1 ( C \na) ( C \x^2 C)   \ok^i  \A_i
  = 
0
\ee
Note also that 
\be
\P_1(   \oC {\ov \na}) (C \x^2 C)   \ok^i  \A_i
=
\P_1  (C \x^2 C)   \ok^i  (\oC \oy_i)
=0
\ee
because there are no terms of the form
\be 
 \oC_{\dot \b}   (C \x^2 C)  
\ee
in $E_1$. 

These follow from the form of $E_1$ in Equation (\ref{E1eq}) and the form of $\na$ and $\ov \na$.  The projection operator $\P_1$ projects the result back onto $E_1$, which is really quite a restricted form.

  Table   \ref{minustwoscalars2} shows the mapping to the cohomology space ${\cal H}_{\rm Fock}$.
We can map this further onto  ${\cal H}_{\rm LIP}$ because of the discussion in Section \ref{isolampsection}.  However the linear terms are ZMT, and we shall return to them later.

\begin{table}[hpbt]
\caption{\Large First Table for Scalars with ${\rm Dimension}  = -2$:   
  $N_{\rm Form}=3,4, 5$.  }
\la{minustwoscalars}
\vspace{.1in}
\framebox{ 
{ \Large $\begin{array}{ccccccc}  
\\N_{\rm Form} = 4 &&
N_{\rm Form} =  5 
\\
(C \x^2 C)   k_i  A^i
  & \stackrel{d_1}{\lra} &  
(C \x^2 C)   k_i  (C \y^i)
\\ 
(C \x^2 C)    {\ov k}^i
 \A_i
   \in 
E_{\infty} 
\\ 

\end{array}$ } }
\end{table}

\begin{table}[hptb]
\caption{\Large Second Table for Scalars with ${\rm Dimension}  = -2$:   The Mapping   $E_{\infty} \leftrightarrow {\cal H}$.}
\vspace{.1in}
\framebox{ 
{ \Large $\begin{array}{ccccccc}  
\la{minustwoscalars2}
\\
E_{\infty} 
&&  {\cal H}_{\rm Fock} 
&&  {\cal H}_{\rm LIP} 
\\
\\
 (C \x^2 C)    {\ov k}^i
 \A_i 
& \Lra 
&
(\x^4) {\ov k}^i
 \oF_{1,i} + \cdots
& \Lra 
&

\int d^4 x \; {\ov k}^i
 \oF_{1,i} 
\\

\end{array}$ } }
\end{table}

\subsection{Scalar Cohomology with Dimension  $-1$}

This is the next to simplest case of diagram chasing.  
The scenario is summarized in Tables 
\ref{minusonescalars3},
\ref{minusonescalars1},
\ref{minusonescalars2},
\ref{minusonescalars4},
\ref{minusonescalars5} and \ref{minusonescalars6}.
Tables 
\ref{minusonescalars1} and \ref{minusonescalars2} eliminate some objects at higher $N_{\rm Ghost}$ in a way that often repeats in these Tables. 

Table 
\ref{minusonescalars3}  shows that 
\ben
\item
  $(C \x \oC) e_i A^i$  survives to $E_{\infty}$  only if it satisfies $
e_i        \og^{ij}  =
e_i \og^{ijk} =0 $,
 and that 
\item
$( C x^2 C) \lt \{
{\ov k}^{i} m  \A_i
+
{\ov k}^{ij}  \A_j  \A_i
\rt \} $ survives to $E_{\infty}$ only if the coefficients  satisfy $  k_i \og^{ik}  + 2
k_{ij} \og^{ijk} =0 $.
\een
These equations are listed in Table \ref{minusonescalars4}. 
Then Table \ref{minusonescalars5} shows the correspondence between  $E_{\infty}$ and  ${\cal H}_{\rm Fock}$, and Table \ref{minusonescalars6} shows the correspondence between 
 ${\cal H}_{\rm Fock}$ and  ${\cal H}_{\rm LIP}$.  The coefficients in ${\cal H}_{\rm LIP}$ must also satisfy the constraints in Table \ref{minusonescalars4}.

This case serves as a paradigm for the more complicated cases, because it shows how constraints arise in this simple case.

\begin{table}[hptb]
\caption{\Large First Table for Scalars with ${\rm Dimension}  = -1$:   
Mappings of $N_{\rm Form}=3 \leftrightarrow N_{\rm Form} =  4$.  }
\la{minusonescalars3}
\vspace{.1in}
\framebox{ 
{\Large $\begin{array}{ccccccccc}  
\\
N_{\rm Form} = 3 &&
N_{\rm Form} = 4  \\

\\

(C \x \oC)  e_i   A^i  
&   \stackrel{d_6}{\lra} &
(C \x^2 C)e_i   \lt \{
  m  \og^{ij} \A_j
+
\og^{ijk}   \A_j  \A_k
\rt \}
\\

\begin{array}{c} (C \x \oC)     A^k 
\\
\lt \{
 g_{ik} {\ov k}^{i}
+
2 g_{ijk} {\ov k}^{ij}
\rt \}
\\
\end{array}  
& \stackrel{d_6^{\dag}}{\longleftarrow} & 
\begin{array}{c}  
(C \x^2 C)  \\
\lt \{
m {\ov k}^i
 \A_i +
 {\ov k}^{ij}\A_j  \A_i
\rt \}
\\
\end{array}  
\\ 
\\

(C \x \oC)   \ove^j  \A_i  
&  \stackrel{d_6}{\lra} &
(\oC \x^2 \oC) 
\ove^j \lt \{
m g_{ij} A^j + 
 g_{ijk} A^j A^k 
\rt \}
\\

\\
\end{array}$ } }
\end{table}

\begin{table}[hptb]
\caption{\Large Second Table for Scalars with ${\rm Dimension}  = -1$:   
Mappings of $N_{\rm Form}=4 \leftrightarrow N_{\rm Form} =  5$.  }
\la{minusonescalars1}
\vspace{.1in}
\framebox{ 
{\Large $\begin{array}{ccccccccc}  
\\
N_{\rm Form} = 4 &&
N_{\rm Form} =  5 
\\

(C \x^2 C)   m  A^i
  & \stackrel{d_1}{\lra} &  
(C \x^2 C)   m  (C \y^i)
\\ 
(C \x^2 C) f_{(ik)}  A^k  A^i
  & \stackrel{d_1}{\lra} &  
(C \x^2 C) f_{(ik)} A^k  (C \y^i)
\\

(C \x^2 C)   \A_j  A^i
 & \stackrel{d_1}{\lra} &  
(C \x^2 C)  \A_j  (C \y^i)
\\ \\

\\
\end{array}$ } }
\end{table}

\begin{table}[hptb]
\caption{\Large Third Table for Scalars with ${\rm Dimension}  = -1$:   
Mappings of $N_{\rm Form}=5 \leftrightarrow N_{\rm Form} =  6$.  }

\vspace{.1in}
\framebox{ 
{\Large $\begin{array}{ccccccc}  
\la{minusonescalars2}
\\
N_{\rm Form} = 5 &&
N_{\rm Form} = 6 
\\
(C \x^2 C) f_{[ik]} ( C \y^i) A^k
 & \stackrel{d_1}{\lra}  &
(C \x^2 C) f_{[ik]} 
\lt \{
( C \y^i)  ( C \y^k)
\rt \}
\\

\end{array}$ } }
\end{table}

\begin{table}[hptb]
\caption{\Large Fourth Table for Scalars with ${\rm Dimension}  = -1$:   
Equations arising from Table \ref{minusonescalars3} }
\vspace{.1in}
\framebox{ 
{ \Large $\begin{array}{ccccccc}  
\la{minusonescalars4}
\\
e_i        \og^{ij} 
&=& 0
\\
e_i \og^{ijk} 
&=& 0
\\
 k_i \og^{ik}  + 2
k_{ij} \og^{ijk} 
&=& 0
\\

\end{array}$ } }
\end{table}

\begin{table}[phtb]
\caption{\Large Fifth Table for Scalars with ${\rm Dimension}  = -1$:   
Mappings of $E_{\infty} \ra {\cal H}$. These are subject to the constraints in Table \ref{minusonescalars4} }

\vspace{.1in}
\framebox{ 
{ \Large $\begin{array}{ccccccc}  
\la{minusonescalars5}
\\
E_{\infty} &&  {\cal H}_{\rm Fock} \\
\\
 (C \x^2 C) \lt \{
  {\ov k}^i
m \A_i + {\ov k}^{ij}
\A_i \A_j \rt \} & \Lra &
(\x^4)
\;  \lt \{
 m {\ov k}^i \oF_{1,i}  +
{\ov k}^{ij}
\oF_{2,ij}  \rt \} 
+ \cdots
\\ 
(C \x \oC)  e_i   A^i  
& \Lra &
(\x^4)
e_i  \oG^i 
+\cdots
\\

\end{array}$ } }
\end{table}

\begin{table}[phtb]
\caption{\Large Sixth Table for Scalars with ${\rm Dimension}  = -1$:   Mappings of ${\cal H}_{\rm Fock} \ra {\cal H}_{\rm LIP}$. These are subject to the constraints in Table \ref{minusonescalars4} }

\vspace{.1in}
\framebox{ 
{ \Large $\begin{array}{ccccccc}  
\la{minusonescalars6}
\\
 {\cal H}_{\rm Fock} 
&&
  {\cal H}_{\rm LIP} \\
\\
(\x^4)
\;  \lt \{
 m {\ov k}^i \oF_{1,i}  +
{\ov k}^{ij}
\oF_{2,ij}  \rt \} 
+ \cdots
& \Lra &
\int d^4 x \;
\;  \lt \{
 m {\ov k}^i \oF_{1,i}  +
{\ov k}^{ij}
\oF_{2,ij}  \rt \} 
\\ 
(\x^4)
e_i  \oG^i 
+\cdots
& \Lra &
\int d^4 x \;
e_i  \oG^i 
\\

\end{array}$ } }
\end{table}

\subsection{Scalars with ${\rm Dimension}  = 0$  }

\la{subseczero}
 The Scalars with ${\rm Dimension}  = 0$  are the next case  of diagram chasing that we will consider. There are now a substantial number of possible objects to consider. The situation  is summarized in seven tables: Tables 
\ref{zeroscalars1}, \ref{zeroscalars4}, \ref{zeroscalars2},
\ref{zeroscalars3}, \ref{zeroscalars5}, \ref{zeroscalars6} and \ref{zeroscalars7}.  

Tables \ref{zeroscalars1}, \ref{zeroscalars2} and \ref{zeroscalars3} 
have a simple effect.  All the objects in them are eliminated by the action of $d_1$ .  
Table \ref{zeroscalars4} is more complicated.  It eliminates the first objects with $d_2$.  For the rest it  gives rise to the constraints listed in Table \ref{zeroscalars5}.  The sixth Table   \ref{zeroscalars6}  shows the correspondence between  $E_{\infty}$ and  ${\cal H}_{\rm Fock}$, and the seventh Table   \ref{zeroscalars7}  shows the correspondence between  ${\cal H}_{\rm Fock}$ and  ${\cal H}_{\rm LIP}$.  The objects in ${\cal H}_{\rm LIP}$ and ${\cal H}_{\rm Fock}$ must satisfy the constraints on the coefficients mentioned in  Table \ref{zeroscalars5}.

\begin{table}[hptb]
\caption{\Large First Table for Scalars with ${\rm Dimension}  = 0$:   
Mappings of $N_{\rm Form}=2 \leftrightarrow N_{\rm Form} =  3$.  }

\vspace{.1in}
\framebox{ 
{\Large  $\begin{array}{ccccccc}  
\la{zeroscalars1}
\\
N_{\rm Form} = 2 &&
N_{\rm Form} = 3 
\\
(\y^i \x \oC)      &\stackrel{d_1}{\lra} &  
(\x \oC)^{\a} \oC^{\dot \b} A^i_{\a \dot \b}   
\\
(C \x \oy_i)       &\stackrel{d_1}{\lra} &  
(\x C)^{\dot \b} C^{\a} \A_{i \dot \b \a}   
\\

\\
\end{array}$ } }
\end{table}

\begin{table}[hptb]
\caption{\Large Second Table for Scalars with ${\rm Dimension}  = 0$:   
Mappings of $N_{\rm Form}=3 \leftrightarrow N_{\rm Form} =  4$.  }
\Large

\vspace{.1in}
\framebox{ 
{   $\begin{array}{ccccccc}  
\la{zeroscalars4}
\\
 N_{\rm Form} = 3 
&&
N_{\rm Form} =   4
\\

(C \x \oC)  f_{ij} A^j A^i  
&   \stackrel{d_2}{\lra} &
(C \x^2 C)  f_{ij} (\y^i \y^j)
\\

(C \x \oC)   \of^{ij}   \A_i \A_j  
&  \stackrel{d_2}{\lra} &
(\oC \x^2 \oC) 
 \of^{ij} (\oy_i \oy_j)
\\
\\
\begin{array}{c}
(C \x \oC)
\\
\lt \{
  e_i  m A^i 
+ \ove^j  m \A_i   
+  e_i^j  A^i   \A_j  
\rt \}
\end{array}
&   \stackrel{d_6}{\lra} &
\lt \{
\begin{array}{c}
(C \x^2 C)e_i   m
\\
\lt \{
  m  \og^{ij} \A_j
+
\og^{ijk}   \A_j  \A_k
\rt \}
\\+
(\oC \x^2 \oC) 
\ove^i m 
\\
\lt \{
m g_{ij} A^j + 
 g_{ijk} A^j A^k 
\rt \}
\\
+(\oC \x^2 \oC) 
e_i^j  A^i   
\\
\lt \{
m g_{jk} A^k + 
 g_{jkl} A^k A^l 
\rt \}
\\
+
(C \x^2 C)  
 e_i^j   \A_j 
\\
\lt \{
  m  \og^{ik} \A_k
+
\og^{ikl}   \A_k  \A_l
\rt \}
\end{array}
\rt.
\\
\\
\lt.
\begin{array}{c}
(C \x \oC) A^i
\\
\lt \{
m \lt (
 2 \ok^j g_{ij} 
+
 2 \ok^{jk} g_{ijk} 
\rt )  
\rt.
\\
+
\lt.
 \lt (
  2  \ok^{jq} g_{ij} 
+
6   \ok^{jkq} g_{ijk} 
\rt )  \A_q \rt \}
\\
+*
\\
\end{array}
\rt \}
&   \stackrel{d_6^{\dag}}{\lla} &

\begin{array}{c}
(C \x^2 C) 
\lt \{
 m^2 \ok^i \A_i
\rt.
\\
\lt.
 + m \ok^{ij}\A_i \A_j
 +  \ok^{ijk}\A_i \A_j
\A_k
\rt \}
\\
+
(\oC \x^2 \oC)
\lt \{
  m^2 k_i A^i
\rt.
\\
\lt.
 + m k_{ij} A^i A^j 
 + m k_{ijk} A^i A^j A^k
\rt \}
\end{array}
\\
\\
\end{array}$ } }
\end{table}

\begin{table}[hptb]
\caption{\Large Third Table for Scalars with ${\rm Dimension}  = 0$:   
Mappings of $N_{\rm Form}=4 \leftrightarrow N_{\rm Form} =  5$.  }

\vspace{.1in}
\framebox{ 
{\Large   $\begin{array}{ccccccc}  
\la{zeroscalars2}
\\
N_{\rm Form} = 4 &&
N_{\rm Form} =  5 
\\

(C \x^2 C)   m^2  A^i 
&\stackrel{d_1}{\lra}   
&(C \x^2 C)   m^2  (C \y^i) 

\\ 
(C \x^2 C)  m A^k  A^i
&\stackrel{d_1}{\lra}   
&(C \x^2 C)    m A^k   (C \y^i) 

\\ 
(C \x^2 C)  A^l A^k  A^i
&\stackrel{d_1}{\lra}   
&(C \x^2 C)    A^l A^k   (C \y^i) 

\\
(C \x^2 C)  m \A_j  A^i
&\stackrel{d_1}{\lra}   
&(C \x^2 C)    m \A_j (C \y^i) 

\\
(C \x^2 C)  A^l \A_j  A^i
&\stackrel{d_1}{\lra}   
&(C \x^2 C)    A^l \A_j (C \y^i) 

\\
(C \x^2 C)  \A_l \A_j  A^i
&\stackrel{d_1}{\lra}   
&(C \x^2 C)    \A_l \A_j (C \y^i) 
\\

(C \x^2 C)  (\oy_i \oy_j )
&\stackrel{d_1}{\lra}   
&(C \x^2 C)    \oy_i^{\dot \b} \A_{j \dot \b \g} C^{\g}
 \\

\end{array}$ } }
\end{table}

\begin{table}[hptb]
\caption{\Large Fourth Table for Scalars with ${\rm Dimension}  = 0$:   
Mappings of $N_{\rm Form}=5 \leftrightarrow N_{\rm Form} =  6$.  }

\vspace{.1in}
\framebox{ 
{\Large $\begin{array}{ccccccc}  
\la{zeroscalars3}
\\
N_{\rm Form} = 5 &&
N_{\rm Form} = 6 
\\
(C \x^2 C) f_{[ijk]} ( C \y^i)( C \y^j) A^k
 & \stackrel{d_1}{\lra}  &
(C \x^2 C) f_{[ijk]} 
\lt \{
( C \y^i)( C \y^j) ( C \y^k)
\rt \}
\\

(C \x^2 C) \of^{[ij]} ( C \pa \A_i \oy_j) 
 & \stackrel{d_1}{\lra}  &
(C \x^2 C) \of^{[ij]} ( C \pa \A_i \pa \A_j C) 
\\

\end{array}$ } }
\end{table}

\begin{table}[hptb]
\caption{\Large Fifth Table for Scalars with ${\rm Dimension}  = 0$:   
Equations arising from Table \ref{zeroscalars4} }

\vspace{.1in}
\framebox{ 
{\Large  $\begin{array}{ccc}  
\la{zeroscalars5}
\\

\\
e_i  \og^{ij} 
&=&0
\\
e_i  \og^{ijk}  
+
 e_i^{(j} 
 \og^{k)i} 
&=&0
\\
e_{(i}^q  
 g_{kl)q}
&=&0
\\
\\
 \lt (
 \ok^j g_{ij} 
+
  \ok^{jk} g_{ijk} 
\rt )  
&=&0

\\
 \lt (
2  \ok^{jq} g_{ij} 
+
6   \ok^{jkq} g_{ijk} 
+*
\rt ) 
&=&0
\\

\end{array}$ } }
\end{table}

\begin{table}[hptb]
\caption{\Large Sixth Table for Scalars with ${\rm Dimension}  = 0$:  
Mapping $E_{\infty} \ra {\cal H}_{\rm Fock}$. These are subject to the constraints in Table  \ref{zeroscalars5} }

\vspace{.1in}
\framebox{ 
{ \Large $\begin{array}{ccccccc}  
\la{zeroscalars6}
\\
E_{\infty} &&  {\cal H}_{\rm Fock}\\
\\
\begin{array}{c}
 (C \x^2 C) \lt \{
 {\ov k}^i
m^2 \A_i 
\rt.
\\
\lt.
+ m{\ov k}^{ij}
\A_i \A_j 
+ {\ov k}^{ijk}
\A_i \A_j \A_k
\rt \} 
\end{array}
& \Lra &
\begin{array}{c}
(\x^4)
 \;  \lt \{
m^2 {\ov k}^i \oF_{1,i}  
\rt.
\\
\lt.
+
m {\ov k}^{ij}
\oF_{2,ij} 
  +
{\ov k}^{ijk}
\oF_{3,ijk}  \rt \} 
\\
+
\cdots
\end{array}
\\ 
\\
\begin{array}{c}
(C \x \oC) \lt \{
 e_i  m A^i  
+ \ove^i  m \A_i  
\rt.
\\
\lt.
+ e_i^j   A^i \A_j  
\rt \}
\end{array}
& \Lra &
\begin{array}{c}
(\x^4)
\lt \{
 e_i m \oG^i -  m
\ove^i \G_i 
\rt.
\\
+  e_i^j \lt (
\oG^i \A_j 
-
\G_j A^i 
\rt.
\\
\lt.
\lt.
-
Y^{\a}_j \y^{i \a} 
+
\oY^{i \dot \a} \oy_{j}^{\dot \a} 
 \rt )
\rt \}
\\
+\cdots
\end{array}

\\
\end{array}$ } }
\end{table}

\begin{table}[hptb]
\caption{\Large Seventh Table for Scalars with ${\rm Dimension}  = 0$:  
Mapping ${\cal H}_{\rm Fock} \ra {\cal H}_{\rm LIP}$. These are subject to the constraints in Table  \ref{zeroscalars5} }

\vspace{.1in}
\framebox{ 
{ \Large $\begin{array}{ccccccc}  
\la{zeroscalars7}
\\
 {\cal H}_{\rm Fock}
&& 
 {\cal H}_{\rm LIP}\\
\\
\begin{array}{c}
(\x^4) \;  \lt \{
m^2 {\ov k}^i \oF_{1,i}  
\rt.
\\
\lt.
+
m {\ov k}^{ij}
\oF_{2,ij} 
  +
{\ov k}^{ijk}
\oF_{3,ijk}  \rt \}
\\
+ \cdots 
\end{array} 
& \Lra &
\begin{array}{c}
\int d^4 x \;  \lt \{
m^2 {\ov k}^i \oF_{1,i}  
\rt.
\\
\lt.
+
m {\ov k}^{ij}
\oF_{2,ij} 
  +
{\ov k}^{ijk}
\oF_{3,ijk}  \rt \} 
\end{array}
\\ 
\\
\begin{array}{c}
(\x^4)
\lt \{
 e_i m \oG^i -  m
\ove^i \G_i 
\rt.
\\
\lt.
+  e_i^j \lt (
\oG^i \A_j 
-
\G_j A^i 
\rt.
\rt.
\\
\lt.
\lt.
-
Y^{\a}_j \y^{i \a} 
+
\oY^{i \dot \a} \oy_{j}^{\dot \a} 
 \rt )
\rt \}
\\
+ \cdots
\end{array}
& \Lra &
\begin{array}{c}
\int d^4 x \;  
\lt \{
 e_i m \oG^i -  m
\ove^i \G_i 
\rt.
\\
\lt.
+  e_i^j \lt (
\oG^i \A_j 
-
\G_j A^i 
\rt.
\rt.
\\
\lt.
\lt.
-
Y^{\a}_j \y^{i \a} 
+
\oY^{i \dot \a} \oy_{j}^{\dot \a} 
 \rt )
\rt \}
\end{array}
\\

\\
\end{array}$ } }
\end{table}

\section{ Spinor Cohomology }

Our next three examples depend on considering objects that have a  spinor index.  To keep track of the spinor index, we  contract the spinor index  with a spinor of $\f_{\a}$.  We will ignore the complex conjugate ${\ov \f}_{\dot \a}$ since information for it is easily constructed from the information below.

\la{SpinorCohomologySection}

\subsection{ Cohomology with Dimension Zero with one spinor index contracted with a spinor  $\f_{\a}$  Dimension $\fr{5}{2}$}

Here we  assume the Dimension of $\f_{\a}$  is $\fr{5}{2}$. The scenario is summarized in Tables 
\ref{fivehalfscalars1} and
\ref{fivehalfscalars2}. The first Table \ref{fivehalfscalars1},
 is quite simple. It summarizes all the action of the $d_1$.   The first object listed here is  eliminated by $d_1$.
  The second Table   \ref{fivehalfscalars2}  shows the correspondence between  $E_{\infty}$ and  
${\cal H}_{\rm LIP}$. Since it is simple and illustrated by the scalar examples, we skip the intermediate mapping to ${\cal H}_{\rm Fock}$.  There are no constraints.

\begin{table}[hptb]
\caption{\Large First Table for  spinors coupled to a constant source $\f_{\a}$ with  ${\rm Dimension} \; \f_{\a}  = \fr{5}{2}$:   
Mappings of $N_{\rm Form}=5 \leftrightarrow N_{\rm Form} =  6$.  }

\vspace{.1in}
\framebox{ 
{\Large $\begin{array}{ccccccc}  
\la{fivehalfscalars1}\\
N_{\rm Form} = 5 &&
N_{\rm Form} =  6 
\\

 (C \x^2 C) (\f C)
f_{j}  A^j 
 & \stackrel{d_1}{\lra} 
&
  (C \x^2 C)  (\f C)
f_{j} (C \y^j) 
\\

 (C \x^2 C) (\f C)
\oh^{j}  \A_j 
 & \in &
E_{\infty}  \\

\\
\end{array}$ } }
\end{table}

\begin{table}[hptb]
\caption{\Large Second Table for  spinors coupled to a constant source $\f_{\a}$ with  ${\rm Dimension} \; \f_{\a}  = \fr{5}{2}$:   
Mapping of   $E_{\infty} \ra {\cal H}_{\rm LIP}$.  There are no constraints. }

\vspace{.1in}
\framebox{ 
{ \Large $\begin{array}{ccccccc}  
\la{fivehalfscalars2}.
\\
E_{\infty} &&  {\cal H}_{\rm LIP}\\
\\
(C \x \oC) \f^{\a} C_{\a} \lt \{
  \oh^i   \A_i   
\rt \}
& \lra &
\int d^4 x \;  
\f^{\a} C_{\a}  \lt \{
 \oh^i   \oF_{1,i}  
\rt \}
\\

  \\

\end{array}$ } }
\end{table}

\subsection{ Cohomology with Dimension Zero with one spinor index contracted with a spinor of $\f_{\a}$  Dimension $\fr{3}{2}$}

Our next example also is based on objects that  have a spinor index  contracted with a spinor  $\f_{\a}$.  But now we now assume the Dimension of $\f_{\a}$  is $\fr{3}{2}$. The scenario is summarized in Tables 
\ref{threehalfscalars2},
\ref{threehalfscalars1},
\ref{threehalfscalars3} and 
\ref{threehalfscalars4}.  Table \ref{threehalfscalars1}
 is quite simple. It summarizes all the action of the $d_1$.   The objects listed here are all eliminated by $d_1$, as are the first objects in Table \ref{threehalfscalars2}.
Table \ref{threehalfscalars2} gives rise to the constraints listed in Table  \ref{threehalfscalars3}. The latter objects in it are not affected by $d_r$ for $r \neq 5$.  The fourth Table   \ref{threehalfscalars4}  shows the correspondence between  $E_{\infty}$ and  
${\cal H}_{
\rm LIP}$. Since it is simple and illustrated by the scalar examples, we skip the intermediate mapping to ${\cal H}_{\rm Fock}$.  The objects in ${\cal H}_{\rm LIP}$ must satisfy the constraints on the coefficients mentioned in   \ref{threehalfscalars3}.

\begin{table}[hptb]
\caption{\Large First Table for  spinors coupled to a constant source $\f_{\a}$ with  ${\rm Dimension} \; \f_{\a}  = \fr{3}{2}$:   
Mappings of $N_{\rm Form}=4 \leftrightarrow N_{\rm Form} =  5$.  }

\vspace{.1in}
\framebox{ 
{ \Large $\begin{array}{ccccccc}  
\la{threehalfscalars2}\\
N_{\rm Form} = 4 &&
N_{\rm Form} =  5 
\\

 (\oC \x^2 \oC)  (\f \y^i)
 & \stackrel{d_1}{\lra} 
&
 (\oC \x^2 \oC)  (\f \pa A^i \oC)
\\
 (C \x^2 C)  f_i (\f \y^i)
 & \stackrel{d_5}{\lra} 
&
\begin{array}{c} 
 (C \x^2 C)  (\f C) f_i
\;
\\
\lt \{ m {\og}^{ij} \A_j + {\og}^{ijk} \A_j \A_k \rt \}
\end{array}
\\

 (C \x^2 C) \lt (
  {\oh}^{j} g_{ij} + 2 {\oh}^{jk} g_{ijk}
\rt )
 (\f \y^i) & \stackrel{d_5^{\dag}}{\longleftarrow} 
&
\begin{array}{c} 
 (C \x^2 C)  (\f C)  
\;
\\
\lt \{ m {\oh}^{j} \A_j + {\oh}^{jk} \A_j \A_k \rt \}
\end{array}
\\

\\
\end{array}$ } }
\end{table}

\begin{table}[hptb]
\caption{\Large Second Table for  spinors coupled to a constant source $\f_{\a}$ with  ${\rm Dimension} \; \f_{\a}  = \fr{3}{2}$:   
Mappings of $N_{\rm Form}=5 \leftrightarrow N_{\rm Form} =  6$.  }

\vspace{.1in}
\framebox{ 
{\Large $\begin{array}{ccccccc}  
\la{threehalfscalars1}\\
N_{\rm Form} = 5 &&
N_{\rm Form} =  6 
\\

 (C \x^2 C) (\f C)
f_{j} m A^j 
 & \stackrel{d_1}{\lra} 
&
  (C \x^2 C)  (\f C)
f_{j}m (C \y^j) 
\\

 (C \x^2 C) (\f C)
f_{ij} A^i A^j 
 & \stackrel{d_1}{\lra} 
&
  (C \x^2 C)  (\f C)
f_{ij} A^i (C \y^j) 
\\

 (C \x^2 C) (\f C)
f_{j}^i \A_i A^j 
 & \stackrel{d_1}{\lra} 
&
  (C \x^2 C)  (\f C)
f_{j}^i \A_i 
 (C \y^j) 
\\

\\
\end{array}$ }}
\end{table}

\begin{table}[hptb]
\caption{\Large Third Table for  spinors coupled to a constant source $\f_{\a}$ with  ${\rm Dimension} \; \f_{\a}  = \fr{3}{2}$:   
Equations arising from Table \ref{threehalfscalars2}. } 

\vspace{.1in}
\framebox{ 
{ \Large $\begin{array}{ccccccc}  
\la{threehalfscalars3}
\\
f_i  \og^{ij} 
&=&0
\\
f_i  \og^{ijk}  
&=&0
\\
\lt (
 \oh^j g_{ij} 
+
2 \oh^{jk} g_{ijk} 
\rt )  
&=&0
\\

\end{array}$ } }
\end{table}

\begin{table}[hptb]
\caption{\Large Fourth Table for  spinors coupled to a constant source $\f_{\a}$ with  ${\rm Dimension} \; \f_{\a}  = \fr{3}{2}$:   
Mapping $E_{\infty} \ra {\cal H}_{\rm LIP}$. These are subject to the constraints in Table  \ref{threehalfscalars3}. } 

\vspace{.1in}
\framebox{ 
{ \Large $\begin{array}{ccccccc}  
\la{threehalfscalars4}.
\\
E_{\infty} &&  {\cal H}_{\rm LIP}\\
\\
 (C \x^2 C) 
   f_i
 \f^{\a} \y^i_{\a}  & \lra &
\int d^4 x \;  \lt \{ f_i
 \f^{\a} C_{\a} \oG^i   \rt \} 
\\ 
\begin{array}{c}
(C \x^2 C) \f^{\a} C_{\a} 
\\
\lt \{
   \oh^i  m \A_i  
+ \oh^{ij}  \A_i \A_j  
\rt \}
\end{array}
& \lra &
\begin{array}{c}
\int d^4 x \;  
\f^{\a} C_{\a} 
\\
 \lt \{
 \oh^i  m \oF_{1,i}  
+ \oh^{ij}  \oF_{2,ij} \rt \}
\end{array}
\\
\end{array}$ } }
\end{table}

\subsection{ Cohomology with Dimension Zero with one spinor index contracted with a spinor of $\f_{\a}$  Dimension $\fr{1}{2}$}

Our final example here also depends on considering objects that have a  spinor index.  To keep track of the spinor index, we  again contract the spinor index  with a spinor  $\f_{\a}$.  But now we will assume the Dimension of $\f_{\a}$  is $\fr{1}{2}$.  The scenario is summarized in Tables 
\ref{onehalfscalars1},\ref{onehalfscalars2}, \ref{onehalfscalars3},  \ref{onehalfscalars4}, \ref{onehalfscalars5},  \ref{onehalfscalars6}, \ref{onehalfscalars7},  \ref{onehalfscalars8}  and \ref{onehalfscalars9}.

The First, Second and Third Tables \ref{onehalfscalars1}, \ref{onehalfscalars2} and
\ref{onehalfscalars3} 
eliminate objects by the action of $d_1$.   

The Fourth  Table   \ref{onehalfscalars4} gives rise to the constraints.

The Fifth, Sixth and Seventh Tables \ref{onehalfscalars5}, \ref{onehalfscalars6} and
\ref{onehalfscalars7} 
eliminate more objects by the action of $d_1$.   

Note that many terms such as
\be
(\oC \x^2 \oC) \of^i  (\f \pa \oy_i)  
\ee
or
 \be
(C \x^2 C) (\f \pa \oy_i)  
\ee
are not present because they represent contractions
\be
 (\f \pa \oy_i) = \f^{\a} \oy_{\dot \b, \a \dot \g}
\e^{\dot \b  \dot \g} 
\ee
which are not in the space $E_1$, following the discussion in Section \ref{delta0disc}.

The Ninth Table   \ref{onehalfscalars9}
 shows the correspondence between  $E_{\infty}$ and  
${\cal H}_{
\rm LIP}$. Since it is simple and illustrated by the scalar examples, we skip the intermediate mapping to ${\cal H}_{\rm Fock}$.  
The objects in ${\cal H}_{\rm LIP}$ must satisfy the constraints on the coefficients mentioned in  Table \ref{onehalfscalars8} which follow from  Table \ref{onehalfscalars4}.

Note that the second  object in the  ${\cal H}_{\rm LIP}$ column in Table    \ref{onehalfscalars9} has a rather surprising form. In detail it is 
\[
{\cal A}_{{\rm Insert}\;\fr{1}{2}} = 
\int d^4 x \; \f^{\a} \F'_{\a}
\]
\[
=
\int d^4 x \; \f^{\a}
f_i^j
\lt \{
\oG^{i} \A_j
C_{\a} 
+
\lt (
\pa_{\a \dot \b} A^i 
+ C_{\a}\oY^{i}_{ \dot \b} \rt )
 \oy_{j}^{\dot \b}
\rt.
\]
\be
\lt.
+
\y^i_{\a} 
\lt ( 
m {g}_{jl} {A}^l  +
{g}_{jlq} A^l A^q
+
{Y}_j^{  \b} {c}_{\b } 
\rt )
\rt \}
\la{insertonehalffromfivechap}
\ee

In this form we have made the replacement  
\be
\oF_{1,j} \ra 
- \lt (
m {g}_{jl} {A}^l  +
{g}_{jlq} A^l A^q
+
{Y}_j^{  \b} {c}_{\b } 
\rt )
\ee
in order to show the form of this object in detail.  This is the simplest operator that looks like it has a reasonable chance of having a supersymmetry anomaly associated with it.  We shall discuss more details about the origin of this form in the next Chapter.

The corresponding anomaly is also in 
Table    \ref{onehalfscalars9}. It has the form
\be
{\cal A}^{(1)}_{\f} = 
\int d^4 x \; \f^{\a} C_{\a}
  \lt \{  m {\oh}^{ij}   \oF_{2,ij}
+  {\oh}^{ijk}  \oF_{3,ijk}
\rt \}
\la{insertonehalfanomfromfivechap}
\ee
where $m$ is the mass parameter, and 
\be
\oF_{1,j}= \lt ( 
m {g}_{jl} {A}^l  +
{g}_{jlq} A^l A^q
+
{Y}_j^{  \b} {c}_{\b } 
\rt )
\ee
\be
\oF_{2,ij}= \A_{(i} \oF_{1,j)} - \oy_{(i} \cdot \oy_{j)}
\ee
\be
\oF_{3,ijk}= \A_{(i} \oF_{2,jk)} - \A_{(i} \oy_{j} \cdot \oy_{k)}
\ee

We have dropped possible terms
$
f_i \oG^{i} m
C_{\a} 
$ and 
$  m^2 \oh^{i} \oF_{1,i} $, because they are ZMT.

\begin{table}[hptb]
\caption{\Large First Table for spinors coupled to a constant source $\f_{\a}$ with  ${\rm Dimension} \; \f_{\a}  = \fr{1}{2}$:   
Mappings of $N_{\rm Form}=2 \leftrightarrow N_{\rm Form} =  3$. }

\vspace{.1in}
\framebox{ 
{ \Large $\begin{array}{ccccccc}  
\la{onehalfscalars1}
\\
N_{\rm Form} = 2 &&
N_{\rm Form} = 3 
\\
(\f \x \oC)     A^i  
& \stackrel{d_1}{\lra} & (C \x \oC)  
 (\f \y^i) 
  \\
(\f \x \oC)  \A_i   &\stackrel{d_1}{\lra} &   
 (\f \x \oC) (\oC  \oy_{j})
\\

\\
\end{array}$ } }
\end{table}

\begin{table}[hptb]
\caption{\Large Second Table for  spinors coupled to a constant source $\f_{\a}$ with  ${\rm Dimension} \; \f_{\a}  = \fr{1}{2}$:   
Mappings of $N_{\rm Form}=3 \leftrightarrow N_{\rm Form} =  4$.  }

\vspace{.1in}
\framebox{ 
{ \Large $\begin{array}{ccccccc}  
\la{onehalfscalars2}
\\
N_{\rm Form} = 3 &&
N_{\rm Form} = 4 
\\
 (C \x \oy_i)  (\f C) & \stackrel{d_1}{\lra} & (C \x )^{\dot \b} \A_{i,\dot \b \a}  C^{\a}(\f C)
\\

\\
\end{array}$ } }
\end{table}

\begin{table}[hptb]
\caption{\Large Third Table for  spinors coupled to a constant source $\f_{\a}$ with  ${\rm Dimension} \; \f_{\a}  = \fr{1}{2}$:   
Mappings of $N_{\rm Form}=4 \leftrightarrow N_{\rm Form} =  5$.  }

\vspace{.1in}
\framebox{ 
{ \Large $\begin{array}{ccccccc}  
\la{onehalfscalars3}\\
N_{\rm Form} = 4 &&
N_{\rm Form} = 5 
\\
 f_{ik} (C \x^2 C) A^k (\f \y^i) & \stackrel{d_1}{\lra} 
&
\begin{array}{c} 
 (C \x^2 C) f_{ik}(\f \y^i) \;
( C \y^k) 
\end{array}
\\
\\

 (\oC \x^2 \oC) f_{ik} A^k (\f \y^i) 
& \stackrel{d_1}{\lra} 
&
\begin{array}{c} 
 (\oC \x^2 \oC) f_{ik} A^k (\f \pa A^i \oC)

\end{array}
\\
\\

 (\oC \x^2 \oC) f_{i}^k \A_k (\f \y^i) & \stackrel{d_1}{\lra} 
&
\begin{array}{c} 

 (\oC \x^2 \oC)
f_{i}^k\lt \{
 \A_k (\f \pa A^i \oC) 
+ 
 (\oC \oy_k) (\f \y^i)
\rt \}

\end{array}
\\
\\

 (\oC \x^2 \oC) m (\f \y^i) & \stackrel{d_1}{\lra} 
&
\begin{array}{c} 
 (\oC \x^2 \oC)
 m (\f \pa A^i \oC) 
\end{array}
\\
\\

 (C \x)^{\dot \b} A^i_{\a \dot \b} C^{\a}
 (\f C) & \stackrel{d_1}{\lra} 
&
\begin{array}{c} 
(C \x)^{\dot \b} \y^i_{\g , \a \dot \b} C^{\a}
 C^{\g}
 (\f C) \end{array}
\\
\\

\\
\end{array}$ } }
\end{table}

\begin{table}[hptb]
\caption{\Large Fourth Table for  spinors coupled to a constant source $\f_{\a}$ with  ${\rm Dimension} \; \f_{\a}  = \fr{1}{2}$:   More  Mappings of $N_{\rm Form}=4 \leftrightarrow N_{\rm Form} =  5$.  }

\vspace{.1in}
\framebox{ 
{ \Large $\begin{array}{ccccccc}  
\la{onehalfscalars4}\\
N_{\rm Form} = 4 &&
N_{\rm Form} = 5 
\\

 (C \x^2 C) (\f \y^i)
\lt \{
m  f_i 
+ f_i^j  \A_j   \rt \}
& \stackrel{d_5}{\lra} 
&
\begin{array}{c} 
 (C \x^2 C)  (\f C) \;
\\
\lt \{
m  f_i 
+ f_i^j  \A_j   \rt \} 
\\
 \lt \{
m {\og}^{ik} \A_k + {\og}^{ikq} \A_k \A_q
\rt \}
\end{array}
\\
\\

\begin{array}{c} 
 (C \x^2 C)
\\
 \lt \{
m (\f \y^i)
\rt.
\\
\lt (2 g_{ij} \oh^j +2 g_{ijk} \oh^{jk} \rt )
\\
\lt.
+ 
 (\f \A_q \y^i)
\rt.
\\
\lt.
\lt (
2 g_{ij} \oh^{jq} +
 6 g_{ijk} \oh^{jkq} \rt )
\rt \}
\end{array} 
& \stackrel{d_5^{\dag}}{\lla} 
&
\begin{array}{c} 
 (C \x^2 C)  (\f C) \;
\\
 \lt \{
m^2 {\oh}^{i} \A_i 
+
m {\oh}^{ij} \A_i 
\A_j 
\rt.
\\
\lt.
+
 {\oh}^{ijk} \A_i 
\A_j \A_k 
\rt \}
\end{array}
\\

\\
\end{array}$ } }
\end{table}

\begin{table}[hptb]
\caption{\Large Fifth Table for  spinors coupled to a constant source $\f_{\a}$ with  ${\rm Dimension} \; \f_{\a}  = \fr{1}{2}$:   
Mappings of $N_{\rm Form}=5 \leftrightarrow N_{\rm Form} =  6$.  }

\vspace{.1in}
\framebox{ 
{\Large $\begin{array}{ccccccc}  
\la{onehalfscalars5}
\\
N_{\rm Form} = 5 &&
N_{\rm Form} = 6 
\\
(C \x^2 C) (\f C)  f_{ijk} A^i A^j A^k  
 & \stackrel{d_1}{\lra}  &
\begin{array}{c}
(C \x^2 C)(\f C)  f_{ijk} 
\\
 A^i A^j   ( C \y^k)
\\
\end{array}
\\

\begin{array}{c} 

 (\oC \x^2 \oC)
f_{i}^k
\\
\lt \{
 \A_k (\f \pa A^i \oC) 
- 
 (\oC \oy_k) (\f \y^i)
\rt \}
\end{array}
 & \stackrel{d_1}{\lra}  &
\begin{array}{c}
(\oC \x^2 \oC) f_{j}^{i}  (\oy_i \oC)  (\f \pa A^j \oC) 
\\
\end{array}
\\

(C \x^2 C) (\f C) (\oy_i \oy_j)
 & \stackrel{d_1}{\lra}  &
\begin{array}{c}
(C \x^2 C) (\f C) (\oy_i \pa \A_j C)
\\
\end{array}
\\

\oC \x^2 \oC) A^i ( \f  \pa   \A_j \oC) 
 & \stackrel{d_1}{\lra}  &
\begin{array}{c}
\oC \x^2 \oC) A^i ( \f  \pa \oC) (\oy_j \oC)
\\
\end{array}
\\

\end{array}$ } }
\end{table}

\begin{table}[hptb]
\caption{\Large Sixth Table for  spinors coupled to a constant source $\f_{\a}$ with  ${\rm Dimension} \; \f_{\a}  = \fr{1}{2}$:   
Mappings of $N_{\rm Form}=6 \leftrightarrow N_{\rm Form} =  7$.  }

\vspace{.1in}
\framebox{ 
{\Large $\begin{array}{ccccccc}  
\la{onehalfscalars6}
\\
N_{\rm Form} = 6 &&
N_{\rm Form} = 7 
\\

(C \x^2 C) (\f C) f_{ijk} A^i A^j ( C \y^k) 
 & \stackrel{d_1}{\lra}  &
\begin{array}{c}
(C \x^2 C)(\f C)  f_{ijk} 
\\
\lt \{ A^i  ( C \y^j) ( C \y^k)
\rt \}
\\
\end{array}
\\

(\oC \x^2 \oC) (\oy_j \oC)  (\f \pa \A_i \oC)  
 & \stackrel{d_1}{\lra}  &
\begin{array}{c}
(\oC \x^2 \oC) (\oy_j \oC)  (\f \pa \oC) (\oC \oy_i )
\\
\end{array}
\\

\end{array}$ } }
\end{table}

\begin{table}[hptb]
\caption{\Large Seventh Table for  spinors coupled to a constant source $\f_{\a}$ with  ${\rm Dimension} \; \f_{\a}  = \fr{1}{2}$:   
Mappings of $N_{\rm Form}=7 \leftrightarrow N_{\rm Form} =  8$.  }

\vspace{.1in}
\framebox{ 
{\Large $\begin{array}{ccccccc}  
\la{onehalfscalars7}
\\
N_{\rm Form} = 7 &&
N_{\rm Form} = 8 
\\
(C \x^2 C) (\f C) f_{[ijk]} ( C \y^i)( C \y^j) A^k
 & \stackrel{d_1}{\lra}  &
\begin{array}{c}
(C \x^2 C)(\f C)  f_{[ijk]} 
\\
\lt \{
( C \y^i)( C \y^j) ( C \y^k)
\rt \}
\\
\end{array}
\\

\end{array}$ } }
\end{table}

\begin{table}[hptb]
\caption{\Large Eighth Table for  spinors coupled to a constant source $\f_{\a}$ with  ${\rm Dimension} \; \f_{\a}  = \fr{1}{2}$:  Equations arising from Table \ref{onehalfscalars4}. }
\vspace{.1in}
\framebox{ 
 { \Large
 $\begin{array}{ccc}  
\la{onehalfscalars8}
\\
   f_{i}^{j} {\og}^{ki}
+   f_{i}^{k} {\og}^{ji}
&=&0
\\
   f_{q}^{(i} {\og}^{jk)q}
&=&0
\\
\lt (
2 g_{ij} \oh^{jq} +
6 g_{ijk} \oh^{jkq} \rt ) 
&=&0
\\

\end{array}$ } }
\end{table}

\begin{table}[hptb]
\caption{\Large Ninth Table for  spinors coupled to a constant source $\f_{\a}$ with  ${\rm Dimension} \; \f_{\a}  = \fr{1}{2}$:   
Mappings  of $E_{\infty} \ra {\cal H}_{\rm LIP}$.  These are subject to the constraints in Table \ref{onehalfscalars8}. } 

\vspace{.1in}
\framebox{ 
{\Large $\begin{array}{ccccccc}  
\la{onehalfscalars9}.
\\
E_{\infty} &&  {\cal H}_{\rm LIP}\\
\\
 
\begin{array}{c}  
(C \x^2 C) \f^{\a} C_{\a} \lt \{
  
h^i  m^2 \A_i  
\rt.
\\
\lt.
+ \oh^{ij} m \A_i \A_j  
+ \oh^{ijk}  \A_i \A_j \A_k 
\rt \}
\end{array}
& \lra &
\begin{array}{c}  
\int d^4 x \;  
\f^{\a} C_{\a}  \lt \{
 \oh^i  m \oF_{1,i}  
\rt.
\\
\lt.
+ \oh^{ij}  \oF_{2,ij} 
+ \oh^{ijk}  \oF_{3,ijk} \rt \}
\end{array}
\\
\\
\begin{array}{c}  
 (C \x^2 C) 
\\
\f^{\a} \y^i_{\a} 
\lt \{
    f_i^j \A_j
 \rt \}
\end{array}
 & \lra &
\begin{array}{c}  
\int d^4 x \;
 \f^{\a} 
\lt \{
f_i^j
\lt [
\oG^{i} \A_j
C_{\a} 
\rt.
\rt.
\\
\lt.
+
\lt (
\pa_{\a \dot \b} A^i 
+ C_{\a}\oY^{i}_{ \dot \b} \rt )
 \oy_{j}^{\dot \b}
\lt.
-
\y^i_{\a} 
\oF_{1,i} \rt ]
\rt \}
\end{array}

\end{array}$ } }
\end{table}



\chapter{ Some Composite and Elementary Supermultiplets}

\la{superfieldschapter}

We can still usefully talk about superfields, even though we are no longer using  superspace to construct our action. The use of the superfields is to facilitate multiplication of multiplets and to pick out components of compound multiplets.  But one needs to use the superfields judiciously.  Really all we have here is supermultiplets, and their composition is facilitated by thinking of them as superfields. 

In Table \ref{Lowcompsup}, we summarize the antichiral superfields discussed in this section and used below in the formulae (\ref{superzero}), (\ref{superone}) and (\ref{supertwo}).

\begin{table}[hptb]
\caption{\Large Some Composite and Elementary Antichiral `Superfield' Multiplets}
\la{Lowcompsup}
\vspace{.1in}
\framebox{ 
{\Large $\begin{array}{cccc}  
\\
{\rm Mult} &
\A \;{\rm Term} &
\oy \;{\rm Term} &
\oF \;{\rm Term} 
\\
{\hat \A}_i
&
\A_i &
\oy_{i \dot \b}
&
\oF_{1,i}
\\
{\hat ?}^i
&
?^i &
\oY^{i}_{ \dot \b}
&
\oG^i
\\
\\
{\hat \f}_{\a}
&
\f_{\a} 
&
W_{\a \dot \b}  
&
\F_{\a} 
\\
\\
\begin{array}{c}
{\hat \f}'_{\a}
\\
= {\widehat \y}^i_{\a}
\end{array}
&
\begin{array}{c}
{  \f}'_{\a}
\\
= { \y}^i_{\a}
\end{array}
&
\begin{array}{c}
{W}'_{\a \dot \b} =
\\
\lt (
\pa_{\a \dot \b} A^i + C_{\a} \oY^i_{\dot \b} 
\rt )
\end{array}
&
\begin{array}{c}
\F'_{\a} 
\\=
\oG^i C_{\a}
\end{array}
\\
\\

\begin{array}{c}
{\hat \f}''_{\a}
\\
= {\widehat \y}^i_{\a} {\widehat \A}_j
\end{array}
&
\begin{array}{c}
{  \f}''_{\a}
\\
=
\y^i_{\a} \A_j

\end{array}
&
\begin{array}{c}
{W}''_{\a \dot \b}
=
\lt \{
\y^i_{\a} \oy_{j \dot \b}
\rt.
\\
\lt.
+
\lt ( 
\pa_{\a \dot \b} A^i + C_{\a} \oY^i_{\dot \b} 
\rt ) \A_j
\rt \}

\end{array}
&
\begin{array}{c}
\F''_{\a} 
= \lt \{
\oG^i C_{\a} \A_j 
+
\y^i_{\a} \oF_{1,j} 
\rt.
\\
\lt.
+ \lt ( 
\pa_{\a \dot \b} A^i + C_{\a} \oY^i_{\dot \b} 
\rt ) \oy^{\dot \b}_j
\rt \}
\end{array}
\\
\\

&
&
{ \begin{array}{c}  

\end{array}} 
&

\\

\end{array}$
}} 
\end{table}

Here is the chiral superfield:
\be
{\hat A}^i = A^i + \q^{\b} \y^i_{\b}
+ \fr{1}{2} \q^2 F_1^i
\ee

It has a composite auxiliary field of the form: 
\be
F_{1}^{i} = -  \lt ( 
m {\ov g}^{ij} {\ov A}_j  +
 {\ov g}^{ijk} {\ov A}_j  {\ov A}_k +
{\ov Y}^{i \dot \b} {\ov c}_{\dot \b } 
\rt )
\ee
Its complex conjugate is the Antichiral Scalar Superfield. 
\be
{\ov {\hat A}}_i = \A_i + \oq^{\dot \b} \oy_{i \dot \b}
+ \fr{1}{2} \oq^2 \oF_{1,i}
\ee
which has the composite auxiliary field: 
\be
\oF_{1,i} = -\lt ( 
 m {g}_{il} {A}^l  +
{g}_{ilq} A^l A^q
+
{Y}_i^{  \b} {c}_{\b } 
\rt ).
\ee
We will use a source which is an antichiral spinor superfield:
\be
{\hat \f}_{\a} = \f_{\a} + \oq^{\dot \b} W_{\a \dot \b} + \fr{1}{2} \oq^2 \F_{\a}
\ee
These components transform as in Table \ref{antichiral}.

 \begin{table}[hptb]
\caption{\Large   Antichiral Spinor Transformations}
\la{antichiral}
\vspace{.1in}
\framebox{ 
{\Large $\begin{array}{lll}  
\\
\d \f_{\a}&= & 
  W_{\a \dot   \b} {\oC}^{\dot   \b} 
+ \x^{\g \dot \d} \partial_{\g \dot \d} \f_{\a}
\\
 \d W_{\a \dot \b} &  =& 
\pa_{ \g \dot \b }  \f_{\a} C^{\g} + \F_{\a} {\ov C}_{\dot \b}  
+ \x^{\g \dot \d} \partial_{\g \dot \d}   W_{\a \dot \b} 
\\
\d
 \F_{\a} &  =&  \pa_{ \g \dot \b }  W_{\a}^{\;\; \dot \b}  C^{\g} 
 + \x^{\g \dot \d} \partial_{\g \dot \d}  \F_{\a}
\\
\end{array}$} }
\end{table} 

There is no `superfield' which corresponds to the notation ${\hat ?}^i$ in Table \ref{Lowcompsup}.  This is a consequence of the formulation of the Physical BRS-ZJ identity that we performed in Section \ref{physzj}. On the other hand, it appears that we get a new `superfield' in compensation, namely ${\hat \f}'_{\a}$ in Table \ref{Lowcompsup}. This new `superfield' 
transforms in the same way as an antichiral spinor superfield. 
We can multiply this superfield with the antichiral scalar superfield to get  ${\hat \f}''_{\a}$ in Table \ref{Lowcompsup}. 
 This is the major new result in this paper.  Most of the new results of this paper depends on its existence:
\be
{\hat \f}'_{\a} = \f'_{\a} + \oq^{\dot \b} W'_{\a \dot \b} + \fr{1}{2} \oq^2 \F'_{\a}
\ee
\be
= f_i 
{\hat \y}^i_{\a } =  
f_i \lt \{
\y^i_{\a} + \oq^{\dot \b} 
\lt (
\pa_{\a \dot \b} A^i
+
 C_{\a} \oY^i_{ \dot \b} 
\rt )
+
\fr{1}{2} \oq^2 \oG^i C_{\a}
\rt \}
\ee
This multiplet contains all the fields and sources in the the `Upper Index Quartet':
\be
A^i, \y^i_{\a}, \oY^i_{\dot \b}, \oG^i
\ee
It transforms as an antichiral spinor multiplet. In other words the transformations in Table \ref{physicaltable} induce the transformations in Table   \ref{primeantichiral} with the above definitions.

{\em However, this transformation  as an antichiral spinor multiplet is true
only  for the case where the tensors $  g_{ij}, g_{ijk}$ are all zero. This is the origin of the constraints.  When any of $ g_{ij}, g_{ijk}$ are not zero, one needs constraints to ensure that this multiplet transforms simply.}

 \begin{table}[thbp]
\caption{\Large   Antichiral Spinor Transformations}
\la{primeantichiral}
\vspace{.1in}
\framebox{ 
{\Large $\begin{array}{lll}  
\\
\d \f'_{\a}&= & 
  W'_{\a \dot   \b} {\oC}^{\dot   \b} 
+ \x^{\g \dot \d} \partial_{\g \dot \d} \f'_{\a}
\\
 \d W'_{\a \dot \b} &  =& 
\pa_{ \g \dot \b }  \f'_{\a} C^{\g} + \F'_{\a} {\ov C}_{\dot \b}  
+ \x^{\g \dot \d} \partial_{\g \dot \d}   W'_{\a \dot \b} 
\\
\d
 \F'_{\a} &  =&  \pa_{ \g \dot \b }  W_{\a}^{'\;\; \dot \b}  C^{\g} 
 + \x^{\g \dot \d} \partial_{\g \dot \d}  \F'_{\a}
\\
\end{array}$}} 
\end{table} 

In some sense, the existence of the multiplet ${\hat \f}'_{\a}$ is a consequence of the integration of the auxiliary in the Physical Formulation of the BRS-ZJ identity.

 The simplest example of a cohomologically non-trivial spinor that could be coupled to the source ${\hat \f}_{\a} $ and then added to the Physical action is:

\be
{\cal A}_{\rm Operator; \fr{3}{2}} 
=
\int d^4 x 
\int d^2 \oq 
{\hat \f}^{\a} 
 f_i {\hat \y}^i_{\a}
\ee

It can be expanded into components:
\be
{\cal A}_{\rm Operator; \fr{3}{2}} 
=
\int d^4 x 
\int d^2 \oq 
{\hat {\f}}^{\a} 
{\hat \f}'_{\a} 
\ee
\be
=
\int d^4 x 
\lt \{
{\f}^{\a} 
\F'_{\a} +
{W}^{\a \dot \b} 
{W'}_{\a \dot \b} 
+
{\F}^{\a} 
\f'_{\a} 
\rt \}
\ee
where
\be
{\hat \f}'_{\a} = \f'_{\a} + \oq^{\dot \b} W'_{\a \dot \b} + \fr{1}{2} \oq^2 \F'_{\a}
\ee
and the components are:
\be
\f'_{\a}
 =
f_i  \y^{i}_{\a} 
\ee
and
\be
W'_{\a \dot \b} =
f_i 
\lt (
\pa_{\a \dot \b} A^i 
+ C_{\a}\oY^{i}_{ \dot \b} \rt )
\ee
and
\be
\F'_{\a}
 =
f_i \oG^{i} 
C_{\a} 
\ee


\chapter{ Insertion of Cohomology Invariants}

\la{insertionchapter}

\section{Renormalization and terms in the action}

\subsection{Non-renormalization Theorem}

The general theory of renormalization of quantum field theories tells us that we should usually expect to require  all possible local polynomial renormalization counterterms  
  so long as those counterterms have the appropriate dimension and quantum numbers.  In the pure chiral supersymmetry theory,  one could expect that the   mass $g_{ij}$ and Yukawa tensors $g_{ijk}$  would get renormalized to new tensors  $ g'_{ij}$ and $g'_{ijk}$.

However the non-renormalization theorem of supersymmetry says that this does not happen.  It says that we should require no counterterms of these types.
This seems to be true in perturbation theory.

\subsection{Cohomology and Physical Effects}

Generally if we have local polynomials that satisfy
\be
{\cal B }^{(0)} = \d {\cal P}^{(-1)},
\ee
this should mean that the relevant ${\cal B }^{(0)}$ has no physical effect because it `vanishes on-shell'.  The variation by $\d$ is the sum of a redefinition of the field and a replacement of the Zinn Sources by the equations of motion.  Neither of these create any on-shell physical matrix element.

An element of the cohomology space is not of this type, and so we expect it to have a physical effect in general.

Tere are several different  kinds of cohomology terms that we have found above.  They can be introduced into the action by multiplying them by a `source', which may, or may not, be spacetime dependent.  If the `source' is spacetime dependent then we need to ensure supersymmetry by using whatever supermultiplets are available.

\subsection{Zero Momentum Terms (ZMT)}

In Section \ref{isolampsection}, we discussed the meaning of this term and its implications for cohomology.  In the present chapter this discussion becomes relevant again.  Because the BRS-ZJ identity is quadratic, the insertion of a new term into the action often forces us to `close the algebra' by adding new terms.  The insertion of a new term immediately changes $\d$.  If the new $\d$ does not satisfy $\d^2=0$ we need to see what is necessary to make that happen, etc. 

In particular, we need to make sure that we do not introduce ZM terms in this way, and that is relevant to the present chapter as we shall see.

\section{Insertion  of Scalar Cohomology generates Global Symmetry or Local Gauge Symmetry, together with spontaneous breaking of those symmetries}

 For the scalar cases above in subsection \ref{ScalarCohomologySection},
 we shall  consider the cases where the source has dimension 2, 1 and 0, and where it has ghost charge minus one ($\W$)  and zero ($U$).  So there are five possible items we could add to the action: 
\be
{\cal A}^{(0)}_{\rm Insert,{\rm Dim} \; W = 2} = 
\int d^4 x \;
\lt \{
 \ok^i \oF_{1,i} U
 +
k_i F^i_{1} {\ov U}
\rt \}
\ee

\be
{\cal A}^{(0)}_{\rm Insert,{\rm Dim} \; U = 1} = 
\int d^4 x \;
\lt \{
\lt ( m \ok^{i} \oF_{1,i} 
 +
 \ok^{ij} \oF_{2,ij} \rt )
U
\rt.
\ee
\be
\lt.
 +
\lt ( m k_{i} F^{i}_{1}  +
k_{ij} F^{ij}_{2} \rt )
{\ov U}
\rt \}
\ee

\be
{\cal A}^{(0)}_{\rm Insert,{\rm Dim} \; U = 0} = 
\int d^4 x \;
\lt \{
 \lt (
m^2 \ok^{i} \oF_{1,i} 
 + m
 \ok^{ij} \oF_{2,ij} 
 + 
 \ok^{ijk} \oF_{3,ijk} 
\rt ) U
\rt.
\ee
\be
\lt.
\lt (
 +
m^2 k_{i} F^{i}_{1}   + m
k_{ij} F^{ij}_{2}   + 
k_{ijk} F^{ijk}_{2} \rt ){\ov U}
\rt \}
\ee

\be
{\cal A}^{(0)}_{\rm Insert,{\rm Dim} \; \W = 1} = 
\int d^4 x \;
\lt \{
 e_i \oG^i \W
 +
\ove^i \G_i {\ov \W}
\rt \}
\ee
\be
{\cal A}^{(0)}_{\rm Insert,{\rm Dim} \; \W = 0} = 
i \int d^4 x \;
\lt \{ e_i m \oG^i \W -  m
\ove^i \G_i {\ov \W}
\rt.
\ee
\be
\lt.
+  e_i^j \lt (
\oG^i \A_j 
-
\G_j A^i 
-
Y^{\a}_j \y^{i \a} 
+
\oY^{i \dot \a} \oy_{j}^{\dot \a} 
 \rt )
\lt ( \W + \ov \W \rt )
\rt \}
\ee

These are five distinct cases.  For each one the insertion gives rise to a new $\d$ to consider. In each case that new  $\d$ may or may not be nilpotent.  In each case there are modifications or constraints that need to be imposed to make the $\d$ nilpotent.  Recall the definitions in subsections \ref{brsinphyspform} and
\ref{boundaryopinphyspform}.   Because of the connection between the $\d{\cal A}$ constructed from an action,  and the action ${\cal A}$, namely:
\be
{\cal A} * {\cal A} = 0  
\la{AstarA}
\ee
\be  \Lra 
\ee
\be   \d^2_{\cal A} =0
\la{deltanilpotent}
\ee
we know that if we can complete $\d_{\cal A}$ so that it satisfies equation (\ref{deltanilpotent}), then  there is a corresponding ${\cal A}$  which satisfies 
equation (\ref{AstarA}), and vice versa.

Remarks:
\ben
\item
The term $U$ is really more like a new mass or constant $(m^2 w, m w, w$ where $w$ is dimensionless) for the three cases ${\rm Dim} = -2,-1,0$.  This is equivalent to starting with a different mass or Yukawa term. It is clear how we need to complete the action or $\d$: we simply revise the constants $g_{ij}, g_{ijk}$.  The term $\int d^4 x  m^2 g_{i} F^i_1 $ is generally present too, and of course it requires a vaccuum expectation value, a shift of $A^i \ra A^i + m v^i$ and the generation of Goldstone bosons etc. in general. But nothing unknown in the literature is to be expected here. 

\item
The term $\W$ is anticommuting.  Starting with it and completing the action ${\cal A}$ so that it satisfies  equation (\ref{AstarA}), or equivalently, completing  $\d$ so that it satisfies equation (\ref{deltanilpotent}), in general gives rise to a  new action.  But we know what to expect for this action! If we leave $\W$ a constant, then we will get the constraint equations that characterize a theory with a spontaneously broken global abelian symmetry.  If we let $\W^a$ have an index then we get a theory with a non-Abelian spontaneously broken global symmetry.  If we consider 
the possibility of letting $\W^a \ra \W^a(x)$ be spacetime dependent, then we will generate a spontaneously broken local supersymmetric non-Abelian gauge symmetry.  All of these are well known of course, and no attempt will be made here to derive them from the above. 

\een

\section{Insertion  of Spinor Cohomology generates 
What??}

\subsection{Scalar Cohomology generates gauge theory.}
From the above remarks we see that the cohomology in Section 
\ref{ScalarCohomologySection} generates the global or local versions of group invariant symmetries, including ultimately the well-known spontaneously broken supersymmetric gauge theories.  All of this develops from the following simple terms which we derive from Table \ref{zeroscalars6}, by adding a ghost source as described above, and then looking for the constraints that will ensure nilpotence:

\be
\begin{array}{c}
(C \x \oC) i \lt \{
 e^a_{i}  m A^i  
- \ove^{a i}  m \A_i  
+ e^{aj}_i   A^i \A_j  
\rt \}
 \w^a
\end{array}
\in E_{\infty} 
\ee
\be
\Lra
\ee
\be
\begin{array}{c}
\int d^4 x \;  
i \lt \{
 m  e^a_{i} \w^a
  \oG^i -  m
\ove^{a i} \w^a \G_i 
\rt.
\\
\lt.
+ \w^a e^{aj}_i \lt (
\lt (
\oG^i \A_j 
-
\G_j A^i 
-
Y^{\a}_j \y^{i \a} 
+
\oY^{i \dot \a} \oy_{j}^{\dot \a} 
 \rt )
\rt )
\rt \}
\end{array} 
\in   {\cal H}
\ee

Of course we also know that there is more cohomology involving $\w^a$ that gives rise to the well known gauge anomalies here. 

\subsection{What does the spinor Cohomology generate?}

So what happens for the rest of the cohomology above in section \ref{SpinorCohomologySection}? It involves spinors 
$\f_{\a}$ with ghost charge zero, rather than scalars $\w^a$ with ghost charge one. 

It appears that this is no longer familiar ground.  It seems to be  something new.
Should we add an index $\f^a_{\a}$ and also consider spacetime dependence $\f^a_{\a}(x)$?  It would appear that both of these seem reasonable generalizations of the above results.  We know from Chapter \ref{superfieldschapter} that adding spacetime dependence to $\f_{\a}$ should make sense.
  The possibility of adding an index will not be considered here, but it is an interesting possibility too.

We already know something about the anomaly possibilities here too from section \ref{SpinorCohomologySection}.  They involve the supersymmetry ghost $C_{\a}$ rather than $\w^a$.  
Even adding $\f^a_{\a}(x)$ does not immediately require us to make  $C_{\a} \ra C_{\a}(x)$, which would give rise to supergravity. 

As we will see below in Chapter \ref{remarkschapter}, this spinor cohomology and its constraints can be expected to exist for an infinite spectrum of spinors $\f^a_{(\a_1 \cdots \a_n ),( \dot \b_1 \cdots \dot \b_m)}$ with all numbers of indices, which reminds us strongly of the spectrum of the superstring.  All of these look like they can have supersymmety anomalies analogous to the above possibilities.

A look at the situation where we take a supermultiplet ${\hat \f}_{\a}$ as a source reveals that
\ben
\item
We can probably build up an action that also has kinetic terms for the $\f_{\a}, W_{\a \dot \b} ,\F_{\a}$ terms too.
\item\
We will need ghosts to complete the action and to close the algebra.  There also appear to be ghosts for ghosts etc.
\item
A new cohomology problem is generated.  It may be that anomalies are added and removed by the new algebras.
\item
The terms $\f_{\a}, W_{\a \dot \b} ,\F_{\a}$  interact in interesting ways with the chiral matter.
\item
We could probably incorporate indices on the fields
$\f^a_{\a}, W^a_{\a \dot \b} ,\F^a_{\a}$
and examine more constraint equations like the ones below.
\item
We could probably also include more spinor indices on the fields
${\hat \f}^a_{\a \b}$ etc. and then
examine more constraint equations like the ones below.
\item
It appears that this might constitute a generalization of supersymmetric gauge theory for example.

\een

 For this paper, to keep things as simple as possible, we will now restrict ourselves to the situation where $\f_{\a}$ is a constant without an index and the fields 
$W_{\a \dot \b} ,\F_{\a}$ are absent.  In this case we do not need to transform $\f_{\a}$ under $\d$ and the new cohomology problem is tractable without much more work.

Let us now look at the  cohomology with inserted cohomologically non-trivial spinor  operators in more detail.

\section{Cohomology with Dimension Zero with Insertion of Spinor Invariant coupled to $\f_{\a}$ with Dimension $\fr{3}{2}$}
\be
{\cal A}_{\rm Total; \fr{3}{2} }=
{\cal A}_{\rm Physical} +
{\cal A}_{\rm Insert; \fr{3}{2} }
\ee
\be
\d_{\rm Total; \fr{3}{2} }=
\d_{\rm Physical} +
\d_{\rm Insert; \fr{3}{2} }
\ee
\be
{\cal A}_{{\rm Insert} ; \fr{3}{2}} =
\int d^4 x 
{\f}^{\a} f_i \oG^i C_{\a} 
\la{inv52}
\ee

We shall ignore this case because it is a ZMOF    term.  One could promote the field $\f_{\a}$ to $\f_{\a}(x)$ to remove the ZMOF   term, but this will not be done here.

\section{Cohomology with Dimension Zero with Insertion of Spinor Invariant coupled to $\f_{\a}$  with Dimension $\fr{1}{2}$}

\[
{\cal A}_{{\rm Insert}\;\fr{1}{2}} = 
\int d^4 x \; \f^{\a} \F'_{\a}
\]
\[
 =
\int d^4 x \; 
f_i^j
 \f^{\a} \lt [
\oG^{i} \A_j
C_{\a} 
\rt.
\]
\[
\lt.
+
\lt (
\pa_{\a \dot \b} A^i 
+ C_{\a}\oY^{i}_{ \dot \b} \rt )
 \oy_{j}^{\dot \b}
\rt.
\]
\be
\lt.
\lt.
+
\y^i_{\a} 
\lt ( m {g}_{jl} {A}^l  +
{g}_{jlq} A^l A^q
+
{Y}_j^{  \b} {c}_{\b } 
\rt )
\rt ]
\rt \}
\la{insertsevenhalves}
\ee

We are tempted to add an additional  term 
\be
{\cal A}_{ZMOZ;\fr{1}{2}}  =
\int d^4 x \; 
\f^{\a} 
f_i \oG^{i} m
C_{\a} 
\ee
but we do not do so because this is a ZMOZ term, in view of the discussion in Section \ref{isolampsection}. We remove it. If we were prepared to let $\f_{\a} \ra \f_{\a} (x)$ then we could replace it in the theory, but that is a change which brings in a a big change in $\d$ and in the action, and in this paper we are trying to keep things simple.  But note that the addition of such a term is very reminiscent of the Higgs-Kibble mechanism in vector-scalar gauge theories.

Even when $\f_{\a}$ is a constant and we ignore ${\ov \f}_{\dot \b}$,  we need more terms to close the algebra:

\be
{\cal A}_{\rm Total; \fr{1}{2} }=
{\cal A}_{\rm Physical} +
{\cal A}_{\rm Insert \; \fr{1}{2} }
+
{\cal A}_{\rm Completion} 
\la{theanomaction}
\ee

We start with the relation
\be
\d 
{\cal A}_{\rm Insert \; \fr{1}{2} }
=0
\ee

For simplicity, we have assumed that ${\hat \f}_{\a} \ra \f_{\a}$ is a constant independent of spacetime, and then we can put $W_{\a \dot \b} = \F_{\a}=0$ and ignore 
the supersymmetry transformations for $\f_{\a}$.  We also ignore the complex conjugate ${\ov \f}_{\dot \a}$ here. 
In this special case we can (apparently) write down the 
${\cal A}_{\rm Completion} 
$ term quite easily:
\be
{\cal A}_{\rm Completion} 
=
\int d^4 x 
\lt \{
{\f}^{\a} 
{\f}_{\a}
f^i_j
f^j_k
A^k
\oF_{1,i}
 \rt \}
\la{sevencomplete}
\ee
and then the identities analogous to \ref{star1PI}
hold
\be
\la{starNew}
{\cal A}_{\rm Total; \fr{1}{2} }
*{\cal A}_{\rm Total; \fr{1}{2} }
=0
\la{sevenhalfastar}
\ee
which implies that the derived $\d$ satisfies
\be
\d^2 =0
\ee

Clearly this term ${\cal A}_{\rm Completion} $ does not have an obvious superspace origin.  But there is no reason to expect it to do so, since even the action we start with does not have a very clear superspace origin itself. 

The relevant nilpotent operator $\d$ is in Table \ref{deltaInsertphionehalf}.    The operator $d_{7,\W,\fr{1}{2} }$ in Table \ref{specseqsum} is the relevant operator for the present case. We shall look at its cohomology below in subsection \ref{onehalfinsertsubsection}.

\begin{table}
\caption{\Large   Field, Composite and Zinn Source Transformations  including a spinor $\f_{\a}$ with dimension  $\fr{1}{2}$: This is free from ZMOF    term. }
\framebox{
{\Large $\begin{array}{lll}  
\la{deltaInsertphionehalf} 
\\
\d A^i&= & 
\fr{\d {\cal A}}{\d \G_i} 
=  \y^{i}_{  \b} {  C}^{  \b} 
+ \x^{\g \dot \d} \partial_{\g \dot \d} A^i
\\
\d {\ov A}_i&= & 
\fr{\d {\cal A}}{\d {\ov \G}^i} 
=  {\ov \y}_{i  \dot \b} {\ov  C}^{ \dot  \b} 
+ \x^{\g \dot \d} \partial_{\g \dot \d} {\ov A}_i
\\
&&
+ \lt (
  f_i^j  \A_j 
\rt )
\f^{\a} C_{\a}  
\\

\d \y_{\a}^i &  =& 
\fr{\d {\cal A}}{\d {  Y}_i^{   \a} } = 
\pa_{ \a \dot \b }  A^{i} {\ov C}^{\dot \b}  
+ 
C_{\a}   
F_1^i
+ \x^{\g \dot \d} \partial_{\g \dot \d}  \y^{i}_{\a  }
\\
&& 
+  f_j^i  \f^{\b} \y^j_{\b} C_{\a}   
\\
&&
+  f_j^k  f_k^i  (\f)^2 A^j  C_{\a} 
  
\\

\d
 {\ov \y}_{i \dot \a} &  =& 
\fr{\d {\cal A}}{\d { {\ov Y}}^{i \dot   \a} } = 
\pa_{ \a \dot \a }  {\ov A}_{i} { c}^{\a}  
+ 
{\ov C}_{\dot \a}   
{\ov F}_{1,i}
+ \x^{\g \dot \d} \partial_{\g \dot \d} 
 {\ov \y}_{i \dot \a} 
\\
&& 
+  f_i^j  \f^{\b} C_{\b}   \oy_{j \dot \b}  
\\
\d \G_i 
&= &
 \fr{\d {\cal A}}{\d A^i} 
=
 - \fr{1}{2} \pa_{ \a \dot \b  }       \pa^{ \a \dot \b  }        {\ov  A}_{i} 
\\
&& +   m {g}_{iq} F_1^q  + g_{ijk} { F}_2^{jk}
 - \pa_{ \a \dot \b } Y_{i}^{ \a}    {\ov C}^{\dot \b}   
+ \x^{\g \dot \d} \partial_{\g \dot \d} \G_i
\\
&& 
+  f_i^j  \lt \{ 
m g_{jk}  \f^{\b} \y^k_{\b}   
+ 2 g_{jkl}  \f^{\b} A^k \y^l_{\b}   
\rt \}
+  f_i^j \f^{\a} \pa_{\a \dot \b  } \oy^{\dot \b}_j   
\\
&&
+ 
\f^{\a}\f_{\a}
\lt \{
f_i^j f_j^k {\ov F}_{1,k} 
+
f_k^j f_j^l A^k \lt ( m g_{il}  + 2 g_{ilq} A^q  
\rt ) 
\rt \} 
\\

\d {\ov \G}^i 
&= & \fr{\d {\cal A}}{\d {\ov A}_i} 
=
- \fr{1}{2} \pa_{ \a \dot \b  }       \pa^{ \a \dot \b  }        { A}^{i} 
\\
&& 
+  m {\ov g}^{ij} {\ov  F}_{1, k}  
+  {\ov g}^{ijk}     {\ov  F}_{2,j k} 
\\
&& - \pa_{ \a \dot \b } {\ov Y}^{ i \dot \b}    { C}^{\a}   
+ \x^{\g \dot \d} \partial_{\g \dot \d} 
 {\ov \G}^i
\\
&&
+  f_j^i \f^{\a} C_{\a} \oG^j 
\\
\d Y_{i}^{ \a} 
&=&\fr{\d {\cal A}}{\d {  \y}^i_{   \a}} 
= 
-
  \pa^{\a \dot \b  }   
{\ov \y}_{i   \dot \b}
+  m {g}_{iq}   
\y^{q \a} 
\\
&&
 +
2 g_{ijk}  \y^{j \a} A^k    
-
\G_i  
 {  C}^{  \a}
+ \x^{\g \dot \d} \partial_{\g \dot \d}  Y_{i}^{ \a}
\\
&& 
+  f_i^j \oF_{1,j}  \f_{\a}  
\\
\d 
{\ov Y}^{i \dot \a} 
&=&\fr{\d {\cal A}}{\d {\ov \y}_i^{ \dot \a} 
} 
= 
-
  \pa^{\b \dot \a  }   
{ \y}^i_{ \b}
+  m {\ov g}^{ik}   
{\ov \y}_{k}^{\dot  \a} 
\\
&& +
2 {\ov g}^{ijk} {\ov \y}_{j}^{\dot  \a} 
{\ov A}_k  
-
{\ov \G}^i  
 { \ov C}^{\dot  \a}
+ \x^{\g \dot \d} \partial_{\g \dot \d}  
{\ov Y}^{i \dot \a} 
\\
&&
+  f_j^i \f^{\a} 
\lt \{
\pa_{\a \dot \a  } A^j  
+ C_{\a} {\ov Y}^j_{ \dot \a  }    
\rt \}
\\
\\
\d C_{\a}
&=&
0
\\
\d  {\ov C}_{\dot \b}
&=&
0

\end{array}$}} 
\end{table}


\chapter{
Diagram Chasing for the Six Easy Examples with inserted operators}

\la{diagramchasewithinsertchapter}

{\Large

\la{insertdiagramchasechapter}

 We will look again at the results of Chapter \ref{diagramchasechapter},  but with the insertions discussed in Chapter 
\ref{insertionchapter}.

\section{Scalar Cohomology with inserted operators}

We shall not look at this again because we can anticipate the results as  explained in Chapter 
\ref{insertionchapter}.

However it must be stressed that the combination of the non-renormalization theorem and the cohomology results is confusing.  It appears that the cohomology is in a sense irrelevant when one is dealing with the possibility of adding new mass or Yukawa superterms.

This becomes important when we consider the spinor invariants, as will be explained below.

\section{ Spinor Cohomology }

The first two examples yield only ZMT terms and so we will not look at them again.

\subsection{ Cohomology with Dimension Zero with one spinor index contracted with a spinor of $\f_{\a}$  Dimension $\fr{1}{2}$}

\la{onehalfinsertsubsection}

The four Tables 
\ref{insertonehalfscalars1},
\ref{insertonehalfscalars2},
\ref{insertonehalfscalars3} and 
\ref{insertonehalfscalars4} 
summarize the effect of the new mapping which results from the inclusion of the inserted operator in the action.  They result from $d_7$ for this case.  We ignore ZMT because it appears that they  cannot appear in perturbation theory with this action.

From Table \ref{insertonehalfscalars1}, we see that the operator $d_7$ maps a `mass-Yukawa' type of term into a spinor anomaly type of term.  In order to actually use this in perturbation theory, we would need to add a new `mass-Yukawa' type of term to the action, which would violate the non-renormalization theorem.  So this part of the cohomology appears to be irrelevant to the real problem,.

On the other hand,  rrom Table \ref{insertonehalfscalars1}, we see that the operator $d_7^{\dag}$ maps a spinor anomaly type of term into  a `mass-Yukawa' type of term.  What this means is that certain spinor anomaly terms can originate from `mass-Yukawa' type terms. In other words, if we put these `mass-Yukawa' type terms into the action, we could eliminate the anomaly.  

We will adopt the attitude that we could do that if we had to.  Of course doing it would violate the non-renormalization theorem. 

So this is quite confusing, as advertised.  

But what is fairly clear is that some spinor anomaly terms cannot be eliminated even if we agree to violate the non-renormalization theorem.  These  `highly anomalous' terms are the ones that we are looking for, at least at first. 

There is another level of confusion available here too.  Recall from the discussion in subsection \ref{subseczero},
that only certain kinds of `mass-Yukawa' type terms are in the cohomology space.  We could try to implement this fact also in our current discussion. I think that is the wrong approach, but the point is not really very clear. 

These `highly anomalous' terms are the solutions of the equations that originate, as shown in Table \ref{insertonehalfscalars3},  from the following generating function:
\be
{\cal T}
\equiv
{\cal T}_{\fr{1}{2}}
=
f_j^i 
\lt ( 2 h_{ik} \og^{jk}  + 6 h_{ikl} \og^{jkl} \rt )
\la{onegenfunc}
\ee

The advantage of choosing this way of looking at the problem is that \bitem
\item
Equations that derive from a generating function are easier to deal with and solve in general.  Once those solutions are better understood, it might be time to review the rest of the issues. 
\item
This seems the most logical approach anyway. 
\eitem

\begin{table}[hptb]
\caption{\Large First Table for  spinors coupled to a constant source $\f_{\a}$ with  ${\rm Dimension} \; \f_{\a}  = \fr{1}{2}$:   
Mappings of $N_{\rm Form}=4 \leftrightarrow N_{\rm Form} =  5$.  }

\vspace{.1in}
\framebox{ 
{ \Large $\begin{array}{ccccccc}  
\la{insertonehalfscalars1}\\
N_{\rm Form} = 4 &&
N_{\rm Form} =  5 

\\

\begin{array}{c} 
 (C \x^2 C) 
\\
\lt \{
m \ok^{ij} \A_i
\A_j
+ {\ok}^{ijk} \A_i \A_j
\A_k
\rt \}
\end{array}
 & \stackrel{d_7}{\lra} 
&
\begin{array}{c} 
 (C \x^2 C)  (\f C)  
\\
  \lt (  f_i^q \A_q
\rt ) 
\\
\lt \{
 2m \ok^{ij}  \A_j
+ 3{\ok}^{ijk}   \A_j
\A_k
\rt \}
\end{array}

\\
\\
\begin{array}{c} 
 (C \x^2 C)
\\
\lt \{ 
m \A_i \A_j \lt (
 2  \oh^{ik} \of^{j}_k  
\rt )
\rt.
\\
\lt.
+\A_i \A_j \A_k
\lt (
 3  \oh^{ijq} \of^{k}_q  
\rt )
\rt \}
\end{array}
 & \stackrel{d_7^{\dag}}{\longleftarrow} 
&
\begin{array}{c} 
 (C \x^2 C)  (\f C)  
\;
\\
  \lt \{ 
 m {\oh}^{ij}   \A_i
 \A_j+  {\oh}^{ijk}   \A_i
 \A_j \A_k \rt \}
\end{array}
\\

\\
\end{array}$ }}
\end{table}

\begin{table}[hptb]
\caption{\Large Second Table for  spinors coupled to a constant source $\f_{\a}$ with  ${\rm Dimension} \; \f_{\a}  = \fr{1}{2}$:  Equations arising from Table \ref{insertonehalfscalars1}. }
\vspace{.1in}
\framebox{ 
 { \Large
 $\begin{array}{ccccc}  
\la{insertonehalfscalars2}
 \\
 f_i^{(j} \ok^{k)i}
&=&
0
& {\rm Ignore?}
\\
\\
 f_i^{(j} \ok^{kl)i}
&=&
0
& {\rm Ignore?}
\\
\\
 2  \oh^{ik} \of^{j}_k  
&=&
0
& {\rm Keep}
\\
 3  \oh^{ijq} \of^{k}_q  
&=&
0
& {\rm Keep}

\end{array}$ }}
\end{table}

\begin{table}[hptb]
\caption{\Large Third Table for  action from inserted spinor coupled to a constant source $\f_{\a}$ with  ${\rm Dimension} \; \f_{\a}  = \fr{1}{2}$:   
Equations arising from 
Table \ref{insertonehalfscalars2} and  Table \ref{onehalfscalars5} assuming non-renormalization. All the equations can be derived from the Generating Function ${\cal T}$ of Equation (\ref{onegenfunc}).} 

\vspace{.1in}
\framebox{ 
 { \Large
 $\begin{array}{ccccc}  
\la{insertonehalfscalars3}

\\
 
   f_{i}^{j} {\og}^{ki}
+   f_{i}^{k} {\og}^{ji}
& =& \fr{\pa {{\cal T}} }{\pa h_{jk} } 
=0
\\
   f_{q}^{(i} {\og}^{jk)q}
& =& \fr{\pa {{\cal T}} }{\pa h_{ijk} } 
=0
\\
\lt (
2 g_{ij} \oh^{jq} +
6 g_{ijk} \oh^{jkq} \rt ) 
& =& \fr{\pa {\ov{\cal T}} }{\pa \of^i_j } 
=0
\\
 2  \oh^{k(i} \of^{j)}_k  
& =& \fr{1}{2} \fr{\pa {\ov{\cal T}} }{\pa g_{ij} } 
=0
\\
 3  \oh^{q(ij} \of^{k)}_q  
& =&  \fr{1}{2} \fr{\pa {\ov{\cal T}} }{\pa g_{ijk} } 
=0
\\

\end{array}$ }}
\end{table}

\begin{table}[hptb]
\caption{\Large Fourth Table for  spinors coupled to a constant source $\f_{\a}$ with  ${\rm Dimension} \; \f_{\a}  = \fr{1}{2}$:   
Mappings  of $E_{\infty} \ra {\cal H}$.  These are subject to the constraints in Table \ref{insertonehalfscalars3}, if we assume non-renormalization of the chiral action terms.
 } 

\vspace{.1in}
\framebox{ 
{\Large $\begin{array}{ccccccc}  
\la{insertonehalfscalars4}.
\\
E_{\infty} &&  {\cal H}\\
\\
 
\begin{array}{c}  
(C \x^2 C) \f^{\a} C_{\a} 
\\
\lt \{
 \oh^{ij} m \A_i \A_j  
+ \oh^{ijk}  \A_i \A_j \A_k 
\rt \}
\end{array}
& \lra &
\begin{array}{c}  
\int d^4 x \;  
\f^{\a} C_{\a}  
\\
\lt \{
 m \oh^{ij}  \oF_{2,ij} 
+ \oh^{ijk}  \oF_{3,ijk} \rt \}
\end{array}
\\
\\
\begin{array}{c}  
 (C \x^2 C) 
\\
\f^{\a} \y^i_{\a} 
\lt \{
 f_i^j \A_j
 \rt \}
\end{array}
 & \lra &
\begin{array}{c}  
\int d^4 x \;
 \f^{\a} 
\lt \{
f_i^j
\lt [
\oG^{i} \A_j
C_{\a} 
\rt.
\rt.
\\
\lt.
+
\lt (
\pa_{\a \dot \b} A^i 
+ C_{\a}\oY^{i}_{ \dot \b} \rt )
 \oy_{j}^{\dot \b}
\lt.
-
\y^i_{\a} 
\oF_{1,i} \rt ]
\rt \}
\end{array}

\end{array}$ }}
\end{table}


\chapter{
Solutions of Constraint Equations}

\la{notsimpleconstraintchapter}

\section{Orthonormal Basis}

To solve the  Constraint equations for the first non-trivial case, when
 $N_{\rm Chiral } = 2 $ ,  let us  use an orthonormal basis for the $N_{\rm Chiral } = 2 $ space :
\be
a^i , b^i
\ee
\be
a_i , b_i
\ee
with 
\be
a^i b_i = 0
\ee
\be
b^i a_i = 0
\ee
\be
a^i a_i = 1
\ee
\be
b^i b_i = 1
\ee
and write all the tensors in terms of these, taking into account the symmetry properties that are appropriate:
\be
f_i^j = 
t_1 a_i a^j   
+
t_2  a_i b^j 
+
t_3 b_i a^j   
+
t_4 b_i b^j   
\ee
and 
\be
\og^{ij} = 
m_1 a^i a^j
+
m_2 \lt ( a^i b^j + b^i a^j \rt )
+
m_3 b^i b^j 
\ee
\be
\og^{ijk} = 
e_1 a^i a^j a^k
+
e_2 \lt ( a^i a^j b^k + b^i a^j a^k + a^i b^j a^k \rt )
\ee
\be+
e_3 \lt ( a^i b^j b^k + b^i a^j b^k + b^i b^j a^k \rt )
+
e_4 b^i b^j b^k 
\ee

\be
h_{ij} = 
q_1 a_i a_j
+
q_2 \lt ( a_i b_j + b_i a_j \rt )
+
q_3 b_i b_j 
\ee
\be
h_{ijk} = 
r_1 a_i a_j a_k
+
r_2 \lt ( a_i a_j b_k + b_i a_j a_k + a_i b_j a_k \rt )
\ee
\be+
r_3 \lt ( a_i b_j b_k + b_i a_j b_k + b_i b_j a_k \rt )
+
r_4 b_i b_j b_k 
\ee

\section{Collection for ${\cal T}_{\fr{1}{2}}$}

It is easy to rewrite the generating function in terms of the new independent variables: 
\[
{\cal T}_{\fr{1}{2}} =
f_i^j 
\lt (
\og^{iq} h_{jq} 
+
h_{jpq} 
\og^{ipq}  
\rt )
\]
\[
=t_1    m_1    q_1  
+
t_1 m_2 q_2 
\]
\[
+
t_2 
m_1
q_2 
+
t_2 m_2 q_3
\]
\[
+
t_3 
m_2
q_1 
+
t_3 
m_3
q_2 
\]
\[
+
t_4 
m_2
q_2 
+
t_4 
m_3
q_3 
\]
\[
+
t_1 e_1 r_1 +
t_1 2 e_2 r_2 +
t_1 e_3 r_3 
\]
\[
+
t_2 e_1 r_2 +
t_2 2 e_2 r_3 +
t_2 e_3 r_4 
\]
\[
+
t_3 e_2 r_1 +
t_3 2 e_3 r_2 +
t_3 e_4 r_3 
\]
\be
+
t_4 e_2 r_2 +
t_4 2 e_3 r_3 +
t_4 e_4 r_4 
\la{geng2}
\ee

\section{Rewritten form for the  Constraints}

\la{rewritten}

The equations then become

\subsubsection{ $f\ra t$ Equations}

 \be
\fr{\pa {\cal T}_{\fr{1}{2}} }{\pa t_i} 
 =0 ; i=1,2,3,4
\ee

\subsubsection{ $ \og \ra   m,e $ Equations}

 \be
\fr{\pa {\cal T}_{\fr{1}{2}} }{\pa m_i} 
 =0  ;i=1,2,3
\ee
 \be
 \fr{\pa {\cal T}_{\fr{1}{2}} }{\pa e_i} 
 =0  ; i=1,2,3,4 
\ee

\subsubsection{ $ h \ra  q,r $ Equations}

  \be \fr{\pa {\cal T}_{\fr{1}{2}} }{\pa q_i} 
 =0 ;i=1,2,3
\ee
\be \fr{\pa {\cal T}_{\fr{1}{2}} }{\pa r_i} 
 =0 ;i=1,2,3,4
\ee

\section{A somewhat Interesting Solution for the ${\rm Dim} \f = \fr{1}{2}$ Case}
\la{intsolonehalf}

It is not proposed in this paper to try to solve the constraint equations in any general way.   It looks fairly easy to solve the above equations for this special case using a simple MAPLE program. There are 18 equations in 18 unknowns.  There appear to be 78 solutions, most of which have plenty of zeros in them.  

We will look at one of these solutions.  Solution  ${\cal S}_{71}$ below has no zeros, but it is equivalent to a solution with some zeros in it.

  It has  six independent independent parameters, namely  $e_3,t_2,
t_3,
t_4,
r_3,
q_2$; 
and the other twelve parameters are expressed in terms of these by rational functions. 

\be {\cal S}_{71} = [E, T,R,M,Q]
\ee
where
\be
E = [e_1,e_2,e_3,e_4]
= [{\frac {{t_{{4}}}^{2}e_{{3}}}{{t_{{2}}}^{2}
}},-{\frac {t_{{4}}e_{{3}}}{t_{{2}}}},e_{{3}},-{\frac {t_{{2}}e_{{3}}}
{t_{{4}}}}] \ee

\be T = [t_1,t_2,t_3,t_4]
 = [ {\frac {t_{{3}}t_{{2}}}{t_{{4}}}},t_{{2}},t_{{3}},t_{{4}} ] \ee \be R = [r_1,r_2,r_3,r_4]
 =[{\frac {{t_{{4}}}^{2}r_{{3}}}{{t_{{3}}}^{2}}},-{\frac {t_{{4}}r_{{3
}}}{t_{{3}}}},r_{{3}},-{\frac {t_{{3}}r_{{3}}}{t_{{4}}}}]\ee 
\[ 
M = [m_1,m_2,m_3]
\]
\be
 = [  {\frac {e_{
{3}}r_{{3}} \left( {t_{{4}}}^{2}+t_{{3}}t_{{2}} \right) t_{{4}}}{t_{{3
}}{t_{{2}}}^{2}q_{{2}}}},-{\frac {e_{{3}}r_{{3}} \left( {t_{{4}}}^{2}+
t_{{3}}t_{{2}} \right) }{t_{{3}}t_{{2}}q_{{2}}}},{\frac {e_{{3}}r_{{3}
} \left( {t_{{4}}}^{2}+t_{{3}}t_{{2}} \right) }{t_{{3}}q_{{2}}t_{{4}}}
}]\ee \be Q = [q_1,q_2,q_3]=
  [-{\frac {t_{{4}}q_{{2}}}{t_{{3}}}},q_{{2}},-{\frac {t_{{3}}q_{{2}}
}{t_{{4}}}}]\ee 

If we take transformations such as:
\be
a'_i = \fr{1}{t_2^2 + t_4^2} \lt ( t_4  a_i  + t_2 b_i \rt ) 
\ee
this reduces to a form with plenty of zeros.  But is it decoupled?
There is little point in looking in detail at this issue for this particular example, since there are so many examples to look at.  There are many questions that need general answers:

\ben
\item
Are all possible solutions decoupled in some sense?  In general we must consider  solutions ( See Chapter \ref{remarkschapter}) 
where we start with more general expressions such as a linear combination (with suitable powers of $m$) of the following 
\be
 \int d^4 x \d^2 \oq  f^{j_1 \cdots j_n}_{a, i}
{\hat \f}^{a \a} {\hat \y}^i_{\a} {\hat \A}_{j_1} \cdots  {\hat \A}_{j_n } \ee
and consider the cases for general  $N_{\rm Chiral}$, and also consider spontaneous breaking by $A^i \ra m v^i + A^i$.  
\item
What if we add supersymmetric gauge theory? 
\item
It would not necessarily be disappointing if `Interesting' solutions ( i.e. solutions that are not `decoupled' and that have supersymmetry anomalies) are rare.  It is not inconceivable they could shed some light on the bizarre spectrum of the elementary particles.

\een

Evidently there is quite a lot to look at.  No further effort in this direction will be made here.

\la{remarksolt}

\section{General Remarks on the Solutions of the Constraint Equations}

\la{genremsolcon}

 Solving these equations for the case where the dimension is higher or where there are more than two chiral fields is a task for the future. Preliminary results indicate that there appear to be plenty of solutions, some of which have non-zero values for all or nearly all of the parameters.  So there are ghost charge zero invariants and ghost charge one invariants as well as non-zero couplings and masses.  Presumably  we can forget about complications such as Spontaneous Gauge Breaking and Goldstone Bosons, at least at first.

It is not clear that there are anomalies however.  These solutions need to be examined carefully, and if a given solution looks promising, the diagrams need to be carefully calculated, the coboundary terms need to be recognized and the BRS identity needs to be checked.

In the event that there are no anomalies in the simplest cases, the next step would be to use $N_{\rm Chiral } = 3$ or ${\rm Dimension} = \fr{9}{2}$ as in Table \ref{threespininv}.   It seems likely that there must be a large number of solutions as the dimensions and number of fields increase.  The equations for $N_{\rm Chiral } = 3$ are easy to write down.  One just adds another orthogonal vector to the set above and proceeds to look for solutions.  However even the massless case with $N_{\rm Chiral } = 3$ generates 29 equations in 29 unknowns. My simple MAPLE program, which works quickly for the 18 variable $N_{\rm Chiral } = 2$ case, mentioned above, looks inadequate to treat this case. 

It is of course possible that no anomalies are present unless we include gauge interactions, which introduce new tensor structures.      It is possible that no anomalies are present even though they seem to be possible.  Calculations and investigations need to be done to see what happens.

\section{Vacuum Expectation Value and Shift of $A^i$}

\la{vevsection}

The above equations bear some resemblance to the problems connected with the case where we start with the assumption that there is a term $\int d^4 x \; m^2 g_i F^i$ in the action.  This implies the existence of a vacuum expectation value, Goldstone bosons and possibly Goldstone fermions etc.   Obviously one needs to analyze this subject if one wants to make contact with the SSSM, so this is only a start. 

The VEV is determined so that the potential is at a minimum.  The value of the potential will be zero or greater.

The potential is
\be
V =  \oF'_{1 i} 
 F_1^{'i} 
\ee
where
\be
 F_1^{'i} 
=
- \lt ( 
m^2 {\ov g}^{i}   +
m {\ov g}^{ij} {\ov A}_j  +
 {\ov g}^{ijk} {\ov A}_j  {\ov A}_k 
\rt )
\ee

One considers the possibility of a  nonzero vacuum expectation value 
\be
<A^i>_0 = m v^i
\ee
One adjusts the value $v^i$ so as to choose a minimum for the potential.

If there exists a solution of 
\be
< F_1^{'i}> 
=
-m^2 \lt ( 
 {\ov g}^{i}   +
{\ov g}^{ij} {\ov v}_j  +
 {\ov g}^{ijk} {\ov v}_j  {\ov v}_k 
\rt ) =0
\ee
then the minimum of the potential is at zero. If no such solution exists, then the vacuum expectation values of the auxiliary fields are not zero, and the Vacuum has a non-zero energy density:
\be
V=< F_1^{'i}> 
< \oF'_{1 i}> 
\ee
 This is the origin of the cosmological constant problem for spontaneous breaking of supersymmetry.
These kinds of equations were considered in  
\ci{oRaif}.


\chapter{Remarks}
\la{remarkschapter}

\Large

\section{Some More General Cohomology Results}

Here are some additional results from the spectral sequence.  The following objects are in the cohomology space of the Physical $\d$ in Table \ref{physicaltable}, but only when the mass $g_{ij}$ and the coupling $g_{ijk}$ and the coefficients $ f^{j_1 \cdots j_m}_{i_1 \cdots i_n}
$ are constrained as discussed below.
The hats indicate that the objects 
${\hat \y}^{i_1}_{\a_1}$ and ${\hat \A}_{j_1}$ are superfields.  Their form was discussed in Chapter  \ref{superfieldschapter}.

Ghost charge zero:
\be 
\F_{\a_1 \cdots \a_n} = 
 f^{j_1 \cdots j_m}_{i_1 \cdots i_n}
\int d^2 {\ov \q}
{\hat \y}^{i_1}_{(\a_1}\cdots  {\hat \y}^{i_n}_{\a_n)} {\hat \A}_{j_1} 
\cdots {\hat \A}_{j_m} \in {\cal H}^0
\la{superzero}
\ee

Ghost charge one
\be 
\oH_{(\a_1 \cdots \a_n} C_{\a_{n+1})}
\ee
\be
= 
 f^{'j_1 \cdots j_m}_{i_1 \cdots i_n}
\int d^2 {\ov \q}
{\hat \y}^{i_1}_{(\a_1}\cdots  {\hat \y}^{i_n}_{\a_n} C_{\a_{n+1})}
{\hat \A}_{j_1} 
\cdots {\hat \A}_{j_m} \in {\cal H}^1
\la{superone}
\ee

Ghost charge two
\be 
{\ov K}_{(\a_1 \cdots \a_n} C_{\a_{n+1}} C_{\a_{n+2})}
\ee
\be
= 
 f^{'' j_1 \cdots j_m}_{i_1 \cdots i_n}
\int d^2 {\ov \q}
{\hat \y}^{i_1}_{(\a_1}\cdots  {\hat \y}^{i_n}_{\a_n} C_{\a_{n+1}} C_{\a_{n+2})}
{\hat \A}_{j_1} 
\cdots {\hat \A}_{j_m} \in {\cal H}^2
\la{supertwo}
\ee
etc. 

\subsection{Constraints}

When we couple the above cohomology generators into the action with a `constant source superfield', as discussed below for some simple cases, then we find that the mass coefficients $g_{i}$, $g_{ij}$ and the coupling coefficients $g_{ijk}$   and the operator and anomaly coefficients $ f^{j_1 \cdots j_m}_{i_1 \cdots i_n}$, $ f^{'j_1 \cdots j_m}_{i_1 \cdots i_n}$, $ f^{''j_1 \cdots j_m}_{i_1 \cdots i_n}$ etc. are all related by constraints.     These constraints can all be derived from a set of ` Generating Functions' which will be described below for some interesting special cases.

 These constraints are most easily derived from the spectral sequence. We shall give a heuristic derivation of them.  The constraints can however be demonstrated in the theory itself for any specific case, and that is a necessary demonstration if one wants to use the results for calculation of anomalies anyway. 

Here is the heuristic derivation.  We already introduced and used the nilpotent operator of the form:
\be
d_{\rm Mass\; Yukawa} \equiv d_{5,\rm MY} = \lt \{
m \og^{ij} \A_j + 
 \og^{ijk} \A_j \A_k  
\rt \} C_{\a}
\y^{i \dag}_{\a}
\la{massYukop} + *
\ee
This operator comes from the terms of the action where we integrate the auxiliary. This is one of the consequences of using the `Physical Form' of the BRS-ZJ identity above. This operator interpolates between objects of the form above where we eliminate the hats and remove the integrals:  

Ghost charge zero:
\[ 
\F'_{\a_1 \cdots \a_n} = 
\]
\be
(C \x^2 C)
f^{j_1 \cdots j_m}_{i_1 \cdots i_n}
{  \y}^{i_1}_{(\a_1}\cdots  {  \y}^{i_n}_{\a_n)} {  \A}_{j_1} 
\cdots {  \A}_{j_m} \in E^0
\la{zerostuff}
\ee

Ghost charge one
\[ 
\oH'_{(\a_1 \cdots \a_n} C_{\a_{n+1})}
\]
\be
= 
(C \x^2 C)
 f^{'j_1 \cdots j_m}_{i_1 \cdots i_n}
{ \y}^{i_1}_{(\a_1}\cdots  { \y}^{i_n}_{\a_n} C_{\a_{n+1})}
{  \A}_{j_1} 
\cdots {  \A}_{j_m} \in {  E}^1
\la{onestuff}
\ee

Ghost charge two
\[ 
{\ov K}'_{(\a_1 \cdots \a_n} C_{\a_{n+1}} C_{\a_{n+2})}
\]
\be
= 
(C \x^2 C)
 f^{'' j_1 \cdots j_m}_{i_1 \cdots i_n}
{\y}^{i_1}_{(\a_1}\cdots  { \y}^{i_n}_{\a_n} C_{\a_{n+1}} C_{\a_{n+2})}
{ \A}_{j_1} 
\cdots { \A}_{j_m} \in E^2
\la{twostuff}
\ee
etc. 

Then there is another operator at the next level which depends on the form of the invariant which we insert.  For the  case that we have looked at in this paper this operator is:
\be
d_{{\rm Insert}; \fr{7}{2}} = 
\f^{\a} C_{\a} \lt \{
 \lt ( 
 f^i_j  \A_i \rt )
\A_j^{\dag} + f^i_j  \y_{i \a}  \y_{j \a}^{\dag} \rt \}
\la{Insertsevenop}
\ee

Finding the cohomology of the operators $d_{\rm Mass\; Yukawa} $ and $d_{{\rm Insert}; \fr{1}{2}} $ in the  space spanned by the forms (\ref{zerostuff}), 
(\ref{onestuff}) and  
(\ref{twostuff})   yields the  Constraints referred to below.

Attempting to derive these results in detail by using the spectral sequence method would make this paper longer without adding much of any use at the present time.

\section{Summary of Spinor Invariants and Anomalies}

In Table \ref{threespininv}, we summarize the first two cohomologically non-trivial spinor objects that could be inserted into the action.  We would couple these to a constant anticommuting  source  $\f^{\a}$ to form the simplest objects discussed above.
In Table \ref{threespininvT}, we summarize the  Generating Function for these two cohomologically non-trivial  spinor objects. 
 
In Table \ref{bispininv}, we do the same for the first cohomologically non-trivial bispinor object that could be inserted into the action.  We would couple this to a constant  commuting  source  $K^{(\a \b)}$ to form the simplest object that could possibly be anomalous with these quantum numbers.  This was not discussed above but is an obvious generalization.   We note that there are two   Generating Functions for this first cohomologically non-trivial  bispinor object. Both of these would have a zero variation for a cohomologically nontrivial composite bispinor object to exist.

The generalization to higher dimensions and higher spins evidently becomes increasingly complex, but looks fairly straightforward to formulate.

 \begin{table}[thbp]
\caption{\Large   First Two Spinor Invariants}
\vspace{.1in}
\framebox{ 
{\Large $\begin{array}{ccccc}  
 \la{threespininv}
{\rm Dim} \;\f = &\F_{\a}=\int d^2 \oq  & \oH C_{\a}  = \int d^2 \oq 
\\
\fr{1}{2} 
&  \lt \{
f_i^j {\widehat \y}^i_{\a} {\widehat \A}_j 
\rt \}
& 
  \lt \{
  
\oh^{ij} m {\widehat \A}_i {\widehat \A}_j
\rt.
\\
&&
\lt.
+
\oh^{ijk} m {\widehat \A}_i {\widehat \A}_j
{\widehat \A}_k
\rt \} C_{\a} 
\\

- \fr{1}{2} 
&  \lt \{
f_i^j m {\widehat \y}^i_{\a} {\widehat \A}_j 
\rt.
& 
  \lt \{
\oh^{ij}  m^2 {\widehat \A}_i {\widehat \A}_j
\rt.
\\
&
\lt.
+
f_i^{jk}  {\widehat \y}^i_{\a} {\widehat \A}_j{\widehat \A}_k
\rt \}
&
\lt.
+
\oh^{ijk} m {\widehat \A}_i {\widehat \A}_j
{\widehat \A}_k
\rt.
\\
&&
\lt.
+
\oh^{ijkl}  {\widehat \A}_i {\widehat \A}_j
{\widehat \A}_k
{\widehat \A}_l
\rt \} C_{\a} 
\\
\end{array}$
}} \end{table}

 \begin{table}[thbp]
\caption{\Large   ${\cal T}_{\rm Dim}$ for First Three  Spinor Invariants }
 \la{threespininvT}
\vspace{.1in}
\framebox{ 
{\Large $\begin{array}{ccccc}  
 {\rm Dim}\;\f =  & {\cal T}
\\
\fr{1}{2} 
  &

f_j^i 
\lt ( 2 h_{ik} \og^{jk}  + 6 h_{ikl} \og^{jkl} \rt )

\\
\\
\fr{-1}{2} 
  &
f_j^i 
\lt ( 2 h_{ik} \og^{jk}  + 6 h_{ikl} \og^{jkl} \rt )

+ f_j^{il} 
\lt ( 2 h_{ikl} \og^{jk}  + 6 h_{iklq} \og^{jkq} \rt )

\\

\end{array}$
} }
\end{table}

 \begin{table}[thbp]
\caption{\Large   First Bispinor}
\vspace{.1in}
\framebox{ 
{\Large $\begin{array}{ccccc}  
 \la{bispininv}
 {\rm Dim} \; K_{\a \b} &Q_{(\a \b)}=\int d^2 \oq  & \F_{(\a} C_{\b)}  = \int d^2 \oq 
& {\ov H} C_{\a}C_{\b}  = \int d^2 \oq 
\\
4
&  k_{[ij]} {\widehat \y}^i_{(\a} {\widehat \y}^j_{\b)} & 
 \lt \{
  f^{i}_j m {\widehat \y}^j_{(\a} 
{\widehat \A}_i
\rt.
& 
 \lt \{
 {\ov h}^{ij} m^2  
{\widehat \A}_i
{\widehat \A}_j
\rt.
\\
&&
+
\lt.  f^{ik}_j  {\widehat \y}^j_{(\a} 
{\widehat \A}_i
{\widehat \A}_k
\rt \} C_{\b)} 
&
\lt.
  {\ov h}^{ijk} m  
{\widehat \A}_i
{\widehat \A}_j
{\widehat \A}_k
\rt.
\\
&&&
\lt.
  {\ov h}^{ijkl}   
{\widehat \A}_i
{\widehat \A}_j
{\widehat \A}_k
{\widehat \A}_l
\rt \}  C_{\a} C_{\b} 
\\

\end{array}$
}} 
\end{table}



\chapter{
Conclusion}
\la{conclusionchapter}

\section{Calculation of Anomalies at one loop}
\la{calcFeyn}

Here is an example of the kind of diagram that we need to evaluate and then test for BRS invariance. The hermitian positive semi-definite mass matrix is
\be
M^i_j = \og^{ik} g_{jk}.
\ee
Note that the diagram is in fact linearly divergent.

 The BRS identity splits up into a number of identities, which we will call Ward identities.  The cohomology tells us that it could happen that we cannot choose momenta in such a way that all the Ward identities are true simultaneously.  The details of that are straightforward but they do require some work.  

\begin{picture}(400,280)
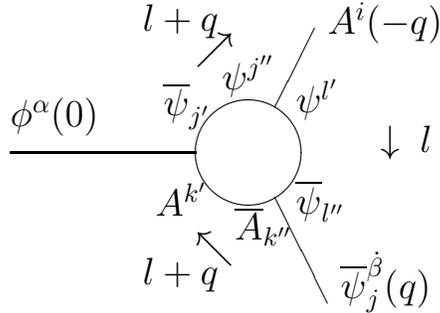
\put(100,02){Feynman Diagram for ${\cal G}_1[\f, A, \oy] $ }
\put(0,250){    ${\cal G}_1[\f, A, \oy]  =
\f^{\a} A^i (-q) {\ov \y}_{j}^{ \dot \b}(q)  
   I_{\a \dot \b ,i}^{j }    $ where}
\put(0,210){ $  I_{\a \dot \b, i}^j  =
{\displaystyle \int d^4 l\;   
\lt \{
f_{k'}^{j'} \lt ( \fr{ (l+q)^2  \;    }{ 
(l+q)^2+ m^2M }  \rt )^{j''}_{ j'}   
g_{i j'' l'} \lt ( \fr{ l_{\a \dot \b}   }{  l^2+ m^2M  } \rt )^{l'}_{l''} \rt. }   $ }
\put(0,170){ $ \lt.
{\displaystyle 
 {\ov g}^{l'' k'' j}
\lt ( \fr{ 1  }{ (l+q)^2+ m^2M}   \rt )^{k'}_{k''} } 
\rt \}
 $ }
%
%
%
\put(210,92){\line(1,2){15}}
\put(210,58){\line(1,-2){20}}
\put(180,75){\line(-1,0){70}}
\put(200,75){\circle{40}}
%
\put(230,120){$A^i(-q)$}
\put(235,20){${\ov \y}_{j }^{\dot \b}(q)$}
\put(110,85){{$\f^{\a}(0)$}}
\put(168,89){${\ov \y}_{j' }$}
\put(190,99){${\y^{j'' }}$}
\put(218,90){${\y^{l'}}$}
\put(218,52){${\ov \y}_{l''}$}
\put(195,42){${\ov A}_{k''}$}
\put(165,51){${A}^{k'}$ }
%
\put(180,110){$\nearrow$}
\put(160,121){$l+q$}
\put(250,75){$\downarrow$}
\put(265,75){$l$}
\put(180,35){$\nwarrow$}
\put(160,25){$l+q$}
%
\end{picture} 

\vspace{1 cm}
It should be noted however that 
\ben
\item
No regularization scheme is necessary to check the Ward identities. The Ward identities are true up to a shift of a linearly (or worse) divergent integral in momentum space, and that always gives a finite result \ci{elias}. 
\item
The right hand side of \ref{brsvar}, which measures the violation of the BRS identity, as calculated from the diagrams, is always a finite local polynomial in the field variables which contains at least one derivative in each monomial.  
\item
The appearance of the anomalies is related to adding terms to convert these derivative terms into non-derivative terms using the equations of motion.  This is of course the point of cohomology--an anomaly is defined only up to a coboundary.
\item
Some solutions of the Constraints will clearly give no anomaly because of some `decoupling' of the action.  Others may not decouple and need a calculation.

\een

\section{Speculations}
\la{speculationsection}

\subsection{The Standard Supersymmetric Model (SSSM)}
\la{SSSMsection}

The  Standard SS model has $N_{\rm Chiral}=53$, or, if we exclude all right handed neutrino superfields,  $N_{\rm Chiral}=50$.  

Here are some questions:
\ben
\item
Is  there any  interesting cohomology in the SSSM?  
\item
What effect do spontaneous breaking of the gauge symmetry and gauged supersymmetry have on the cohomology?  Considerable progress has been made on this problem and a paper is in preparation.  There is a complicated structure of cohomology here too.  The basic answer is that the   Constraints are supplemented by further constraints relating to the gauge invariance and the Goldstone matrix which characterizes the spontaneous gauge symmetry breaking.
\item
Do three families help to create interesting solutions to the   Constraints?
\item
Are there supersymmetry anomalies  in the SSSM?
\item
How does the pattern of supersymmetry breaking relate to the solutions of the   Constraints?
\item
Do the anomalies generate physically realistic supersymmetry breaking in the SSSM?
\een

Here are some promising aspects:
\ben
\item
The relevant Feynman diagrams are linearly (or more) divergent, so the ambiguity needed for anomalies is in fact present.
\item
There seem to be plenty of solutions of the  Constraints, particularly when there are lots of particles 
\item
So the cohomology is also present to give rise to anomalies
\item
It is not preposterous to note that three families of particles will help in any given sector to ensure that there are plenty of solutions to the Constraints.  With only one family, it might be more difficult to get interesting cohomology.

\item
Because of the Legendre transform problem for the auxiliary fields, there is some doubt that the theory can really be formulated correctly in superspace, particularly including the spinor cohomology and the constraints found here, and particularly when there are massless particles present.
\item
This means that the likelihood of supersymmetry anomalies is increased, because there is no obviously correct argument from `manifest supersymmetry' to show that they should not be present.

\item
This pattern continues when we add gauge theory supersymmetry, which has plenty of cohomology too. 
\item
The bizarre solutions to the  Constraints give some promise of yielding interesting patterns of supersymmetry breaking, and maybe some other surprises too. 
\item
The supercurrent and the supercharge do not seem to play much of a role in these potential anomalies. Issues relating to these were discussed in \ci {witten1} and \ci{witten2}. The resulting conundrum might be related to the fact that the definitions of the supercurrent and the supercharge naturally bring in the consideration of supergravity, and in that context it is tricky to define a conserved supercharge.
\item
Also in that context, it is probable that in the local supersymmetric theory with supergravity, these anomalies are not present.  They only show up in composite operators, which may well include all the observable particles, since all of them are formed in a sense from operators that include the Higgs fields and their vacuum expectation value.  Supergravity couples only to the elementary particles however, which are supersymmetric and unobservable. 

Because one must not couple gravity both to the quarks and to the protons, for example, it is clear that an anomaly which affects bound states might not affect the elementary particles.  One must avoid `double counting', and this might provide a way of making these apparently inconsistent approaches consistent.

\een

\subsection{Speculations on some of the Ten Problems for the new Millenium from the 2000 Superstring Conference at Ann Arbour, Michigan}

Are these ideas doomed from the start?  Is the notion that  supersymmetry breaking occurs through anomalies in composite operators obviously wrong in some way? 

To test whether one can show that the ideas are obviously wrong, by {\em reductio ad absurdum},  it is tempting to see what these ideas imply about some of the outstanding problems, to see if there is an obvious contradiction. 

 One way to review this issue is to look at the following list of problems, and see what this mechanism for supersymmetry breaking would imply about them. Comments about the problems are made after each problem where a comment seems possible.

TEN PROBLEMS

1.  {\em Are all the (measurable) dimensionless parameters that characterize the physical universe calculable in principle or are some merely determined by historical or quantum mechanical accident and uncalculable? } 

{ The  Constraints appear to give us relations among some of the dimensionless parameters.  If supersymmetry anomalies exist, these relations are likely to be experimentally relevant.

For the standard model, the  constraints will involve and possibly constrain the parameters of the Cabibbo-Kobayashi-Maskawa matrix and its leptonic analog.}

2.  {\em How can quantum gravity help explain the origin of the universe?}

3.  {\em What is the lifetime of the proton and how do we understand it?}

{ This is a special case of problem 1 above.}

4. {\em  Is Nature supersymmetric, and if so, how is supersymmetry broken? }

{ If there are supersymmetry anomalies, then necessarily they create some form of supersymmetry breaking--the big question is whether that is an experimentally viable form. }

5. {\em Why does the universe appear to have one time and three space dimensions?}

{ The cohomology and anomalies are very closely linked to the dimension and signature of spacetime, as is always true for all features of supersymmetric theories }

6. {\em Why does the cosmological constant have the value that it has, is it zero and is it really constant? }

{ In this context the following remarks gleaned from 
\ci{internetofficialstringtheory} set the stage}:

\hspace{.5in} \parbox{4.in}{  When supersymmetry is not broken, it's easy to get a zero cosmological constant in string theory.  And although a zero cosmological constant might not be the truth, it's incredibly close to the truth.  If you break supersymmetry, if you do it the wrong way, you're going to get a cosmological constant that's much too big, and then you may well get associated problems, such as instabilities, runaways and so on.  So it's easy to find ways that string theory could break supersymmetry, but they all have bad consequences. }

\vspace{.2in}

{ As indicated in these remarks, supersymmetry naturally yields a zero value for the vaccuum energy, even when the gauge symmetry is spontaneously broken. This is only violated if supersymmetry itself is spontaneously broken. Breaking of supersymmetry by supersymmetry anomalies in composite operators would (apparently) leave the vacuum with zero energy, yielding a natural value of zero for the cosmological constant, even after supersymmetry breaking}

7.  {\em What are the fundamental degrees of freedom of M-theory (the theory whose low-energy limit is eleven-dimensional supergravity and which subsumes the five consistent superstring theories) and does the theory describe Nature?}

{ The argument here indicates that manifest supersymmetry cannot be preserved in quantum field theory, because we must integrate the auxiliaries to correctly formulate the BRS identities.  The same argument should be  true for higher dimensional supersymmetric theories.  That means that one should look carefully at their cohomology to see if there are, say, {\rm local} supersymmetry anomalies, that might kill theories with spacetime dimension greater than 3+1, bringing the possible theories down to, or nearer to, the four-dimensional world we inhabit.}

8.  {\em What is the resolution of the black hole information paradox?}

9.  {\em What physics explains the enormous disparity between the gravitational scale and the typical mass scale of the elementary particles?}

{Supersymmetry breaking by supersymmetry anomalies in composite operators might avoid the problem of fine tuning, because of the non-renormalization theorems in 
Supersymmetry.  But they do not appear likely to account for the vast difference in scale}

10.  {\em Can we quantitatively understand quark and gluon confinement in Quantum Chromodynamics and the existence of a mass gap? }

{ If this scheme for supersymmetry breaking is on the right track to some extent, then supersymmetry might be effectively a perfect symmetry for QCD. The calculations would be quite different and possibly even easier because of the greater symmetry.}

\section{Non-Renomalization and the Cohomology}

  The terms in certain Tables above, such as Table \ref{insertonehalfscalars1},
involve new tensors  $k_{ij}$ and $k_{ijk}$ which are just like the mass and Yukawa terms.

In a normal theory, one would want to add all the cohomologically non-trivial operators with the right quantum numbers to form the action in the first place.  These terms are of that form.   But the conventional understanding of Supersymmetry tells us that we should not add any counterterms that are of the chiral type, and these are exactly that.

 Moreover, if we do add them, then we are changing the original theory, since the new terms are of the same kind as the originals.  The resulting theory is really now a theory with the symmetry breaking term 
 $g_{i}+ k_{i}$ and the 
mass matrix $g_{ij}+ k_{ij}$ and the Yukawa tensor $g_{ijk}+ k_{ijk}$, and not what we started with.  And so on. 

It appears that the sensible course is to recognize that there are such solutions  $  k_{ij}$ and   $ k_{ijk}$, and then to refrain from  adding them to the action.

But then what do we do with the equations $d_7$ in Table \ref{specseqsum}?  My guess is that we pretend that the cohomology  that restricts   $ k_{i}$, $ k_{ij}$ and   $  k_{ijk}$ is not really there, so that this equation acts as if there were no projection onto these states, and gives us the Constraints as though there were only half of the $d_7$ operator acting. This is what is done in Chapter \ref{diagramchasewithinsertchapter}.

This guess might be on the wrong track. It is a working hypothesis to calculate the anomalies, which are likely to shed light on this problem if they do indeed exist.

\section{Rules for the calculation of Supersymmetry Anomalies}
\la{envoisection}

All of these speculations are of course empty unless there really are supersymmetry anomalies in composite operators.  We have seen that the simplest possible such anomalies can occur in one-loop diagrams as follows:
\bitem
\item
 The action is of the form in equation (\ref{theanomaction}).
where the subterms are found in (\ref{Aphysical}), (\ref{insertsevenhalves})  and (\ref{sevencomplete}).  I suspect that the term (\ref{sevencomplete}) plays no role since it is of order $\f^{\b} \f_{\b}$ and the anomaly is of order $\f^{\b} $.

\item
This action 
gives rise to the transformations in Table 
\ref{deltaInsertphionehalf}. 
\item
The anomaly appears in the form of equation (\ref{brsvar}):
\be
\d {\cal G}_{\rm 1PI-One \; Loop,\f}
 = 
{\cal A}^1_{\f}
\ee
\item
The anomaly ${\cal A}^1_{\f}
$ is of the form
\ref{insertonehalfanomfromfivechap},
 where the $\oF$ are as defined in 
Table 
\ref{compterms}
\item

The coefficient tensors in the action and the anomaly and the mass and Yukawa coefficients must all be constrained to be a  solution of the constraint equations in
Table \ref{insertonehalfscalars3},
 which can be rewritten and solved as explained in subsection \ref{rewritten}
and \ref{remarksolt}.
\item
The computation involves an analysis of the Ward identities in the context of Feynman diagrams like the example in Section \ref{calcFeyn}.
\eitem

The first questions which must be resolved are: 
\ben
\item
Are there `interesting' solutions to the Constraints?  
\item
`Interesting' solutions could be defined as solutions  which are not, in some sense, the direct sum of 
\ben
\item
 a free sector with cohomology at $N_{\rm Ghost}=0,1$, plus 
\item
 a non-free sector without cohomology at $N_{\rm Ghost}=0,1$. 
\een
\item
Assuming that there are solutions which are `interesting',
 do the corresponding potential  supersymmetry anomalies exist in perturbation theory?
\item
Supposing that anomalies exist, what does this mean when the inserted operators are taken at non-zero momentum: what is the physical interpretation that results from the generated kinetic terms for the sources? and the new cohomology that results from that?
\item
What role does the shift $A^i \ra mv^i + A^i$ play in the invariants
$\int d^4 x \d^2 \oq  f^{j_1 \cdots j_n}_{a, i}
{\hat \f}^{a \a} {\hat \y}^i_{\a} {\hat \A}_{j_1} \cdots  {\hat \A}_{j_n }$  and their cohomology and the solutions of the constraints?

\een


\begin{center}
 Acknowledgments
\end{center}

I am grateful for useful conversations with many colleagues, including F. Brandt, D. Baxter, M. Duff, M. Einhorn, V. Elias, M. Forbes, M. Henneaux, R. Minasian, J. Plefka, L. O'Raiffertaigh, C. Pope, J. Rahmfeld, E. Sezgin, K. Stelle, J. C. Taylor, P. van Nieuwenhuizen, P. West,  S. Weinberg  and E. Witten.  I have benefited greatly  over the years from the insights of  C. Becchi and R. Stora.  I thank my wife Marilyn and my brother Gordon for their patience and encouragement during this work.

\tableofcontents
\listoftables

\end{document}